\documentclass[12pt]{article}%

\usepackage{heck}
\usepackage{sectsty}
\sectionfont{\large\bfseries}
\subsectionfont{\normalsize\sc}
\subsubsectionfont{\normalsize\it}
\usepackage{epsfig}
\usepackage{graphicx}
\usepackage{makeidx}
\usepackage{multicol}
\usepackage{amsfonts}
\usepackage{amssymb}
\usepackage{amsmath}
\providecommand{\U}[1]{\protect\rule{.1in}{.1in}}
\begin{document}

\date{November, 2007}

\preprint{arXiv:0711.0387 \\ PUPT-2246 \\ AEI-2007-149}

\institution{HarvardU}{\centerline{${}^{1}$Jefferson Physical Laboratory, Harvard University, Cambridge,
MA 02138, USA}}%

%TCIMACRO{\TeXButton{Institution}{\institution{PRINCETON}{\centerline{${}%
%^{2}$Department of Physics, Princeton University, Princeton, NJ 08544, USA}}%
%}}%
%BeginExpansion
\institution{PRINCETON}{\centerline{${}^{2}%
$Department of Physics, Princeton University, Princeton, NJ 08544, USA}}%
%EndExpansion
%

%TCIMACRO{\TeXButton{INSTITUTION}{\institution{PLANCK}{\centerline{${}%
%^{3}$Max Planck Institute (Albert Einstein Institute),} \cr\centerline
%{Am M\"uhlenberg 1, D-14476 Potsdam-Golm, Germany}}}}%
%BeginExpansion
\institution{PLANCK}{\centerline{${}^{3}%
$Max Planck Institute (Albert Einstein Institute),} \cr\centerline
{Am M\"uhlenberg 1, D-14476 Potsdam-Golm, Germany}}%
%EndExpansion
%

%TCIMACRO{\TeXButton{Title}{\title{A Cascading Quiver \\ and the MSSM}}}%
%BeginExpansion
\title{Cascading to the MSSM}%
%EndExpansion
%

%TCIMACRO{\TeXButton{Authors}{\authors{Jonathan J. Heckman\worksat{\HarvardU
%}\footnote{e-mail: {\tt jheckman@fas.harvard.edu}},
%Cumrun Vafa\worksat{\HarvardU}\footnote{e-mail: {\tt vafa@physics.harvard.edu}%
%},
%Herman L. Verlinde\worksat{\PRINCETON}\footnote{e-mail: {\tt
%verlinde@princeton.edu}} \\
%and Martijn Wijnholt\worksat{\PLANCK}\footnote{e-mail: {\tt
%wijnholt@aei.mpg.de}}}}}%
%BeginExpansion
\authors{Jonathan J. Heckman\worksat{\HarvardU}\footnote{e-mail: {\tt
jheckman@fas.harvard.edu}},
Cumrun Vafa\worksat{\HarvardU}\footnote{e-mail: {\tt vafa@physics.harvard.edu}%
},
Herman L. Verlinde\worksat{\PRINCETON}\footnote{e-mail: {\tt
verlinde@princeton.edu}} \\[2mm]
and Martijn Wijnholt\worksat{\PLANCK}\footnote{e-mail: {\tt
wijnholt@aei.mpg.de}}}%
%EndExpansion

\abstract{The MSSM can arise as an orientifold of a pyramid-like quiver in the context of intersecting D-branes.
Here we consider quiver realizations of the MSSM which can emerge at the bottom of a duality cascade.
We classify all possible minimal ways this can be done by allowing only one extra node.
It turns out that this requires extending the geometry of the pyramid to an octahedron.
The MSSM at the bottom of the cascade arises in one of two possible ways,
with the extra node disappearing either via Higgsing or confinement.  Remarkably, the quiver of the Higgsing scenario turns out to be nothing but the quiver version of the left-right symmetric extension of the MSSM.
In the minimal confining scenario the duality cascade can proceed if and only if
there is exactly one up/down Higgs pair.  Moreover, the symmetries of the octahedron naturally admit
an automorphism of the quiver which solves a version of the $\mu$ problem precisely when there are an odd number of
generations.}%

%TCIMACRO{\TeXButton{Maketitle}{\maketitle}}%
%BeginExpansion
\maketitle
%EndExpansion
%\renewcommand{\Large}{\large}
%\renewcommand{\large}{\sc}

%%%%%%%%%
%
% Section definitions
%
%%%%%%%%%

%\newcommand{\newsection}[1]{
%\addtocounter{section}{1} %\setcounter{equation}{0}
%\setcounter{subsection}{0} \addcontentsline{toc}{section}{\protect
%\numberline{\arabic{section}}{{\rm #1}}} \vglue .0cm \pagebreak[3]
%\noindent{\large \bf  \thesection. #1}\nopagebreak[4]\par\vskip .3cm}
%%
%\newcommand{\newsubsubsection}[1]{
%\addtocounter{subsubsection}{1}
%\addcontentsline{toc}{subsubsection}{\protect
%\numberline{\arabic{section}.\arabic{subsection}.\arabic{subsubsection}}{ #1}} \vglue .4cm
%\pagebreak[3] \noindent{\it \thesubsubsection.
%#1}\nopagebreak[4]\par\vskip .3cm}
%
%\newcommand{\newsubsection}[1]{
%\setcounter{subsubsection}{0}
%\addtocounter{subsection}{1}
%\addcontentsline{toc}{subsection}{\protect
%\numberline{\arabic{section}.\arabic{subsection}}{ #1}} \vglue .4cm
%\pagebreak[3] \noindent{\sc \thesubsection.
%#1}\nopagebreak[4]\par\vskip .3cm}

%%%%%%%%%%%
%
% Section labelling
%
%%%%%%%%%%%

%\makeatletter
%\newcommand{\seclabel}[1]{%
%  \@bsphack
%  \protected@write\@auxout{}%
%     {\string\newlabel{#1}{{\thesection}{\thepage}}}
%  \@esphack
%  }
%\newcommand{\subseclabel}[1]{%
%  \@bsphack
%  \protected@write\@auxout{}%
%     {\string\newlabel{#1}{{\thesubsection}{\thepage}}}
%  \@esphack
%  }
%\makeatother

%%%%%%%%%%%%%%%%%%%%%%%%%%%%%%%%%%%%%%%%%%%%%%%%%%%%%%%%%%%%%%%%%%%%%%%%%%%%%%%%%
%
% Main file
%
%%%%%%%%%%%%%%%%%%%%%%%%%%%%%%%%%%%%%%%%%%%%%%%%%%%%%%%%%%%%%%%%%%%%%%%%%%%%%%%%%%

\tableofcontents

${}$

\pagebreak

\section{Introduction}\label{Introduction}
\renewcommand{\footnotesize}{\small}
\newcommand{\eol}{\nonumber \\[2mm]}
\newcommand{\eoll}{\nonumber \\[0mm]}

At present, no experiment has verified any signature unique to string theory.
\ Given that the LHC will soon directly probe physics beyond the Standard
Model, it is natural to ask whether string theory can produce \textit{any}
concrete prediction.
On first inspection, even one specific low energy prediction would appear
unlikely because string theory is a theory of quantum gravity and as such its
most novel predictions will involve energy scales close to the string or
the fundamental Planck scale. \ In order to avoid conflict with observation, this scale is
typically taken to be far above the TeV scale.

In the context of grand unified field theories, however, the advent of
Heterotic string theory \cite{GrossHarveyHeterotic} and the 4d gauge theory
structure resulting from its Calabi-Yau compactifications
\cite{CandelasVacuumConfigs} suggested rather limited ways that the matter
content of the Standard Model could emerge from a consistent theory of quantum
gravity. \ This raised theoretical hopes that perhaps string theory would
predict a relatively robust structure for gauge theory and matter structure
from simple topological and representation theoretic criteria.

But this hope has been dashed with the advent of the duality era in string theory
which demonstrates rather convincingly the existence of a large class of
non-perturbative string compactifications with a diverse range of gauge groups
and matter content. Thus an exponentially large number of solutions can be
constructed --the string landscape-- which look more or less like the Standard
Model (for a review see \cite{KachruDouglasFluxCompact}). Of course this is
not to say that any consistent quantum field theory will embed in a consistent
fashion inside a quantum theory of gravity. \ It is therefore important to
delineate the boundary between field theories which possess such an embedding
and those which simply fall in a more general swampland of effective field
theories (see e.g. \cite{VafaSwamp}). These constraints typically arise because
not every local geometry leading to a 4d QFT can appear as part
of a compact internal geometry. \ Nevertheless, such constraints appear to be
difficult to narrow down with our present understanding of string theory. \ It
is thus natural to ask whether \textit{local} aspects of string theory impose
any further constraints on observed QFT's and further, whether symmetries
natural from the perspective of string theory possess a low energy remnant in
the effective field theory. \ More broadly, we ask: Are there any field
theories which are \textit{more distinguished} among others, from a stringy viewpoint?

There seems to be one regard in which considerations following from local
geometry produce a robust prediction. \ Indeed, it does not appear possible to
engineer arbitrary matter content from a configuration of D-branes.
\ For example, for $U(N)$ gauge groups, no representation of rank higher than
two appears for generic $N$. \ Another example is in the context of $U(N)\times
U(k)$ type groups: \ In oriented string theories the matter fields which are
charged under both groups \textit{always} appear to transform as
bifundamentals $(N,\overline{k})$ or $(\overline{N},k)$. \ This is the case,
for example, in the context of intersecting D-branes and the matter localized
at their intersection \cite{BerkoozDouglasLeigh}.

Quite remarkably, the matter content of the Standard Model
compactly fits in such rank two representations. \ It is therefore
natural to expect a local string realization of this type of theory.
\ Representative papers on realizing the Standard Model from a D-brane probe
of a Calabi-Yau singularity may be found in
\cite{FirstBottomUp,BeresnteinJejjalaLeigh,WijnholtVerlindeDPeight}. \ There
is also a large body of work on intersecting D-brane models. \ For recent reviews and a more
complete list of references in the context of intersecting brane
configurations, see \cite{CveticShiuReview,BlumenhagenReview,MarchesanoReview}%
. \ Modulo issues pertaining to tadpole cancellation from orientifold planes,
any such D-brane realization should also possess a smooth large $N$ limit.

In some cases, this large $N$ limit connects directly to a low energy gauge
theory with much smaller rank. \ Indeed, this connection is known in a number
of cases in string theory and involves the concept of a duality cascade. The
aim of this paper is to investigate this question in the context of minimal
realizations of the Standard Model in terms of quiver theories.  We note, however, that
this approach has one disadvantage because it presupposes that the unification
of the gauge coupling constants in the supersymmetric context is an accident.  As we shall explain later,
GUTs do not appear to naturally arise from D-brane constructions.\footnote{In more general string theory constructions, E-type gauge groups can appear in Heterotic and F-theory compactifications.}

We now explain why the concept of a duality cascade is particularly appealing in
the context of D-brane realizations of the Standard Model. \ An ubiquitous
theme in string (motivated) phenomenology is the translation of field
theoretic data into geometry. \ Prominent examples are the engineering of
Standard Model-like gauge theories via singular geometries and D-branes, and
the dual representation of the gauge hierarchy in terms of warped extra
dimensions. % \cite{WijnholtVerlindeDPeight}.
The holographic interplay between gauge and gravitational degrees of freedom
underscores the crucial r\^{o}le D-branes play in establishing a string-theoretic
link between gauge theory and gravity. \ Indeed, a large number of D-branes
will melt into geometry. This process can in fact be done continuously
\cite{KlebanovStrassler}. \ Starting from a configuration with a large number
of D-branes which is captured by the geometry, at distances closer to the tip
of the cone, the dual description in terms of a stack of branes will cause the initially
large number of branes to sequentially decrease until only a finite number of
branes are left at the `bottom' of the geometry. In the dual
gauge theory this corresponds to a duality cascade whereby a series of
Seiberg dualities sequentially decreases the ranks of the gauge group as the
RG flow proceeds from the UV to IR so that deep in the IR the resulting gauge
groups have small finite rank.

Indeed, this is very natural for phenomenology: In the IR, which is the scale at which the Standard Model has been directly probed, we observe $SU(3)\times SU(2)\times U(1)$ gauge symmetry. Could it be
that at higher energy scales, the ranks of the gauge groups increase,
leading ultimately near the Planck scale simply to gravity?

We will see that this scenario can be realized. In fact, quite surprisingly we
find that there is very little choice in the minimal realization of a
cascading structure leading to the Standard Model at the bottom of the
cascade. \ The minimal quiver realization of the
Standard Model involves an orientifold of a \textquotedblleft
covering\textquotedblright\ quiver theory which has the shape of a pyramid
with the lift to the cover of the weak group $SU(2)_{L}$ at the apex of the pyramid.
\ Just as in the Klebanov Strassler cascade, we expect additional nodes to disappear near the end of the cascade.
\ Indeed, it turns out that in order to get a cascading structure, one must minimally add one extra node
which enhances the already symmetric pyramid to an octahedron. \ In this case,
orientifolding the covering theory by a 180 degree rotation along its symmetry
axis leads to an essentially unique cascading model for the Standard Model.
\ Depending on how the cascade terminates, there are two further possible refinements.

The most direct analogue of the Klebanov Strassler cascade corresponds to the
case where the rank of the gauge group factor on the extra node depletes to
zero at the bottom of the cascade. \ In this case it turns out quite
surprisingly that \textit{we need exactly one up/down Higgs pair} for the
cascade to proceed! \ Rather than requiring that the extra node confine at the
bottom of the cascade, it is also possible to Higgs the quiver theory before
the ranks of any node completely depletes so that two of the quiver nodes
collapse to a single quiver site.  Prior to Higgsing, the corresponding quiver theory is in fact nothing but the quiver of the left-right symmetric extension of the MSSM.

We note that the original idea that the Standard Model may lie at the bottom
of a cascade is not new and has been advocated, for example, in
\cite{KlebanovStrassler,StrasslerNagoya}. \ Although the explicit models we
shall consider do not have a conformal limit, more generally, a cascade will
proceed by perturbing the ranks of a given conformal field theory. \ Indeed,
the idea of approaching a conformal limit in the ultraviolet in the context of
quiver realizations of the Standard Model has been studied as a potential
solution to the hierarchy problem in \cite{FramptonVafa}. Furthermore, string
theory constructions with semi-realistic Standard Model vacua at the bottom of
a cascade with a holographic dual have been realized in
\cite{UrangaSMthroat}.

The rest of this paper is organized as follows. \ In section
\ref{DbranesquiversMSSM} we review the relation between D-branes and
the associated $U$-type quivers, as well as their orientifolds and
the resulting $U/USp/SO$-type quivers. In this same section we
summarize how the Standard Model embeds in a minimal quiver
consistent with D-brane constructions in string theory. We call this
minimal quiver realization of the Standard Model the MQSM.
 \ We next review in section
\ref{DualizingCascades} how to Seiberg dualize quiver gauge theories
and how sequences of Seiberg dualities give rise to cascading
gauge theories. %\ To this end, we  some examples of cascades which are well
%understood in string theory.
\ Beginning in section \ref{Realizingone} we commence our analysis of ways
in which the MQSM\ can sit at the bottom of a periodic duality
cascade.
We identify two minimal scenarios by which a %possible ways that any candidate minimal
cascading quiver theory with four quiver nodes can Higgs or confine
down to the three node MQSM. \ In section \ref{alrsymmcascade} we
present a left-right symmetric cascading gauge theory that
reaches
the MQSM via Higgsing. \ %Indeed, just prior to Higgsing the gauge group of the extra
%node is $USp(2)\simeq SU(2)_{R}$.
%\ Deferring further issues related to the bottom of the cascade to later in the paper,
In section \ref{multigen} we present a partial classification
of cascading four node quivers that terminate at the MQSM when the
extra node confines. \ Some technical parts of this section
are deferred to Appendices \ref{CovercascadeA} and \ref{DUALPOT}. \ The
combinatorics of possible cascades in this latter approach is more
intricate and places strong constraints on candidate cascades. \ In
fact, we find that the requirement of a repeating cascade structure
\textit{requires} the presence of exactly one light up/down Higgs
pair! \ Quite unexpectedly, we find that at intermediate stages of the cascade,
 the structure of the quiver theories in
the Higgsing and confining scenarios is nearly identical. \ Indeed,
in section \ref{Unification}
 we explain how the two cascades can be mapped to each other by adding or integrating out
vector-like pairs. \ This is reassuring in that it suggests a robust
extension of the MQSM which admits a periodic cascade structure. \
With the intermediate stages of the cascade analyzed, we next
describe in greater detail the behavior of the two cascading
scenarios near the bottom of the cascade in section  \ref{Bottom}. \
Proceeding from the IR\ to the UV, we find that whereas the scales
of dualization for the Higgsing scenario quite flexibly accommodate
a range of possibilities, in the minimal version of the confining
scenario much of the cascade would be forced to dualize above the
Planck scale. \ We present a minimal modification designed to
accelerate the running of couplings and comment on ways in which
supersymmetry breaking can arise in the low energy theory. \
Combining the above analysis, in section \ref{RUNNING} we discuss in
more detail the structure of the running of the coupling constants
and the fact that in either approach we hit a duality wall at finite
scale. \ In section \ref{DBRANECONFIGS} we discuss the possibility
of realizing the above cascade using a D-brane construction. \
Section \ref{DISTINGUISH} discusses in what sense the cascade distinguishes the matter content and gauge groups of the
MSSM from other possible cascade endpoints.  Finally, in section \ref{DISCUSSION} we conclude and discuss
directions for future research.

\section{D-branes, Quivers and the MQSM}\label{DbranesquiversMSSM}

In order to make the discussion to follow more self-contained,
 and hopefully of interest to a broader range of readers,
in this section we review the types of
gauge theories which can in principle arise from supersymmetric D-brane
constructions.  \ For more detailed reviews we refer to
\cite{CveticShiuReview,BlumenhagenReview,MarchesanoReview}.
Although the restrictions on
the matter content and gauge group types we discuss will continue to hold in
the context of non-supersymmetric gauge theories, in order to apply the above
considerations to cascading gauge theories we shall always restrict our
analysis to theories which preserve at least $\mathcal{N}=1$ supersymmetry.

\subsection{Oriented Theories}\label{oriented}

We first recall how gauge symmetry arises in the context of D-brane
constructions. \ In a sigma-model description, a D-brane is a defect
in spacetime where an open string can end. \ In the presence of $N$
D-branes, each endpoint of an oriented open string is labelled by an
additional Chan-Paton index $i=1,...,N$ indicating which brane a
string can end on. \ When all of the branes form a single stack, the
resulting open string modes transform in the adjoint representation
of a $U(N)$ gauge group. \ Quantizing such an open string yields a
vector multiplet and possibly additional adjoint chiral fields at
the massless level.  \ In all cases, we consider D-branes which fill our four
dimensional spacetime and wrap some internal cycle of a compactified six
dimensional manifold in the extra dimensions. \ Assuming that the internal cycle has volume
$V$, the gauge coupling of the four dimensional effective theory is:%
\begin{equation}
\frac{1}{g_{YM}^{2}}\sim\frac{V}{g_{s}}\text{.}%
\end{equation}

Chiral matter in general arises from topologically stable intersections between two stacks of
branes in the internal dimensions. Both stacks of branes fill space-time,
and intersect at a finite number of points in the internal six-manifold.  As an example,
in type IIA string theory, consider a stack of $N_{1}$ D6-branes and another stack
of $N_{2}$ D6-branes which both fill the spacetime $%
%TCIMACRO{\U{211d} }%
%BeginExpansion
\mathbb{R}
%EndExpansion
^{3,1}$ but intersect at some number of points in the internal
directions of the theory. \ Localized at each transversal intersection
there exists a chiral superfield in the 4d effective theory which is
charged under the bifundamental representation
$(N_{1},\overline{N_{2}})$ of $U(N_{1})\times U(N_{2})$ when the
intersection pairing is positive, and the conjugate when it is
negative. In IIB string theory, chiral matter similarly arises from intersections
between D5- and D7-branes.

More generally, it is possible that D-branes form bound states with lower dimensional
branes. For example, a D7-brane in IIB string theory can carry induced
$D5$- and $D3$-brane charges due to curvature and topologically stable
magnetic fluxes on its worldvolume. If we consider two stacks with $N_1$
and $N_2$ bound state branes, then for each intersection between the D7 and D5-brane
components, an appropriate index theorem computes the number of chiral
superfields charged under $(N_{1},\overline{N_{2}})$ of
$U(N_{1})\times U(N_{2})$ or its conjugate depending on the sign of
the intersection number. \

The matter content of gauge theories on intersecting brane configurations
can thus be summarized in terms of a quiver diagram, with the following rules.
A gauge group is represented by a node of the quiver. A $U(N)$ node
corresponds to a stack of $N$ D-branes. A chiral field transforming
under the bifundamental representation $(N_{1},\overline{N_{2}})$ corresponds to
an arrow which points away from the $U(N_{1})$ node and into the $U(N_{2})$ node.
The number of lines connecting two nodes is determined by the intersection number
between the corresponding branes.

A well-defined gauge theory cannot contain any anomalies. This imposes a number of
constraints on the quiver diagram which reflect specific
consistency conditions on the D-brane configuration inside
a given string background. The cancellation
of non-abelian gauge anomalies requires the number of fundamentals
and anti-fundamentals for every node to be equal. Geometrically, this
corresponds to the fact that the branes are sources for RR-flux and
by Gauss' law the total flux through any compact cycle must vanish so that all tadpoles cancel
\cite{CachazoVafaGeomUnif}. \ In addition, there are
mixed anomalies given by a triangle diagram with one
$U(1)$ factor and two non-abelian factors. In fact these do not need
to vanish, because in any string theory setting there is another
contribution to the anomaly due to the coupling of open strings to
certain closed string axions. This is the generalized Green-Schwarz
mechanism.

\newcommand{\Tr}{{\rm Tr}}

Let us summarize the mechanism for the case of intersecting stacks of D6-branes.
The Chern-Simons (CS) terms for the D6-brane worldvolume action contains couplings
of the form
\begin{equation}
\label{cs} \Tr(F)\wedge C_{5} + \Tr(F\wedge F) \wedge C_{3}
\end{equation}
%
%in the 6+1-dimensional worldvolume action for the brane,
where $F$
denotes the field strength and $C_p$ the p-form RR\ potential. \ In
the  4-d theory, $C_{5}$ reduces to a 2-form
potential $B_2$ with an equation of motion of the form
$d {}^* dB_2 + \Tr(F) = 0$, while $C_{3}$ reduces to an axion
field $a$.
The 10d self-duality relation $\ast_{10}\, d C_5 = d C_3$ relates the
2-form $B_2$ to the axion $a$ via (taking into account the
CS coupling): $da - \Tr(A) = \ast_{4}dB_{2}$.
The axion $a$ thus transforms under the anomalous $U(1)$ gauge symmetry.
The CS coupling of equation (\ref{cs}), combined with the kinetic term for the RR-forms
then leads to a 4d term of the form:%
\begin{equation}
\mu^2 (\Tr(A) - da)^2
%\wedge \ast_4 (Tr(A)-da)
+ a\; \Tr (F\wedge F)  \text{.}%
\end{equation}
%Because this term is not gauge invariant, the corresponding local counterterm
%will cancel the mixed anomaly. \ Together with the kinetic term for $a$, this
% for the gauge field via
%the St\"{u}ckelberg mechanism.
The
second term cancels the mixed anomaly.
The first term represents a St\"uckelberg mass $\mu$ (of order string scale) for the anomalous
$U(1)$ vector boson.
\ Further details of this St\"uckelberg mechanism may be found in
\cite{CveticShiuReview,BlumenhagenReview,BuicanVerlinde}.

In addition to the gauge interactions, the chiral superfields will
also interact via the superpotential of the effective theory. \ In
type IIA string theory, these terms are generated by worldsheet disk
instantons.
%\ See figure \ref{WSINST} for a depiction of how cubic
%terms are generated by worldsheet disc instanton amplitudes.
In type IIB string theory the situation is simpler because the superpotential
can be completely recovered from the classical geometry.
%
%TCIMACRO{\FRAME{ftbpFU}{4.2289in}{1.4295in}{0pt}{\Qcb{In an intersecting
%D-brane configuration, worldsheet disc instanton amplitudes generate
%contributions to the superpotential in the low energy effective theory.}%
%}{\Qlb{WSINST}}{worldsheetinst.eps}{\special{ language "Scientific Word";
%type "GRAPHIC";  maintain-aspect-ratio TRUE;  display "USEDEF";
%valid_file "F";  width 4.2289in;  height 1.4295in;  depth 0pt;
%original-width 6.6245in;  original-height 2.2208in;  cropleft "0";
%filename 'WorldsheetInst.eps';file-properties "XNPEU";}} }%
%BeginExpansion
%\begin{figure} [ptb] \begin{center} \includegraphics[ height=1.4295in, width=4.2289in ]% {WorldsheetInst.eps}% \caption{In an intersecting D-brane configuration, worldsheet disc instanton amplitudes generate contributions to the superpotential in the low energy effective theory.}% \label{WSINST}% \end{center} \end{figure}
%EndExpansion

Before closing this section, we note that anomalous $U(1)$'s leave
behind global symmetries in the low energy theory. \ Even so,
instanton effects in both gauge and string theories will typically
violate these symmetries. \ Although it appears that the specific
contributions from instantons are sensitive to the geometry of the
compactification manifold, we will assume that these effects are
sufficiently small so that an analysis in terms of perturbatively
generated effects will remain reliable.

\subsection{Unoriented Theories}\label{unoriented}

Up to now, all of the open strings which we have discussed have a well-defined
orientation. \ More generally, we can also consider theories where the
orientation of the string worldsheet is not preserved. \ From the perspective
of the worldsheet, such theories correspond to modding out by the $%
%TCIMACRO{\U{2124} }%
%BeginExpansion
\mathbb{Z}
%EndExpansion
_{2}$ group action which acts on all states by $O=\sigma\Omega(-1)^{F_{L}}$,
where $\Omega$ denotes worldsheet parity reversal, $\sigma$ denotes a $%
%TCIMACRO{\U{2124} }%
%BeginExpansion
\mathbb{Z}
%EndExpansion
_{2}$ action in the target spacetime, and $F_{L}$ denotes the spacetime
fermion number of left-moving fields on the worldsheet. \ From the perspective
of the target spacetime, the resulting theory will contain non-dynamical
spacetime defects called orientifold planes which can reverse the orientation
of a string passing through such a plane. \ Because this spacetime defect
corresponds to the fixed point locus of a $%
%TCIMACRO{\U{2124} }%
%BeginExpansion
\mathbb{Z}
%EndExpansion
_{2}$ action, each stack of D-branes also has an image under the $%
%TCIMACRO{\U{2124} }%
%BeginExpansion
\mathbb{Z}
%EndExpansion
_{2}$ action. \ We shall refer to the theory obtained prior to the $%
%TCIMACRO{\U{2124} }%
%BeginExpansion
\mathbb{Z}
%EndExpansion
_{2}$ identification as the \textit{covering theory} of the orientifold
theory. \ We shall label the nodes of a covering theory by capital letters
and those of the orientifold theory by lower case letters. \ In general, for each orientifold plane, there are two possible
associated projections of the oriented theory Hilbert space onto $%
%TCIMACRO{\U{2124} }%
%BeginExpansion
\mathbb{Z}
%EndExpansion
_{2}$ invariant subspaces. \ We now explain how the matter content of the
covering theory determines that of the orientifold theory. \ General rules
for extracting the orientifold of a quiver gauge theory have been given in
\cite{WijnholtGeometry} and for gauge theories derived from brane probes of
toric singularities in \cite{UrangaDimers}.

First consider a stack of branes with a distinct image under the orientifold
action. \ In the covering theory this corresponds to a pair of distinct $U(N)$
gauge group factors. \ In this case, the resulting gauge bosons will also
transform in the adjoint representation of $U(N)$. \ Next consider a stack of
branes which are fixed by the orientifold action. \ The two possible
projections on the Hilbert space correspond to two possible restrictions on
the gauge bosons of the fixed branes:%
\begin{equation}
A=-\gamma A^{T}\gamma^{-1}%
\end{equation}
where $\gamma$ denotes an $N\times N$ matrix which is either symmetric or
anti-symmetric depending on the choice of orientifold projection. \ In the
former case, the resulting gauge group is of $SO$ type, whereas in the latter
case the resulting gauge group is of $USp$ type. \ In what follows we shall
use the convention that $SU(2)\simeq USp(2)$.

Due to the fact that a brane as well as its image must now participate in the
corresponding covering theory, the intersection pairing of a brane with a
possible image will now give rise to additional possible matter
representations. \ Assuming that the open string in question does not attach a
brane to its own image under the orientifold action, in addition to matter
charged in the $(N_{1},\overline{N_{2}})$ (or conjugate) of $U(N_{1})\times
U(N_{2})$, the reversal of string orientation associated with strings
attaching to an image brane also allows bifundamental fields charged in the
$(N_{1},N_{2})$ (or conjugate) representation. \ There is one final
possibility corresponding to bifundamentals which connect a brane to its image
brane in the covering theory. \ Depending on the orientifold projection, the
resulting matter will belong to either the two index symmetric ($S$) or
anti-symmetric ($A$) representation (or conjugates) of the resulting gauge
group. \ In the resulting orientifold quiver theory, we may therefore also
allow arrows which point inwards/outwards to each quiver node.\footnote{Further note
that whereas the non-abelian anomalies must still cancel for each $U(N)$
theory, the presence of two index matter now implies a slightly different
cancellation condition. \ Indeed, recall that the anomaly coefficient of a
fundamental is $+1$ whereas that of the $S$ and $A$ representations of $U(N)$
are respectively $N+4$ and $N-4$ with signs reversed for all anomaly
coefficients in the conjugate representations. \ Because nearly all
representations of $SO$ and $USp$ gauge group factors are real or pseudo-real,
the resulting gauge groups cannot contain any non-abelian anomalies in a
consistent string theory construction. \ On the other hand, there is a well
known restriction on the matter content of $USp$ gauge theories which requires
an even number of chiral fermions charged under the fundamental
\cite{WittenAnomaly}. \ In fact, while Gauss' law enforces the constraint that
all non-abelian anomalies must vanish, the requirement that this same
cancellation take place in K-theory also enforces the perhaps less obvious
constraint that the number of arrows (counted with appropriate multiplicity)
is always even. \ Further discussion of the relation between K-theory charges
and global anomalies may be found in \cite{UrangaKtheory}.}

As in the oriented type IIA theory, worldsheet instantons will generate terms in the
superpotential. \ Due to the lack of orientation of string diagrams,
worldsheet diagrams with the topology of $%
%TCIMACRO{\U{211d} }%
%BeginExpansion
\mathbb{R}
%EndExpansion
\mathbb{P}^{2}$ will also contribute to the resulting superpotential.
\ Nevertheless, the form of individual contributions to the superpotential is
qualitatively similar to the oriented open string theory construction
discussed in the previous section.

\subsection{The Minimal Quiver Standard Model}\label{MQSM}

In this section we review the minimal possible ways in which the MSSM could
embed inside a D-brane construction. By a minimal embedding we shall mean a
quiver gauge theory with a minimal amount of extra gauge group factors and matter.

We now argue that if we restrict to oriented quivers the minimal
quiver will have more than three nodes. \ In order to keep our
discussion as general as possible, we shall allow some of the nodes
to be weakly gauged or even flavor groups. \ Now suppose to the
contrary that an \textit{oriented} three node quiver can accommodate
all of the Standard Model fields. \ In order to account for the
non-abelian groups of the Standard Model, the oriented quiver theory
will at the very least have two gauge group factors $U(3)$ and
$U(2)$. \ Because the lepton doublets do not transform under the
$U(3)$ factor, we conclude that the corresponding arrow must attach
to a distinct quiver node with gauge group $U(N)$ for some $N$. \
This in itself is not a problem because we may view the non-abelian
factor of $U(N)$ as a flavor group symmetry. \ Next consider the
right-handed leptons of the Standard Model. \ These fields transform
as singlets under the two non-abelian factors, but have non-trivial
hypercharge. \ For a three node quiver this would imply that the
right-handed leptons can only attach to $U(N)$ and must therefore
transform in the adjoint representation. \ This implies a
contradiction because the adjoint representation of $U(1)$ is
trivial so that the right-handed leptons would necessarily have zero
hypercharge.

\begin{figure}
[ptb]
\begin{center}
\includegraphics[
height=2.386in,
width=4.6449in
]%
{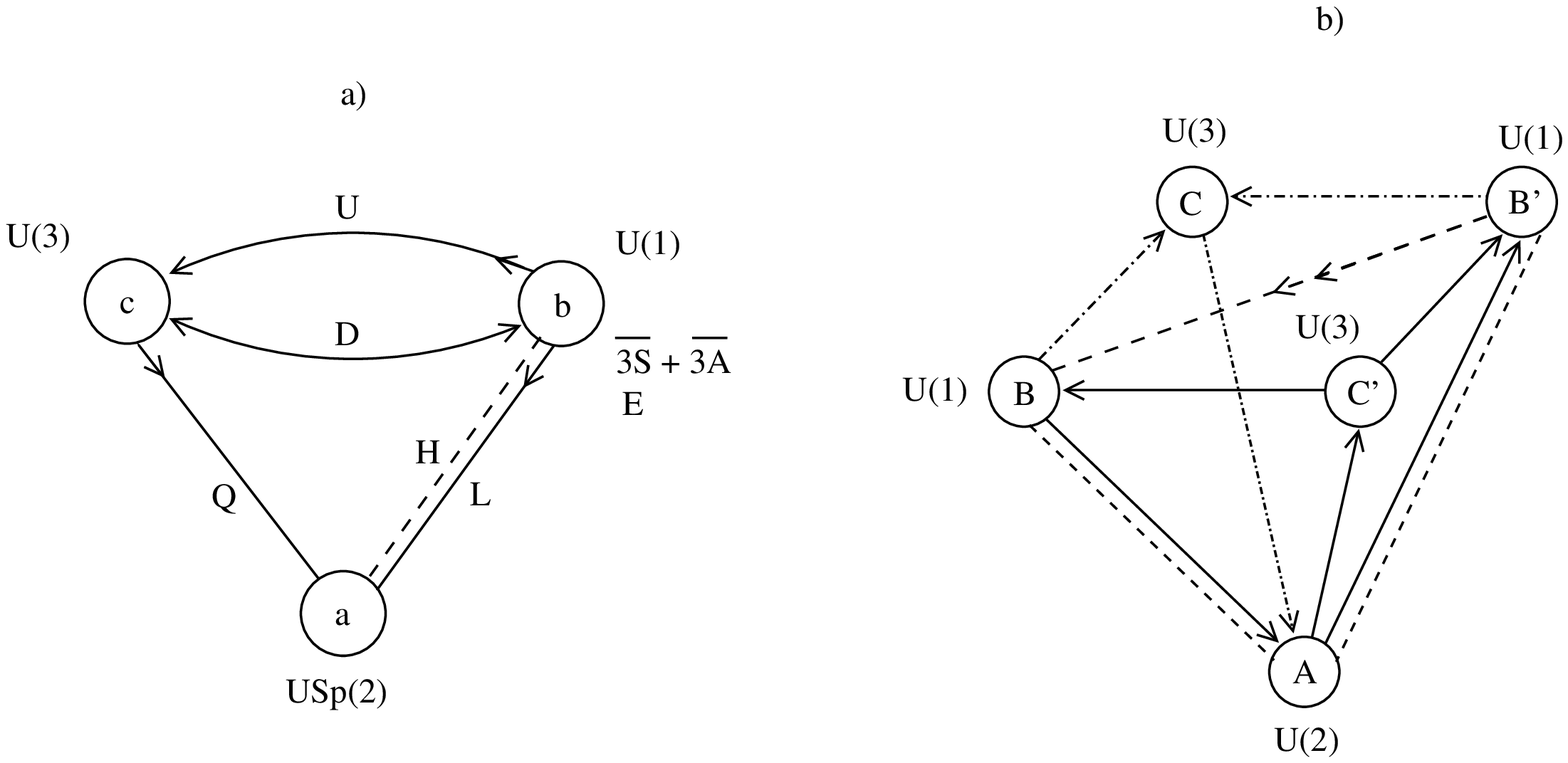}%
\caption{Depiction of the MQSM (a) and the corresponding covering theory (b).
\ In figure a), each directed line denotes three generations of left-handed
chiral fermions. \ The lines with two arrows attached determine the
representation of the fermion under the $U(3)$ and $U(1)$ factors and do not
indicate the multiplicity of the corresponding line. \ The dashed line
indicates the Higgs doublet of the Standard Model.  In the supersymmetric version of this quiver theory, each oriented line denotes a chiral superfield and each dashed line now denotes a vector-like pair of fields.  The field content of the MSSM is also labelled.}%
\label{mqsmandcover}%
\end{center}
\vspace{-2mm}
\end{figure}

A more minimal embedding of the Standard Model can be achieved in
the case of unoriented quiver theories. \ As in the oriented case,
the requirement that the lepton doublets remain charged under the
$SU(2)_{L}$ factor but transform as singlets under $SU(3)_{QCD}$
implies that any minimal embedding will possess at least three
quiver nodes. \ As discussed in
\cite{SchellekensDijkstra,BerensteinPinansky}, the minimal
quiver Standard Model (MQSM) with $\mathcal{N}=0$ supersymmetry
corresponds to the three node quiver shown in figure
\ref{mqsmandcover}a. \ A supersymmetric version of this quiver has
recently been constructed in string theory by partially Higgsing the
brane probe theory of a del Pezzo 5 Calabi-Yau singularity
\cite{WijnholtGeometry}. \ The gauge group of the MQSM is
$U(3)\times USp(2)\times U(1)$. \ The hypercharge of the Standard
Model is given (up to rescaling) by the unique linear combination of
$U(1)$ charges which does not suffer from mixed
anomalies\footnote{We normalize
$\mathcal{Q}_{U(1)_{c}}$ so that fundamentals of $U(3)$ have charge
$\pm1/3$.}:
\begin{equation}
\mathcal{Q}_{Y}=\frac{1}{2}\mathcal{Q}_{U(1)_c}-\frac{1}{2}\mathcal{Q}%
_{U(1)_b}\text{.}%
\end{equation}
The other linearly independent combination of $U(1)$ charges suffers
from mixed anomalies which are cancelled by the generalized
Green-Schwarz mechanism described previously. \ In addition to the
matter content corresponding to the Standard Model, we have also
included three additional fields which transform in the conjugate
two index anti-symmetric tensor representation $(\overline {A})$ of
$U(N)$ corresponding to the $U(1)$ factor. \ For $N=1$, the massless
modes of these fields are automatically trivial, however the stringy
tower of modes includes fields in the symmetric representation which are
non-trivial. Inclusion of these fields in a string theoretic
realization is crucial for tadpole cancellation even for $N=1$
(reflected in the gauge theory in the vanishing of the 1-loop FI
term \cite{LawrenceMcGreevy}). For $N>1$, the massless modes of these
anti-symmetric tensor fields are non-trivial and necessary for
$SU(N)$ anomaly cancellation.

Any candidate D-brane construction consistent with the above matter content
must also include orientifold planes of some type. \ This follows from the
presence of $USp$ gauge factors and matter charged under general two index
representations. \ The covering quiver of the MQSM\ is shown in figure
\ref{mqsmandcover}b.%
%TCIMACRO{\FRAME{ftbpFU}{4.6449in}{2.386in}{0pt}{\Qcb{Depiction of the MQSM (a)
%and the corresponding covering theory (b). \ In figure a), each directed line
%denotes three generations of left-handed chiral fermions. \ The lines with two
%arrows attached indicate the representation of the fermion under the $U(3)$
%and $U(1)$ factors and do not indicate the multiplicity of the corresponding
%line. \ The dashed line indicates the Higgs doublet of the standard model.}%
%}{\Qlb{mqsmandcover}}{mqsmandcover.eps}{\special{ language "Scientific Word";
%type "GRAPHIC";  maintain-aspect-ratio TRUE;  display "USEDEF";
%valid_file "F";  width 4.6449in;  height 2.386in;  depth 0pt;
%original-width 8.2806in;  original-height 4.2402in;  cropleft "0";
%filename 'MQSMandCOVER.eps';file-properties "XNPEU";}} }%
%BeginExpansion
%EndExpansion

It is important to note that D-brane realizations of quivers and
especially potential realizations of the cascade idea are in some
sense orthogonal to the idea of grand unification. However, quite
independently of cascades, one could ask if
D-branes can be used to realize GUTs. As it turns out, quiver
realizations of GUTs are more problematic. For example the chiral
matter content of the $SO(10)$ GUT transforms in the spinor
representation, which cannot arise from D-branes. \ Possible quiver
realizations of $SU(5)$ GUTs have been discussed for example in
\cite{Lykkenbraneguts}. \ Some general issues with D-brane
realizations of GUTs have been discussed in \cite{BerensteinGUTS}.
In the context of type IIB compactifications, more general constructions based on $F$-theory naturally avoid
 many of the above restrictions.  Further details on model building
 in $F$-theory constructions may be found in \cite{DonagiWijnholt,BHV}.

\section{Dualizing Quivers and Duality Cascades}\label{DualizingCascades}

In this section we review how Seiberg duality acts locally on the nodes of a
quiver gauge theory. \ In the context of renormalization group flows of quiver
gauge theories, the gauge couplings of an asymptotically free gauge group
factor will flow to strong coupling in the infrared (IR) of the theory.
\ Assuming that the other gauge group factors are sufficiently weakly coupled,
it is then appropriate to apply a Seiberg duality so that the resulting description of the theory
becomes weakly coupled. \ By performing a sequence of Seiberg dualities
consistent with RG flow, we arrive at a duality cascade. \ After presenting
the Klebanov Strassler duality cascade as well as how the cascade operates in
orientifolds of the theory, we next discuss general criteria which we shall
impose on any candidate duality cascade which flows to the MQSM\ in the deep IR.

\newcommand{\fracc}{\frac{\mbox{\small 1}}{\mbox{\small 2}}}

\subsection{Dualizing Quivers}\label{Dualizing}

Seiberg duality for oriented quiver gauge theories has been studied in
\cite{CachazoVafaGeomUnif,BerensteinDouglasSeiberg,WijnholtLargeVolume,HerzogSeiberg}%
. \ In this section we review the salient features for the analysis to follow
and extend these results to unoriented quiver gauge theories. \ In particular,
we shall explain how Seiberg duality in the oriented covering theory
determines the Seiberg dual in the orientifold theory. \ While this is a
straightforward extension of results from the oriented case, to the best of
our knowledge, this analysis has not appeared in the literature.

To begin our analysis, we first review how Seiberg duality acts on
supersymmetric $SU(N)$ QCD\ with $F$ flavors \cite{SeibergSeibergDual}. \ In
addition to the gauge symmetry, the model contains an $SU(F_{1})\times
SU(F_{2})$ flavor group symmetry. \ Note that in order to cancel non-abelian
cubic anomalies we must take $F=F_{1}=F_{2}$.\ \ Labeling the representation
content of the chiral fields by a triple, the field content of this theory
consists of a chiral superfield charged under the $(N,\overline{F},1)$ and
a chiral superfields charged under the $(\overline{N},1,F)$. \ Treating the
flavor groups as weakly coupled gauge theories such that the resulting
anomalies cancel by adding additional flavors uncharged under the $SU(N)$
factor, the associated quiver has three nodes with chiral superfields
$X_{i\overline{f_{1}}}$ and $Y_{f_{2}\overline{i}}$ where the subscripts on
the fields indicate the two gauge groups a bifundamental is charged under, and
the presence (resp. absence) of a bar indicates it transforms in the
anti-fundamental (resp. fundamental) of the corresponding quiver node. \ We
note that the distinction between fundamental and anti-fundamental is
superfluous when the gauge group is of $SO$ or $USp$ type. \ When the number
of flavors $0<F<N$, non-perturbative effects generate an ADS\ superpotential
which destabilizes the vacuum and lead to runaway behavior. \ When $3N>F>N+1$,
the\ Seiberg dual description of the same gauge theory is given by a gauge
theory with gauge group $SU(F-N)$, the same flavor group and $F$ dual quark
superfields charged under the $(\overline{F-N},F,1)$ and $F$ dual quarks
charged under the $(F-N,1,\overline{F})$. \ We note that this same description
also extends to the cases $F=N+1$ and $F=N$ in which case the gauge theory
confines. \ In order to enforce the constraint that the dimensions of the
(quantum) moduli spaces in the original description and dual descriptions
match, we must also introduce a meson field $M_{f_{2}\overline{f_{1}}}%
=\mu^{-1}Y_{f_{2}\overline{i}}X_{i\overline{f_{1}}}$ and superpotential:%
\begin{equation}
W_{mag}=\widetilde{Y}_{i\overline{f_{2}}}M_{f_{2}\overline{f_{1}}}%
\widetilde{X}_{f_{1}\overline{i}}%
\end{equation}
where $\mu$ denotes an energy scale associated with the meson field and
$\widetilde{X}$ and $\widetilde{Y}$ denote the flavors of the dual
theory. \ Note that the quiver theory now contains a closed triangle and that the
magnetic dual superpotential corresponds to the minimal closed path obtained
from such a triangle. \ From the perspective of the quiver gauge theory,
Seiberg duality corresponds to reversing the orientation of the arrows
attached to the dualized node. \ In addition, the dual theory contains a meson
field corresponding to the bifundamental $(\overline{F_{1}},F_{2})$ attached
between the two flavor group factors. \ In the context of unoriented theories,
a similar description would persist if the representation content of the
chiral superfields under the flavor groups had been changed so that the fields
transformed in the fundamental (resp. anti-fundamental) of both gauge groups.
\ Indeed, the only change in the corresponding theories is the transformation
properties of the dual meson field. \ In all cases the resulting closed
triangle term in the dual theory would still be present.

\begin{figure}
[ptb]
\begin{center}
\includegraphics[
height=3in,
width=4.1in
]%
{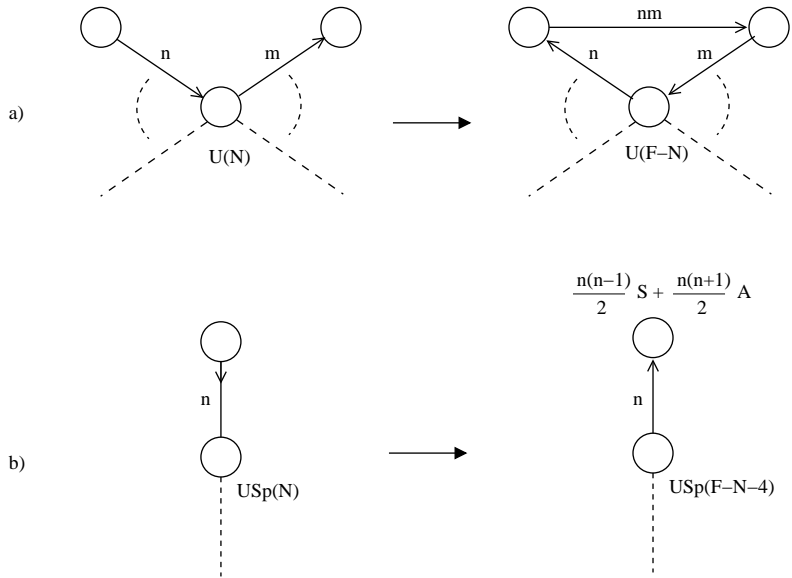}%
\caption{Depiction of Seiberg duality for oriented quiver theories with $SU$
type gauge group (a) and $USp$ type gauge groups which descend from an
orientifold action (b). \ In both cases, dualizing the electric theory (left)
reverses the orientation of all attached arrows and produces a number of dual
meson fields corresponding to fundamental fields in the dual description
(right). \ In the $USp$ case both $N$ and $F$ are even.}%
\label{seibergduals}%
\end{center}
\vspace{-2mm}
\end{figure}

The above results generalize to three node quivers with $n$ incoming arrows
and $m$ outgoing arrows. \ To determine the dual description, we view the two
flavor group symmetries as~$SU(nF_{1})\times SU(mF_{2})$ which are broken to
the diagonal $SU(F_{i})$ flavor symmetry. \ It now follows that in the Seiberg
dual description, the orientation of all arrows attached to the $SU(N)$ node
reverse direction and that $nm$ meson fields charged in the $(\overline{F_{1}%
},F_{2})$ now attach between the two flavor group factors. \ The resulting
superpotential terms can also be determined by a similar analysis of how the
flavor group breaks and so we shall omit the details. \ In the quiver theories
of interest we will always dualize a $U(N)=SU(N)\times U(1)/%
%TCIMACRO{\U{2124} }%
%BeginExpansion
\mathbb{Z}
%EndExpansion
_{N}$ factor, where a similar analysis applies.  See figure \ref{seibergduals} for a depiction of the local action
of Seiberg duality in both oriented and unoriented quivers.
%TCIMACRO{\FRAME{ftbpFU}{3.685in}{2.7086in}{0pt}{\Qcb{Depiction of Seiberg
%duality for oriented quiver theories with $SU$ type gauge group (a) and $USp$
%type gauge groups which descend from an orientifold action (b). \ In both
%cases, dualizing the electric theory (left) reverses the orientation of all
%attached arrows and produces a number of dual meson fields corresponding to
%fundamental fields in the dual description (right). \ In the $USp$ case both
%$N$ and $F$ are even.}}{\Qlb{seibergduals}}{seibergduals.eps}%
%width 3.685in;  height 2.7086in;  depth 0pt;  original-width 7.7089in;
%original-height 5.6585in;  cropleft "0";  croptop "1";  cropright "1";
%cropbottom "0";  filename 'SEIBERGDuals.eps';file-properties "XNPEU";}} }%
%BeginExpansion
%EndExpansion

Next consider $USp(2N)$ gauge theory with a single bifundamental\footnote{In
order for the global anomalies of the $USp$ theory to cancel, there must be an
even number of such quark fields.} transforming in the representation
$(2N,2F)$ of $USp(2N)\times SU(2F)$. \ The gauge group of the dual description
is $USp(2F-2N-4)$ and the flavor symmetry is again $SU(2F)$
\cite{SeibergSeibergDual,PouliotIntriligatorSeibergDual}. \ As a quiver gauge
theory, this corresponds to a two node quiver with a field $X_{if}$. \ Due to
the fact that the fundamental representation is pseudoreal, the dual $2F$
quarks $\widetilde{X}_{i\overline{f}}$ transform in the representation
$(\overline{2F-2N-4},\overline{2F})=(2F-2N-4,\overline{2F})$. \ In addition to
the dual quarks, the theory also contains a meson field:%
\begin{equation}
M_{[ff^{\prime}]}=\mu^{-1}\gamma^{ii^{\prime}}X_{if}X_{i^{\prime}%
f^{\prime}}%
\end{equation}
which transforms in the two index anti-symmetric representation $A_{2F}$ of
$SU(2F)$. \ In the above, $\gamma$ denotes an anti-symmetric matrix. \ In order to enforce the usual moduli space condition, the dual
theory also contains a superpotential term:%
\begin{equation}
W_{mag}=M_{[ff^{\prime}]}\gamma^{ii^{\prime}}\widetilde{X}_{i\overline{f}}\widetilde{X}_{i^{\prime
}\overline{f^{\prime}}}\text{.}%
\end{equation}
Hence, Seiberg duality again acts by reversing the orientation of all arrows
attached to the $USp$ factor and including an additional meson field in the
$SU(N)$ factor.

It is instructive to also treat the action of Seiberg duality in the covering
theory. \ In this case, we again obtain a three node quiver of the type
studied above in the context of $SU(N)$ gauge theories. \ Dualizing the
$SU(N)$ factor which descends to a $USp$ factor, we find that the dual meson
fields correspond to a bifundamental charged under the two $SU(2F)$ nodes of
the covering theory. \ Performing the requisite identification of the two
nodes in the orientifold, it follows from the general analysis of section
\ref{DbranesquiversMSSM} that the resulting meson field descends to a two
index anti-symmetric tensor representation of $SU(2F)$, as expected.

To conclude our presentation of dualization for $USp$ factors, we next treat
the case of $g$ fields transforming in the representation $(2N,2F)$ of
$USp(2N)\times SU(2F)$. \ In order to determine the meson field content in the
dual theory, we first note that we may view the $g$ fields as transforming in
the diagonal subgroup of $SU(2gF)$. \ Decomposing $A_{2gF}$ into irreducible
representations of $SU(2F)$ yields:%
\begin{align}
SU(2gF) &  \rightarrow SU(2F)%\label{antione}
\eoll
\textstyle
A_{2gF} &  \rightarrow gA_{2F}+\fracc{g\left(  g-1\right)  }\left(
2F\otimes2F\right) \label{antithree} \\
\textstyle
&  =\fracc{g\left(  g+1\right)  }A_{2F}+\fracc{g\left(  g-1\right)  }%
S_{2F}
\nonumber
\text{.}
\end{align}
The resulting superpotential can also be determined by analyzing the breaking
pattern of the flavor symmetry.

Much of the structure of the dualized $USp$ factor with $g$ arrows attached
can also be seen by dualizing the gauge group of the covering theory.
\ Indeed, it follows from the general analysis of $SU(N)$ factors discussed
above that in the resulting covering theory there will be precisely $g^{2}$
meson fields between the $SU(2F)$ node and its image. \ As expected,
the number of tensor matter fields in the orientifold theory is precisely
$g^{2}$.

We conclude this section by presenting a similar analysis for the Seiberg dual
of $SO$ type quiver nodes. \ Because the analysis is quite similar, we shall
omit all details unnecessary for the analysis of the rest of the paper.
\ Starting with $g$ chiral superfields in the $(N,F)$ of $SO(N)\times SU(F)$,
dualizing the $SO$ gauge group factor yields a gauge group $SO(F-N+4)\times
SU(F)$ with dual quarks given by reversing all arrows in the corresponding
quiver theory. \ For $g=1$, the dual meson field is given by the two index
symmetric representation $S_{F}$ of $SU(F)$. $\ $More generally, the
representation $S_{gF}$ of $SU(gF)$ decomposes as:%
\begin{align}
SU(gF) &  \rightarrow SU(F)%\label{symmone}
\eoll
\textstyle
S_{gF} &  \rightarrow gS_{F}+\fracc{g\left(  g-1\right)  }\left(  F\otimes
F\right)
\label{symmthree} \\
\textstyle
&  =\fracc{g\left(  g+1\right)  }S_{F}+\fracc{g\left(  g-1\right)  }%
A_{F}\text{.}  \nonumber
\end{align}

\subsubsection{Mass Terms and Vector-Like Pairs}

At the level of the connectivity of the quiver theory, dualizing a quiver node
reverses the orientation of arrows attached to a given node and also adds a
number of additional bifundamental meson fields to the dual quiver theory.
\ The presence of additional chiral matter will sometimes produce vector-like
pairs between different quiver nodes. \ Unless a symmetry explicitly forbids
the presence of a quadratic term in the dual superpotential, general arguments
from effective field theory imply that such a vector-like pair will develop a
mass and lift from the low energy spectrum. \ We now
explain how quadratic terms can arise in the dual magnetic theory.

Although strictly speaking the mass of the corresponding superfield is
controlled by the normalization of the K\"{a}hler potential, unless otherwise
stated we shall assume that the masses have developed
an appropriate value so that they may be integrated out of the low energy theory.
\ For simplicity we shall also assume that the
FI\ terms have been set to zero and that the K\"{a}hler metric is non-singular
near the origin of field space.

A vector-like pair of fields $X$ and $Y$ in a dual magnetic theory will
develop a mass when the superpotential contains a term of the form:%
\begin{equation}
W_{mag}\supset mX_{ac}Y_{\overline{c}b} \label{Wmag}%
\end{equation}
where the indices $a$ and $b$ label general flavor indices and $c$ indicates a
gauge group index. \ We note that when $a$ and $b$ denote gauge group factors,
they must contract to form a gauge invariant operator.

Because Seiberg duality only adds terms of cubic order to the magnetic dual
superpotential, it is enough to study the transformation under dualization of
composite operators in the superpotential of the electric theory. \ Assuming
that all massive vector-like pairs have already been integrated out in the
original electric theory, we may assume that at least one of the fundamental
fields $X$ or $Y$ of the dual magnetic theory corresponds to a composite
operator in the original electric theory variables. \ Letting $\alpha$ denote
the index of the fundamental representation for the gauge group to be
dualized, if $X$ is a meson field in the original electric theory, it must
take the form:%
\begin{equation}
X_{ac}\sim\frac{1}{\mu_{\alpha}}U_{a\alpha}V_{\overline{\alpha}c}\text{ or
}\frac{1}{\mu_{\alpha}}U_{a\overline{\alpha}}V_{\alpha c}%
\end{equation}
for some fields $U$ and $V$. \ Assuming $Y_{\overline{c}b}$ denotes a fundamental
field in both the electric and magnetic theories, the electric theory must
contain a term of the form:%
\begin{equation}
W_{elec}\supset\frac{m}{\mu_{\alpha}}U_{a\alpha}V_{\overline{\alpha}%
c}Y_{\overline{c}b}\text{ or }\frac{m}{\mu_{\alpha}}U_{a\overline{\alpha}%
}V_{\alpha c}Y_{\overline{c}b}%
\end{equation}
where we shall assume that the indices $a$ and $b$ contract in an appropriate
fashion. \ From the perspective of paths in the quiver theory, this requires
the presence of an oriented triangle which passes through the quiver nodes
$a,\alpha$ and $c$. \ A similar analysis applies with $X$ and $Y$ interchanged.

Next suppose that both $X$ and $Y$ correspond to meson fields in the electric
theory. \ A similar analysis now implies that the electric theory must contain
a term of the form:%
\begin{equation}
W_{elec}\supset\frac{m}{\mu_{\alpha}^{2}}U_{a\alpha}V_{\overline{\alpha}%
c}S_{\overline{c}\alpha}T_{\overline{\alpha}b}%
\end{equation}
where schematically, the mesons of the electric theory $UV$ and $ST$
respectively map to $X$ and $Y$ in the dual magnetic theory.

The above arguments help to explain why such quartic terms in the superpotential correspond to dangerous irrelevant operators.
Indeed, so long as a given operator contains sufficiently large anomalous dimensions, such terms can significantly alter the IR dynamics of the resulting theory.  As will be evident in later sections, this is especially important in the context of duality cascades where terms of quartic order can play a particularly prominent r\^{o}le.

Before concluding this section, we note that when the newly created meson
fields of the dual quiver theory only connects between gauge group factors of
type $SO$ and $USp$, it is always possible to add an appropriate quartic term
to the original electric theory so that the dual superpotential contains a
term quadratic in the dual meson field. \ Similar reasoning also applies when
the meson field transforms in the adjoint of a $U$ type factor.

\subsection{Duality Cascades}\label{Cascades}

Perhaps the most intriguing feature of Seiberg duality is that in the dual
description of the theory, the gauge group changes. \ In the context of
renormalization group flows, the change in rank implies that when an
asymptotically free gauge theory flows to strong coupling in the IR, if
the resulting theory does not flow to a conformal fixed point, the Seiberg
dual theory will correspond to a theory which instead flows to weak coupling
in the IR. \ While strictly speaking a gauge group factor should be dualized
prior to the gauge group reaching infinite coupling, much of the analysis we
discuss will not be sensitive to whether we dualize a gauge group when the
perturbative expansion parameter of the gauge theory becomes infinite or
merely an order one number. \ Matching the two theories at the scale of
dualization $\Lambda$, it follows that while $\Lambda$ corresponds to the
intrinsic scale of the asymptotically free theory, in the dual description
this same scale corresponds to the Landau pole of the dual gauge theory. \ The
general phenomenon whereby a quiver gauge theory undergoes a sequence of
Seiberg dualities as the theory flows from the UV to the deep IR is known as a
duality cascade.

A well known example of the above construction is given by the duality cascade
originally studied in \cite{KlebanovStrassler}. \ To illustrate the above
concepts in an explicit string theory construction, consider first the theory
given by $N$ D3-branes probing the resolved conifold:%
\begin{equation}
xy=uv
\end{equation}
where $x,y,u,v\in%
%TCIMACRO{\U{2102} }%
%BeginExpansion
\mathbb{C}
%EndExpansion
$ and the singularity at the origin has been replaced by a finite size
$S^{2}=\mathbb{P}^{1}$. \ The resulting quiver gauge theory is given by a two
node quiver with gauge group factors $SU(N_{1})$ and $SU(N_{2})$ with
$N_{1}=N_{2}=N$, and two bifundamentals $X^{1}$ and $X^{2}$ in the
representation $(N_{1},\overline{N_{2}})$ and two bifundamentals $Y^{1}$ and
$Y^{2}$ in the representation $(\overline{N_{1}},N_{2})$ \cite{KlebanovWitten}%
. \ This gauge theory has a strongly interacting conformal fixed point. \ At
large $N$, this conformal fixed point has a holographic dual corresponding to
a space of the form $AdS_{5}\times X_{5}$ where $X_{5}$ is a Sasaki-Einstein manifold.
%TCIMACRO{\FRAME{ftbpFU}{5.3843in}{0.4454in}{0pt}{\Qcb{Depiction of the
%Klebanov Witten quiver theory with fractional branes and the corresponding
%Klebanov Strassler duality cascade. \ At each stage of the cascade, either
%node $1$ or node $2$ Seiberg dualizes and subsequently flows to weak
%coupling.}}{\Qlb{klebstrass}}{klebanovstrassler.eps}%
%width 5.3843in;  height 0.4454in;  depth 0pt;  original-width 9.4316in;
%original-height 0.7541in;  cropleft "0";  croptop "1";  cropright "1";
%cropbottom "0";  filename 'KlebanovStrassler.eps';file-properties "XNPEU";}}
%}%
%BeginExpansion
\begin{figure}
[ptb]
\begin{center}
\includegraphics[
height=0.5in,
width=5.5in
]%
{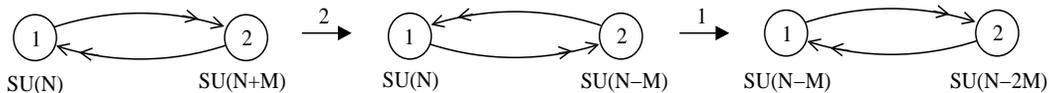}%
\caption{Depiction of the Klebanov Witten quiver theory with fractional branes
and the corresponding Klebanov Strassler duality cascade. \ At each stage of
the cascade, either node $1$ or node $2$ Seiberg dualizes and subsequently
flows to weak coupling.}%
\label{klebstrass}%
\end{center}
\vspace{-2mm}
\end{figure}
%EndExpansion
\

By introducing an imbalance in the numbers $N_{1}$ and $N_{2}$, so that
$N_{1}=N$ and $N_{2}=N+M$, it was shown in \cite{KlebanovStrassler} that as
the theory flows to the IR, the gauge group with larger rank flows to strong
coupling and the gauge group with smaller rank flows to weak coupling.
\ Geometrically, such an assignment of gauge group ranks corresponds to $N$ D3
branes and $M$ D5-branes wrapping the collapsing $S^{2}$. \ Indeed, in this
language each successive Seiberg duality corresponds to a flop of the geometry
which reduces the ranks of the gauge groups \cite{CachazoVafaGeomUnif}.
\ Although the connectivity of the quiver remains the same after dualization,
the ranks of the gauge group factors deplete according to the sequence:%
\begin{equation}
SU(N)\times SU(N+M)\rightarrow SU(N)\times SU(N-M)\rightarrow...
\end{equation}
See figure \ref{klebstrass} for a depiction of the quiver theory as it
undergoes a sequence of dualizations. \ Near the bottom of the cascade, one of
the gauge group factors confines.%

We note that it is also possible to study orientifolds of the above model
which produce $SO\times USp$ type gauge groups \cite{ImaiCascade} as well as
$USp\times USp$ type gauge groups with additional number of flavors added to
cancel all tadpoles locally \cite{SchnitzerCascade}. \ In all cases, the
resulting cascade proceeds in parallel fashion to the case of the covering
theory given by the $SU\times SU$ theory. \ For example, in the $SO\times USp$
cascade, there are two bifundamentals in the corresponding quiver theory. \ In
this case, the ranks deplete via the sequence:%
\begin{equation}
SO(2N+2M+2)\times USp(2N)\rightarrow SO(2N-2M+2)\times USp(2N)\rightarrow...
\end{equation}
As in the covering theory, while the number of bifundamentals remains constant
throughout the entire cascade, the ranks continue to change. \ Indeed, the
existence of a holographic dual description in both the conifold and its
orientifold \textit{guarantees} that the cascade of the covering quiver theory
descends correctly to the orientifold theory.

\begin{figure}
[t]
\begin{center}
\includegraphics[
height=2.9in,
width=3.4in
]%
{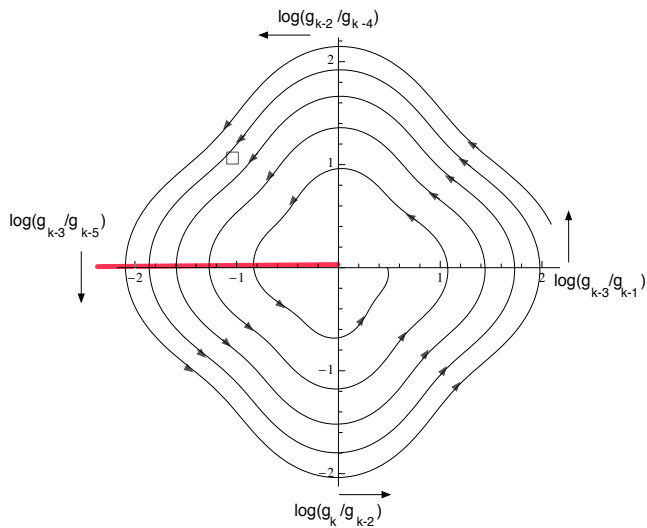}%
\caption{A sketch of the running of couplings during
an accelerated duality cascade. In each successive RG segment
one gauge groups flows to strong coupling while the other
flows to weak coupling. The turns between segments correspond to switching to a
Seiberg dual description. The thick red line along the left horizontal axis denotes a branch cut.
  Each time the RG flow passes this cut the flow switches to the sheet
parameterized by the couplings of the next RG cycle \cite{StrasslerReview}. The flow spirals
inward as the cascade accelerates towards the UV.  In this region, the theory becomes
trapped in a regime of strong coupling.
}\label{rgspiral}%
\end{center}
\vspace{-2mm}
\end{figure}

Extensions of the Klebanov Strassler cascade to more general quiver theories
with vector-like matter have been studied in
\cite{CachazoVafaGeomUnif,FiolWalls}. \ Cascades of gauge theories with chiral
matter have been studied in
\cite{HananyWalcher,HananyFrancoDualityTrees,HerzogCascade}. \ We note that as
opposed to the Klebanov Strassler cascade, many of these latter cases do not
exhibit the same repeating structure for the quiver theory. \ In this regard,
the brane probe theories studied in \cite{UrangaSMthroat} are
particularly interesting in that they have a roughly periodic structure and
flow to semi-realistic Standard Model-like gauge theories at the bottom of the cascade.

A special feature of the Klebanov Strassler cascade is that the rank of the gauge
groups grows only logarithmically with RG scale. This is due to the fact
that higher up in the cascade, the number of flavors of both nodes is
roughly equal to twice the number of colors. This is not
true for most generalizations. In particular, cascading quiver theories
with more than two nodes and two or more generations of chiral matter typically do not maintain this balance.
As a result, the ranks of the gauge groups may start to grow
much faster with scale, and the duality sequence will accelerate accordingly.
The theory gradually gets trapped in a strong coupling regime, as indicated
in figure \ref{rgspiral}. Eventually, the cascade steps accumulate and the
system reaches a so-called duality wall at some finite UV scale
 \cite{StrasslerNagoya}.

\section{Realizing the MQSM at the Bottom of a Cascade}\label{Realizingone}

\begin{figure}
[t]
\begin{center}
\includegraphics[
height=2.3in,
width=4in
]%
{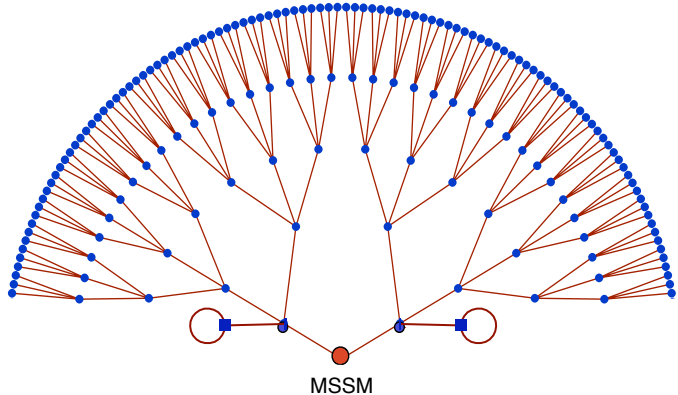}%
\caption{There is most likely a large landscape of UV gauge theories which
connect via a duality cascade to the MSSM in the IR. Starting at the MSSM and proceeding up towards the UV,
the duality cascade may encounter a
succession of bifurcations which represent different possible sequences of
Seiberg dualities.  In this paper, we will identify two ways of connecting the MSSM
to a special cascade which both follow regular periodic duality paths (as indicated by the
two closed circles) rather than a chaotic sequence of dualities.
}\label{tree}%
\end{center}
\vspace{-2mm}
\end{figure}

Starting from the MQSM, there are many ways in which
this theory may connect to a duality cascade in the UV.
Typically, these duality sequences are very irregular, and involve
a succession of gauge theories with rapidly growing gauge group ranks and
number of generations. The RG flow may even display a
chaotic structure due to the fact that small variations in the relative
size of gauge couplings may affect the order in which different quiver
nodes undergo Seiberg dualities
\cite{HananyFrancoDualityTrees,HerzogCascade}.  As a result, a
sequence of dualities will encounter multiple bifurcation points where a
given `magnetic' theory can connect to two or more different `electric'
theories.  Our expectation is that there is likely a large
landscape of UV theories which flow via a duality cascade to the
MSSM in the IR.  See figure \ref{tree} for a schematic depiction of this branching process.  Rather than study a random cascade
which may `accidentally' connect to the MQSM, we shall instead focus on a minimal class of distinguished UV theories which exhibit a periodic duality cascade that terminates at the MQSM. \ As a matter of notation, we shall denote a sequence of dualities as a string of labels indicating which node has been dualized.

\subsection{Cascade Criteria}\label{criteria}

We now abstract from the above example of the Klebanov Strassler cascade
 to provide a stringent set of criteria
which we shall require any candidate cascade which UV completes the MQSM\ to
satisfy. \ Due to the fact that we do not at present possess a string theory
construction of the above cascade, we shall only consider cascades such that
the resulting quiver theory has the maximal chance of possessing a candidate
holographic dual. \ As is evident in the example of the conifold, the
appearance of a well-behaved holographic dual requires that all gauge group
factors with finite gauge couplings must have a smooth large $N$ limit. \ To
reach the bottom of a more general cascade with low rank gauge groups, it
therefore follows that each gauge group factor must dualize repeatedly over
the course of the cascade. \ We shall therefore require that in any candidate cascade:

\begin{itemize}
\item Each node with finite gauge coupling must possess a smooth large
N\ limit. \ As a consequence, each such node must dualize repeatedly during
the cascade.
\end{itemize}

Whereas the ranks of the conifold theory continue to deplete during the entire
cascade, the amount of bifundamental matter remains constant. \ From the perspective
of the associated intersection
pairing, it follows that during the entire cascade the intersection pairing of
the cycles wrapped by the branes remains constant. \ On the other hand, it is
well known that in more general geometric realizations of Seiberg duality in
the context of type IIB brane probe theories, dualization corresponds to a
flop within a K\"{a}hler cone. \ When the quiver theory returns to its
original connectivity, it implies that the geometry has returned to the
original K\"{a}hler cone up to some monodromy at large radius. \ While it is
conceivable that a holographic dual may exist even when the intersection
pairing of the branes does not return to the original connectivity of the
quiver theory, it is reasonable to suppose that cascading brane configurations
which repeatedly return to the same K\"{a}hler cone are most likely to possess
a holographic dual description. \ For this reason, we also require that in any
candidate cascade:

\begin{itemize}
\item The number of bifundamentals must remain roughly constant over the span
of the entire cascade.
\end{itemize}

Practical experience shows that this condition appears particularly difficult to satisfy for a
generic chiral quiver gauge theory.\footnote{For example, by repeatedly Seiberg dualizing
nodes of the quiver theory associated to the D3-brane probe of the geometry $%
%TCIMACRO{\U{2102} }%
%BeginExpansion
\mathbb{C}
%EndExpansion
^{3}/%
%TCIMACRO{\U{2124} }%
%BeginExpansion
\mathbb{Z}
%EndExpansion
_{3}$, the resulting number of bifundamentals rapidly increases after each
successive Seiberg duality. } Indeed, the above condition greatly restricts the number of available candidate cascades.
%\ See figure \ref{cthreeexample} for a sequence of
%Seiberg dualities exhibiting this growth in bifundamental matter for the brane
%probe theory of $%
%TCIMACRO{\U{2102} }%
%BeginExpansion
%\mathbb{C}
%EndExpansion
%^{3}/%
%TCIMACRO{\U{2124} }%
%BeginExpansion
%\mathbb{Z}
%EndExpansion
%_{3}$. \
While it is in principle possible to relax the criteria we propose,
\textit{it is intriguing that even under these stringent conditions, the
chiral matter content of the MQSM\ is arranged in such a way that the
resulting candidate cascades satisfy the above criteria.}%
%TCIMACRO{\FRAME{ftbpFU}{5.591in}{1.1078in}{0pt}{\Qcb{Depiction of a formal
%sequence of Seiberg dualities for the quiver theory given by a D-brane probe
%of $\U{2102} ^{3}/\U{2124} _{3}$. \ As shown, the number of bifundamentals
%rapidly increases during each successive dualization.}}{\Qlb{cthreeexample}%
%}{cthreeexample.eps}{\special{ language "Scientific Word";  type "GRAPHIC";
%width 5.591in;  height 1.1078in;  depth 0pt;  original-width 10.1131in;
%original-height 1.9804in;  cropleft "0";  croptop "1";  cropright "1";
%cropbottom "0";  filename 'CTHREEEXAMPLE.eps';file-properties "XNPEU";}} }%
%BeginExpansion

%\begin{figure} [ptb] \begin{center} \includegraphics[ height=1.1078in, width=5.591in ]% {CTHREEEXAMPLE.eps}% \caption{Depiction of a formal sequence of Seiberg dualities for the quiver theory given by a D-brane probe of $\mathbb{C} ^{3}/\mathbb{Z} _{3}$. \ As shown, the number of bifundamentals rapidly increases during each successive dualization.}% \label{cthreeexample}% \end{center} \end{figure}
%EndExpansion

\subsection{RG Node Locking} \label{NODELOCK}

In addition to the above restrictions, we must also require that any proposed
sequence of Seiberg dualities remains compatible with RG\ flow from the UV\ to
the IR. \ In particular, assuming that the Seiberg dual of any strongly
coupled quiver node subsequently flows to weak coupling in the IR, the next
stage of the cascade must dualize another node of the quiver theory. \ We now
argue that the presence of too much tensor matter can force a quiver node to
flow to weak coupling.

At various stages of the covering theory cascade,
bifundamental matter may appear between a node and its image under the
orientifold action. \ In the orientifold theory these bifundamentals descend
to matter transforming in either the two index symmetric $(S)$, anti-symmetric
$(A)$ or conjugate representations. \ As will be explicit in all of the the
cases studied below, the number of arrows between a quiver node and its image
in the covering theory will always be either zero or at least $2g$. \ These
fields then descend to at least $g$ fields in the $S$ and $g$ in the $A$
representation of $U(N)$. \ The one loop coefficient of the beta function of
the non-abelian gauge coupling in the orientifold theory is therefore:%
\begin{align}
b_{SU(N)}  &  =3N- \fracc{\left(  N+2\right) g }-\fracc{\left(
N-2\right) g
}-F\label{betaone}\\[2mm]
&  =\left(  3-g\right)  N-F \nonumber %\label{betatwo}%
\end{align}
where $F>0$ denotes all additional contributions from fields charged under the
$SU(N)$ gauge group.  The above argument is self-consistent because at weak coupling the one loop approximation to the running of the couplings should be an accurate description.  Nevertheless, it is conceivable that contributions to the anomalous dimensions of fields
can significantly alter the RG flow of coupling constants during a given cascade.  The presence of such a large amount
of tensor matter, however, implies that the effective number of flavors $F_{eff}$ will always be greater than $3N$ so that the above analysis should remain valid beyond the one loop approximation. \ When $g\geq3$, we therefore conclude that any quiver node with a sufficient amount of tensor matter will always flow to weak
coupling. \ As should now be evident, this severely restricts possible
candidate cascade paths.

\subsection{Connecting to the MQSM: Higgsing or Confinement}
%\ at the Bottom of a Cascade
\label{Realizing}

In this subsection we discuss
%Let us now discuss
the possible ways in which the gauge groups and
matter content of the MQSM\ may appear at the bottom of a duality cascade.
\ Note that the field content of the MQSM alone cannot realize a cascade which
satisfies the above stringent criteria.  \ Indeed, suppose
to the contrary that a large $N$ generalization of the MQSM eventually
cascades to the MQSM. \ It follows at once from the above discussion of RG\ node
locking that once the $USp$ factor at node $a$
dualizes, too much tensor matter will be present at the other nodes. \ Hence,
once the $USp$ factor dualizes to a weakly coupled description, all three
gauge group factors will flow to weak coupling in the IR. \ We therefore
conclude that any candidate cascade which realizes the MQSM\ at the bottom of
the cascade must contain at least four quiver nodes. We will label the three
MQSM nodes by $a$, $b$, $c$, and the extra node by $d$.

There are in general two ways in which the MQSM\ can embed inside such a four
node quiver. \ Attaching an extra node to the MQSM, the most direct analogue
of the Klebanov Strassler cascade corresponds to the case where the rank of
the gauge group factor on the extra node depletes to zero at the bottom of the
cascade. \ Rather than requiring that the extra node confine at the bottom of
the cascade, it is also possible to Higgs the quiver theory before the ranks
of any node completely depletes so that two of the quiver nodes collapse to a
single node, thereby reproducing the three node MQSM quiver.

In either scenario, any candidate single node extension of the MQSM must
satisfy the requirements described in section \ref{criteria}. \ Indeed, as
will be evident from the discussion below, the qualitative requirement that
the cascade repeat in an appropriate sense will impose tight restrictions on
the connectivity of four node cascading quiver theories which eventually
terminate at the MQSM.

\section{Cascade in the Higgsing Scenario} \label{alrsymmcascade}

%(XXX Classify the possible ways of Higgsing down ?XXX).

In this section we discuss the cascade of a four node quiver theory which
realizes the MQSM\ at the bottom of the cascade via the Higgsing scenario.
\ In order to satisfy the criteria of section \ref{criteria}, we require that
the connectivity and number generations of the corresponding four node quiver
remain stable throughout nearly all of the cascade.
\ We note that in the explicit example that we shall now consider,
there is -- besides the fact that it is needed for establishing a
periodic cascade --  an additional physical motivation for adding a
fourth node to the MQSM: such an extension naturally restores left-right
symmetry in the UV.

\subsection{A Left-Right Symmetric Extension of the MQSM}

\begin{figure}[t]
\begin{center}
%\resizebox{\textwidth}{!}{
\scalebox{.55}{
\includegraphics{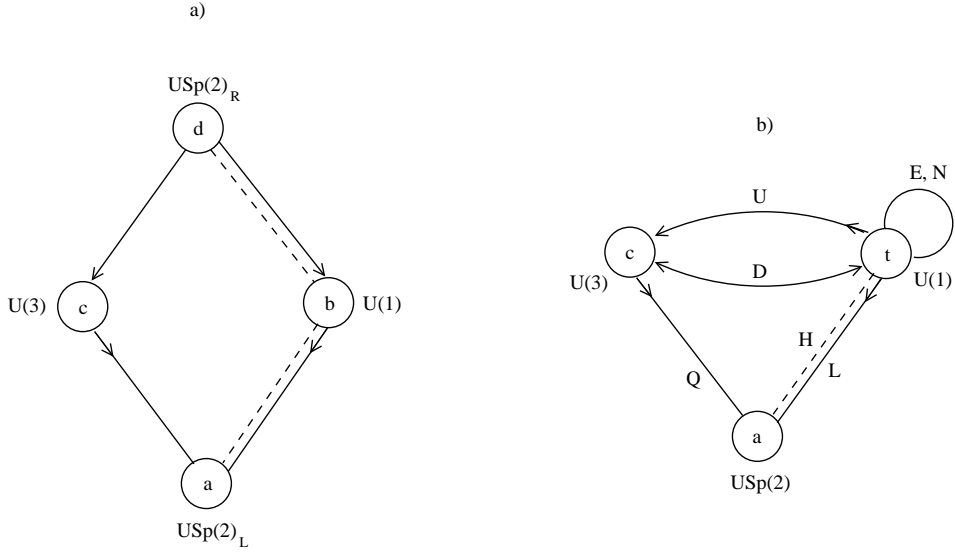}
}
%\scalebox{.8}{
%\includegraphics[width=\textwidth]{LRQuiver.pdf}
\end{center}
\par
\caption{ {A minimal left-right symmetric extension of
the MSSM (a). Each oriented line represents $g=3$ generations of chiral matter and
each dashed line represents a single vector-like pair. Left-right symmetry breaking
reproduces the matter content of the MQSM with three additional fields transforming in the adjoint of node $t$ (b).  We interpret these additional fields as right-handed neutrinos.}}%
\label{LRQuiver}%
\vspace{-2mm}
\end{figure}

Given that node $a$ of the MQSM quiver only couples to left-handed chiral matter,
it is natural to identify the extra node $d$ with its right-handed partner. The $%
%TCIMACRO{\U{2124} }%
%BeginExpansion
\mathbb{Z}
%EndExpansion
_{2}$ symmetry that interchanges node $a$ and $d$ then becomes identified with
left-right symmetry. To reduce the number of unknown parameters in the
construction, we require that this left-right symmetry is restored further up
the cascade. \ The minimal quiver realization of a left-right model is shown
in figure \ref{LRQuiver}. As before, oriented lines represent $g=3$
generations of chiral matter. The dashed lines each represent a single vector
pair of chiral superfields corresponding to the up/down Higgs pair of
$USp(2)_{L}$ and $USp(2)_{R}$. Note that this LR model has no matter lines
between node $b$ and $c$.

We now briefly explain how to Higgs this model to the MQSM. \ The LR quiver
model has two $U(1)$ symmetries, of which only the
combination:
\begin{equation}
\mathcal{Q}_{B-L}={\frac{1}{2}}(\mathcal{Q}_{U(1)_{c}}-\mathcal{Q}_{U(1)_{b}})
\end{equation}
is non-anomalous. The other $U(1)$ symmetry has mixed anomalies,
that we assume are cancelled via a Green-Schwarz mechanism of the
type discussed in section \ref{oriented}. The same GS mechanism also
produces a large mass for the abelian vector boson of the anomalous
$U(1)$.

The right-handed quarks and leptons combine into $USp(2)_{R}$ doublets
\begin{equation}
Q_{R}=\left(
\begin{array}
[c]{c}%
U\\
D
\end{array}
\right)  ,\qquad L_{R}=\left(
\begin{array}
[c]{c}%
N\\
E
\end{array}
\right)  \,.
\end{equation}
There is a vector pair of doublets $H_{3},H_{4}$, which are the right-handed
mirrors of the MSSM Higgs. Left-right-symmetry is broken via the assumption
that $H_{3},H_{4}$ both acquire a vev that is much larger than the
electro-weak symmetry breaking scale
\begin{equation}
H_{3}=\left(
\begin{array}
[c]{c}%
a_{1}\\
a_{2}%
\end{array}
\right)\, ,   \qquad H_{4}=\left(
\begin{array}
[c]{c}%
b_{1}\\
b_{2}%
\end{array}
\right)\, .
\end{equation}
We may use the $USp(2)_{R}$ symmetry to set $a_{2}=0$. The D-term
constraints impose $b_{1}=0$ and $a_{1}=b_{2}$. \
The most economical mechanism for generating this Higgs vev is to assume
that the superpotential takes the form
\begin{equation}
\label{higgsright}
W(H_3,H_4) =   \mu_R (H_3 H_4 - {1\over 2 a_1^2} (H_3 H_4)^2).
\end{equation}
Minimizing $W$ allows for two supersymmetric minima, one of which
breaks the left-right symmetry.\footnote{Alternatively, it is also possible that
left-right symmetry breaking occurs in tandem with supersymmetry
breaking.  While we can consider such scenarios when the LR scale is comparatively low, we shall also later consider models where
the LR scale is quite high (around $10^{10}$ TeV).  In this case it is important to preserve low energy supersymmetry
to protect the large hierarchy with the electro-weak scale.}
The fields $H_{3}$
and $H_{4}$ are neutral under the $U(1)$ subgroup of
$USp(2)_{R}\times U(1)_{b}$ generated by the linear combination
\
\begin{equation}
\mathcal{Q}_{U(1)_{t}}= \mathcal{Q}_{U(1)_{b}}+2\,T_{\!R}{}_{3}. \label{ut}%
\end{equation}
Thus the quiver nodes $b$ and $d$ collapse to a single node $t$, representing
an anomalous $U(1)_{t}$ symmetry. The hypercharge symmetry $U(1)_{Y}$ is
generated by the non-anomalous combination
\begin{equation}
Y={\frac{1}{2}}(\mathcal{Q}_{U(1)_{c}}-\mathcal{Q}_{U(1)_{t}})={%
}\mathcal{Q}_{{B-L}}-T_{\!R}{}_{3}.
\end{equation}
The Higgsing of $USp(2)_{R}\times U(1)_{b}\rightarrow U(1)_{t}$ produces 3
massive gauge bosons with a mass of order the gauge coupling times $a_{1}$. We
will denote this mass scale by $\Lambda_{LR}$. Traditionally the scale
$\Lambda_{LR}$ is taken to be much larger than the electro-weak scale.
Further discussion on the various energy scales in this left-right model appear
in section \ref{RUNNING}.
Of the original fields between the $U(1)$ and $USp(2)_R$ node,
in the low energy theory
we are left over with three right-handed lepton superfields $E$
and
three sterile neutrinos denoted $N$.
The effective
theory far below the scale $\Lambda_{LR}$ is described by the
MQSM quiver in figure \ref{LRQuiver}b.

The Yukawa couplings in the MQSM descend from quartic superpotential
terms in the unbroken LR theory.
The same type of quartic couplings also give a contribution to the $\mu$-term
$\lambda H_{u}H_{d}H_{3}H_{4}\rightarrow\lambda {a_{1}^{2}}H_{u}%
H_{d}$. %+{\frac{a_{1}}{M}}H_{u}H_{d}S
To avoid a $\mu$-term that is too large,
one needs to assume that the corresponding coupling
$\lambda$ is small, or that this
contribution is approximately cancelled via the `bare' mu-term
$\mu_{L}H_{u}H_{d}$.
Reducing the complete quartic superpotential of the LR model, a
priori also produces various undesirable $R$-parity violating couplings.
However, these are easily eliminated if we require the
LR superpotential to be invariant under
a $\mathbb{Z}_{2}$ symmetry which
acts as
\begin{equation}
R(Q)=R(L)=-1,\qquad R(H)=+1.
\end{equation}
Where by abuse of notation, $R$ denotes the matter parity of a given superfield.

\subsection{Left-Right Symmetric Duality Cascade}

\begin{figure}[t]
\begin{center}
%\resizebox{\textwidth}{!}{
\scalebox{.55}{
\includegraphics{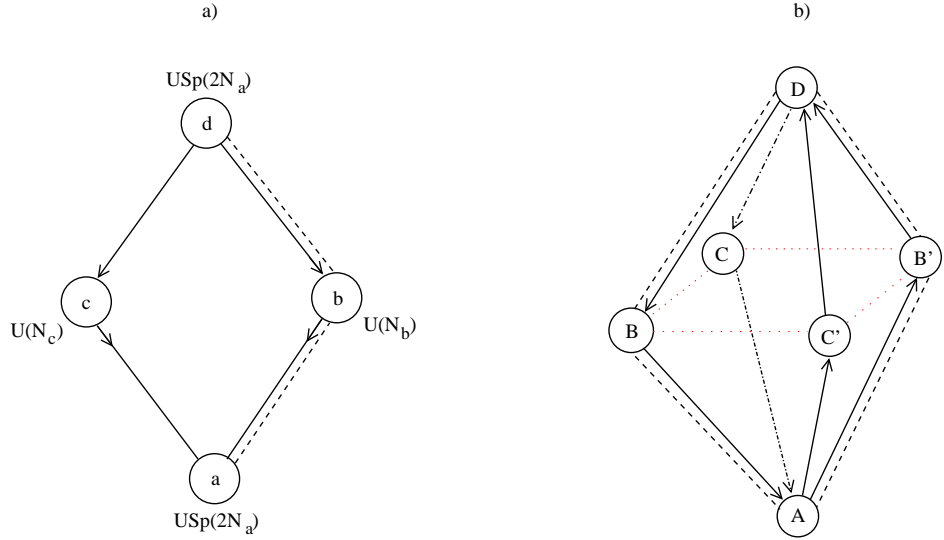}
}
%\scalebox{.8}{
%\includegraphics[width=\textwidth]{LRQuiver.pdf}
\end{center}
\par
\caption{ {The left-right model with general gauge group
ranks (a). Each oriented line indicates $g=3$ chiral matter fields, and the dashed
lines each represent a single vector-like pair of Higgs fields.
The covering theory (b) has the shape of an octahedron.}}%
\label{LRcascade}%
\vspace{-2mm}
\end{figure}

We now turn to describe the LR duality cascade.
On general grounds, the LR model can be viewed as an orientifold of a quiver
gauge theory. As shown in figure \ref{LRcascade}, the covering theory has the
shape of an octahedron. As we will see in the next section, the
cascading quiver that connects to the MQSM via the confining scenario
will exhibit the same octahedral symmetry.\footnote{The symmetry group of the
octahedron is identical to that of the cube, its dual platonic solid. \ The
symmetry group of the cube is $S_{4}\times S_{2}$, corresponding to
permutations of the four primary diagonals and point reflection about the
center of mass. %\ We note in passing that the exceptional Dynkin diagrams make
%a somewhat tangential appearance via the McKay correspondence. \ Indeed, the
%tetrahedral symmetry group corresponds to $E_{6}$ and the octahedral symmetry
%group corresponds to $E_{7}$.
}
\ In order for holography to remain compatible with
the orientifold projection, our expectation is that all cascades of the
covering theory must descend in an appropriate fashion to the orientifold
theory. \

Consider the LR quiver gauge theory with general ranks as depicted in figure
\ref{LRcascade}. First, suppose that node $b$ or $c$ reaches strong coupling
first, and undergoes a Seiberg duality. The $b$ node has $2(g+2)N_{a}$
flavors, so the duality acts via
\begin{equation}
N_{b}\rightarrow2(g+2)N_{a}-N_{b}\,.
\end{equation}
Node $c$ has $2gN_{a}$ flavors, and its duality maps
\begin{equation}
N_{c}\rightarrow2gN_{a}-N_{c}\,.
\end{equation}
Since neither node has tensor matter, or connects to the other node, each
Seiberg duality has a very simple action on the chiral matter content of the
quiver. Each existing chiral line is reversed, indicating the replacement of
quarks and leptons by dual quarks and leptons. The same is true for the vector
lines, connecting nodes $b$ with nodes $a$ and $d$. There are extra mesons
produced in each duality which connect nodes $a$ and $d$. But since $a$ and
$d$ are $USp$ nodes, and as seen from the cover quiver, these mesons
automatically come in vector pairs. Assuming generic quartic superpotential
couplings, these mesons are all massive, and can be omitted from the low
energy theory.

Next consider the case that node $a$ and/or $d$ reach strong coupling and
dualize. For simplicity, we will mostly assume that left-right symmetry is
completely restored during the UV part of the cascade, and that $a$ and $d$
have the same coupling. The two nodes then dualize simultaneously at the same
scale. Each node has $gN_{c}+(g+2)N_{b}$ flavors,
so that in the Seiberg dual theory the ranks change to:
\begin{equation}
2N_{a}\rightarrow gN_{c}+(g+2)N_{b}-2N_{a}-4.
\end{equation}
Again, all matter lines reverse orientation in accord with their replacement by
dual matter. While dual meson fields will also be created which connect nodes $b$ and $c$,
the left-right symmetry of the Higgsing scenario implies that these fields come in vector-like pairs.
We shall therefore assume that these acquire a mass due to quartic superpotential couplings in the UV theory.
A more detailed analysis of how such mass terms arise in the context of the confining scenario cascade is given in Appendix \ref{DUALPOT}.

A priori, we could assume that $a$ and $d$ have different couplings
and that the duality happens at different scales. As long as the two
dualities happen in direct sequence, the above discussion remains
unaltered. However, one could worry then that either node $b$ or $c$
would dualize at an energy scale between the energy scales where $a$ or $d$
dualizes. However, it is easy to see that this
can not happen: whenever $a$ or $d$ dualizes, this creates extra
tensor like matter for nodes $b$ and $c$, that prevents them from
dualizing. This is the RG node locking mechanism discussed in section \ref{NODELOCK}.
In other words, the cascade can proceed only after both
$a$ and $d$ have dualized, so that all tensor and chiral matter
lines between $b$ and $c$ pair up and become massive. The
Seiberg dualities proceed in an alternating sequence where the
duality $(ad)$ is followed by duality map on nodes $b$ and/or $c$,
and then again $(ad)$. It may happen that in between
two $(ad)$ dualities, either both nodes $b$ and $c$ dualize or
just one of the two. Since neither  duality alters the connectivity
of the quiver, the periodicity of the cascade is preserved either way.

It is important to note that throughout this duality cascade, there are never
any massive vector-like meson pairs created that connect nodes $a$ with $b$ or $c$, or node
$d$ with $b$ or $c$. This is important for the following reason. Even if such
meson pairs are massive, and thereby decouple from the low energy theory, they
could still mix with the vector-like Higgs fields that connect node $b$ with
nodes $a$ or $d$. This mixing would produce a large mass for the Higgs scalars,
which would lift them from the low energy spectrum. Thus the absence of this
class of meson pairs is an important fact, which helps secure the stability
of the left-right quiver and preserve the presence of light Higgs scalars.

As a somewhat related point, we note that the superpotential
 obtained through Seiberg duality is always
invariant under matter parity if the initial superpotential is, where
matter parity acts as $\pm1$ on a magnetic edge of the quiver if and only if it acts as
$\pm1$ on the dual electric edge. The reason is that the potential in the
dual theory is of the schematic form $W \sim W_{\mathrm{micro}} + m
x y$ where $x,y$ are dual quarks, and $m$ is a meson which inherits
its parity from the electric theory. With our assignment of matter parity
to $x$ and $y$, $m$ is even if and only if the product $xy$ is even.
From this we conclude that the Seiberg
dual superpotential is also invariant under matter parity.

This concludes our first description of the left-right symmetric
cascade. In section \ref{Bottom} we will describe the IR region of the cascade,
and how it connects with the MQSM.

\section{Cascade in the Confining Scenario}\label{multigen}

The analysis of the previous section establishes that the MQSM\ can in
principle lie at the bottom of a cascade which terminates by partially
Higgsing a four node quiver. \ In this section we begin our analysis of the
other candidate scenario whereby a cascade terminates when the extra node
confines. \ As opposed to the relatively simple combinatorics of candidate
cascades in the LR model, the combinatorics of the confining scenario requires
a much longer sequence of steps before the resulting quiver returns to its
original connectivity.

Because the classification of candidate cascades is more involved than in the
LR\ cascade, we first provide a summary of the analysis to follow. \ In order
to classify candidate cascades in the orientifold theory, we study in Appendix
\ref{CovercascadeA} the admissible cascades of the covering theory which
preserve the $%
%TCIMACRO{\U{2124} }%
%BeginExpansion
\mathbb{Z}
%EndExpansion
_{2}$ orientifold action and properly descend to the orientifold theory. \ We
note that while it is in principle possible to perform this field theoretic
classification of cascades purely in the orientifold theory, in the context of
holography it is important to establish that any candidate cascade correctly
descend from the covering theory to its orientifold.

The combinatorics of the cascade significantly limit possible single node
extensions of the covering theory. \ Although we shall integrate out nearly
all vector-like pairs generated by the cascade, we shall at first allow
$n\geq0$ massless vector-like Higgs pairs of bifundamentals connecting nodes
$b$ and $a$. \ It follows from the covering theory analogue of RG\ node locking
discussed in section \ref{NODELOCK} that when a sufficient amount of chiral
matter connects a quiver node to its image, the resulting pair of quiver nodes
can never dualize. \ In fact, when the extra node attaches to the MQSM\ by
purely chiral matter, we find that all of the \textquotedblleft locking
matter\textquotedblright\ lifts if and only if the quiver theory has
\textit{precisely one Higgs pair}. \ More generally, although we have not
completely classified all possible ways in which vector-like matter can be
added to a single node extension of the MQSM, we present some further examples
where the extra node also attaches by vector-like matter to the MQSM.

\subsection{Cascade Classification}\label{Covercascadeone}

\begin{figure}[t]
\begin{center}
\includegraphics[
height=2.6411in,
width=3.9851in
]{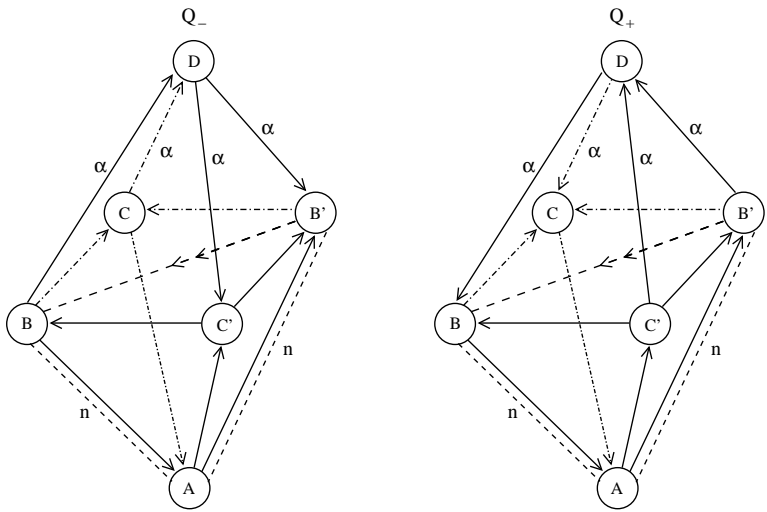}
\end{center}
\par
\vspace{-2mm}\caption{The quiver theories $Q_{-}$ and $Q_{+}$ correspond to
the two distinct ways to attach a single additional node to the covering
theory of the $g$ generation \ MQSM\ by purely chiral matter. \ The label
$\alpha$ by each arrow indicates $\alpha$ bifundamentals. \ All other oriented
lines denote $g$ bifundamentals so that there are $2g$ arrows between
$B^{\prime}$ and $B$. The dashed line denotes a vector-like pair and $n$
denotes the number of such pairs. In this section we show that only $Q_{+}$
supports a periodic cascade which descends to its orientifold when $\alpha=g$ and $n=1$. }%
\label{qminusandqplus}%
\end{figure}

In Appendix \ref{CovercascadeA} we determine necessary conditions which any
single node extension of the MQSM\ by purely chiral matter must satisfy in
order to admit a periodically repeating cascade structure. \ We begin this
classification by studying cascades in the covering theory.\ \ Assigning gauge
group ranks compatible with the $%
%TCIMACRO{\U{2124} }%
%BeginExpansion
\mathbb{Z}
%EndExpansion
_{2}$ orientifold symmetry, we find that cancelling all non-abelian anomalies
greatly restricts the ways in which the extra node can attach to
the MQSM. \ Restricting to this class of quiver topologies, we next consider
candidate cascades in both the covering theory and its orientifold, and find
that unless the arrows of the covering theory have multiplicity $g$, the
cascade cannot proceed. \ The end result of the classification argument as
outlined in Appendix \ref{CovercascadeA} is that only the second quiver in
figure \ref{qminusandqplus}, denoted by $Q_{+}$, can give rise to a periodic
cascade which descends to the orientifold theory denoted by $q_{+}$.
\ Moreover, this cascade properly descends only when $\alpha$, the number of
bifundamentals that connect the extra node $D$ to nodes $B$, $B^{\prime}$, $C$
and $C^{\prime}$ equals the number of generations: $\alpha=g$. \ The analysis
of Appendix A also establishes that the cascade can only proceed when the
gauge group factor of the extra node in the orientifold theory is of $USp$
type.

\subsection{Number of Higgs Pairs}

The analysis of Appendix \ref{CovercascadeA} demonstrates that there is at
most one way to attach one additional node by purely chiral matter to the
covering quiver of the MQSM so that the resulting quiver admits a cascade
which properly descends to a repeating cascade in the orientifold theory. \ We
note that apart from the lines connecting $B$ and $B^{\prime}$ and the
presence of additional vector-like matter, the connectivity of this quiver is
identical to that of an octahedron. \ In this section we show that the pair of
nodes $BB^{\prime}$ can only dualize when the number of Higgs pairs is exactly
one. \ On the other hand, we also show that when only a subset of quiver nodes
are dualized, the number of Higgs pairs in the orientifold theory is unconstrained.

\begin{figure}
[ptb]
\begin{center}
\includegraphics[
height=2.8029in,
width=4.2211in
]%
{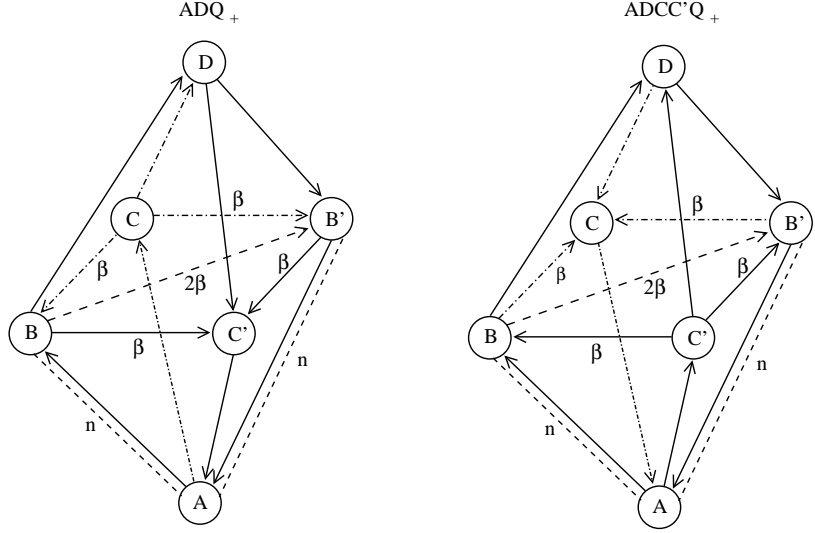}%
\caption{Starting from the quiver theory $Q_{+}$ depicted in figure
\ref{qminusandqplus}, the two candidate sequences of Seiberg dualities
consistent with the $\mathbb{Z} _{2}$ action of the orientifold theory lead to
the quiver theories $ADQ_{+}$ and $ADCC^{\prime}Q_{+}$. \ In the figure,
$\beta=g(n-1)$. \ When $n=1$ all arrows in the common plane of $B,B^{\prime
},C$ and $C^{\prime}$ vanish.}%
\label{twopaths}%
\end{center}
\end{figure}

A cascade in which all nodes are dualized at least once necessarily dualizes
the pair corresponding to $B$ and $B^{\prime}$. \ In order to simultaneously
preserve the $%
%TCIMACRO{\U{2124} }%
%BeginExpansion
\mathbb{Z}
%EndExpansion
_{2}$ symmetry of the orientifold action while dualizing this pair, the number
of bifundamentals connecting $B$ and $B^{\prime}$ must vanish. \ Beginning with the quiver theory $Q_{+}$, we denote
the dualized quiver by a string of letters indicating which nodes have dualized.
A candidate cascade will lead to one of two candidate quiver theories
$ADQ_{+}$ or $ADCC^{\prime}Q_{+}$. \ See figure
\ref{twopaths} for a depiction of these two quiver theories for general $n$.
\ In both cases, while the number of bifundamentals connecting $C$ and
$C^{\prime}$ vanishes, the number connecting $B$ and $B^{\prime}$ is:%
\begin{equation}
n_{BB^{\prime}}=2g(n-1)\text{.}\label{HiggsEquations}%
\end{equation}
In order for the next stage of the cascade to dualize $B$ and $B^{\prime}$ it
follows that $n=1$. \ Assuming that neither cascade proceeds by this route,
the next stage of the two cascades lead to the quiver theories $CC^{\prime
}ADQ_{+}$ and $CC^{\prime}ADCC^{\prime}Q_{+}$ for $ADQ_{+}$ and $ADCC^{\prime
}Q_{+}$, respectively. \ We note, however, that although the amount of
vector-like matter which must be integrated out depends on whether $n$ equals
zero, in all cases the connectivity of the quiver $CC^{\prime}ADQ_{+}$ is
identical to $ADCC^{\prime}Q_{+}$ and that of $CC^{\prime}ADCC^{\prime}Q_{+}$
to $ADQ_{+}$. \ We therefore conclude that if the cascade never proceeds
through the dualization of the pair $B$ and $B^{\prime}$, there are exactly
four distinct connectivities for the quivers given by $Q_{+}$, $CC^{\prime
}Q_{+}$, $ADQ_{+}$ and $ADCC^{\prime}Q_{+}$. \ Because the number of
bifundamentals connecting $B$ and $B^{\prime}$ is $2g$ for the first two
theories and $2g(n-1)$ for the latter two, a cascade which dualizes the pair
$B$ and $B^{\prime}$ can only proceed provided $n=1$.

To show that this result descends to the orientifold theory we return to
lines (\ref{SSIMP}) and (\ref{ASIMP}) of Appendix A. \ When $\alpha=g$ and
$\varepsilon=-1$, the total amount of tensor matter at node $b$ vanishes when
$n=1$. \ A similar analysis holds for the other candidate paths.%
%TCIMACRO{\FRAME{ftbpFU}{4.2211in}{2.8029in}{0pt}{\Qcb{Starting from the quiver
%theory $Q_{+}$ depicted in figure \ref{qminusandqplus}, the two candidate
%sequences of Seiberg dualities consistent with the $\U{2124} _{2}$ action of
%the orientifold theory lead to the quiver theories $ADQ_{+}$ and
%$ADCC^{\prime}Q_{+}$. \ In the figure, $\beta=g(n-1)$. \ When $n=1$ all arrows
%in the common plane of $B,B^{\prime},C$ and $C^{\prime}$ vanish.}}%
%{\Qlb{twopaths}}{twopaths.eps}{\special{ language "Scientific Word";
%type "GRAPHIC";  maintain-aspect-ratio TRUE;  display "USEDEF";
%valid_file "F";  width 4.2211in;  height 2.8029in;  depth 0pt;
%original-width 7.4858in;  original-height 4.9623in;  cropleft "0";
%filename 'TWOPATHS.eps';file-properties "NPEU";}} }%
%BeginExpansion
%EndExpansion

On the other hand, there is no restriction on $n$ when only the quiver nodes
$A$, $D$, $C$ and $C^{\prime}$ participate in the cascade. \ Indeed, tracing
through the discussion given above, we find that such quiver theories also
periodically repeat. \ In fact, up to permutations in the order of dualization
for the nodes $A$ and $D$, there are only two sequences of dualization
consistent with the assumption that a dualized node subsequently flows to weak
coupling. \ The two possible sequences are given by alternately dualizing the
pairs $CC^{\prime}$ and $AD$:%
\begin{align}
(CC^{\prime})(AD)(CC^{\prime})(AD)Q_{+}  &  =Q_{+}\eol (AD)(CC^{\prime
})(AD)(CC^{\prime})Q_{+} & =Q_{+}%
\end{align}
where a pair of dualizations enclosed by brackets commute.

In comparison with the covering theory of the LR cascade, we note that in the
case of the quiver theory $Q_{+}$, there are now matter lines between nodes
$B$ and $B^{\prime}$, as well as between $B$ and $C$. \ Indeed, this
additional connectivity is responsible for the far tighter restrictions on the
matter content of the cascading quivers of the confining scenario.

\begin{figure}
[hbtb]
\begin{center}
\includegraphics[
height=1.7in,
width=4.8in
]%
{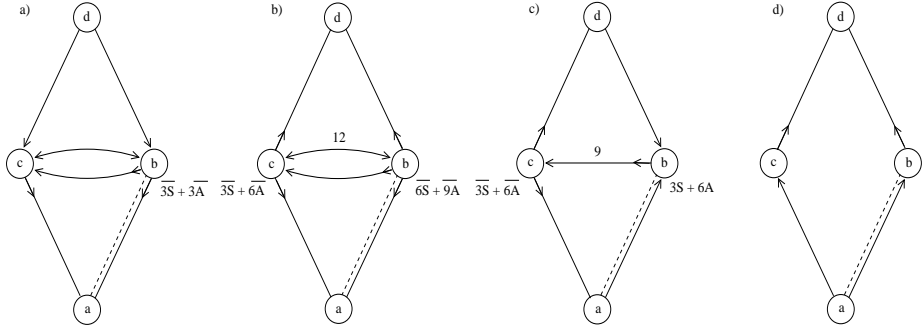}%
\caption{Up to re-orienting the directions of arrows or complex conjugating
tensor matter representations, when the Higgs pair remains exactly massless,
there are four types of quiver theories which can arise from a cascade which
begins with the quiver theory $q_{+}$. \ Unless otherwise indicated, each
oriented line denotes $g=3$ bifundamentals. \ The dashed line denotes a single
Higgs-up/Higgs-down pair. \ A similar classification into four quiver types
also holds for more general values of $g$.}%
\label{fourtypes}%
\end{center}
\vspace{-2mm}
\end{figure}

Assuming that the number of Higgs pairs remains constant during the entire
cascade, up to complex conjugation of some fields and interchanging the r\^{o}les
of nodes $a$ and $d$, there are four types of quivers which can appear in the
process of cascading down from $q_{+}$ (see figure \ref{fourtypes}). \ As the
above qualifications suggest, in section \ref{Superpot}
we show that once the specific form of the superpotential is taken into
account, the number of Higgs pairs can sometimes change as the cascade
proceeds. \ An example of this behavior is shown in figure \ref{higgsphoenix}
which demonstrates that the r\^{o}les of the $USp$ groups at nodes $a$ and $d$ can also
interchange r\^{o}les as the cascade proceeds.%
%TCIMACRO{\FRAME{ftbpFU}{3.9115in}{4.7651in}{0pt}{\Qcb{Up to re-orienting the
%directions of arrows or complex conjugating tensor matter representations,
%when the Higgs pair remains exactly massless, there are four types of quiver
%theories which can arise from a cascade which begins with the quiver theory
%$q_{+}$. \ Unless otherwise indicated, each oriented line denotes $g=3$
%bifundamentals. \ The dashed line denotes a single Higgs-up/Higgs-down pair.
%\ A similar classification into four quiver types also holds for more general
%values of $g$.}}{\Qlb{fourtypes}}{fourtypes.eps}%
%width 3.9115in;  height 4.7651in;  depth 0pt;  original-width 7.4236in;
%original-height 9.0555in;  cropleft "0";  croptop "1";  cropright "1";
%cropbottom "0";  filename 'FOURTYPES.eps';file-properties "XNPEU";}} }%
%BeginExpansion
%EndExpansion

\subsection{More General Vector-Like Matter}\label{MOREGENERAL}

It is intriguing that in its minimal form, the single node extension by purely
chiral matter requires precisely one Higgs pair in order for other stages of
the cascade to proceed. \ For more general single node extensions, however,
there may also be a number of additional massless vector-like pairs present.
\ Letting $v_{GH}$ denote the number of vector-like pairs between a pair of
nodes $G$ and $H$, if we assume that all other vector-like pairs develop a
mass and can be integrated out, a similar cascade will proceed when the number
of added vector-like pairs obey the conditions:%
\begin{align}
n &  =v_{AB}=v_{AB^{\prime}}=v_{DB}+1=v_{DB^{\prime}}+1 \eol
m &  =v_{DC}=v_{DC^{\prime}}=v_{AC}=v_{AC^{\prime}}\\[2mm]
p &
=v_{BC}=v_{B^{\prime}C^{\prime}}=v_{BC^{\prime}}=v_{B^{\prime}C}\nonumber
\text{.}%
\end{align}
In section \ref{Bottom} we shall consider one such extension which serves to
accelerate the first stage of dualization in proceeding from the IR\ to the
UV. \ To simplify some of the analysis to follow, in the rest of this section
we shall restrict to the case of the single node extension by purely chiral
matter.%
%TCIMACRO{\FRAME{ftbpFU}{1.9752in}{2.9611in}{0pt}{\Qcb{Depiction of the single
%node extension of the cover of the MQSM\ which attaches by both chiral matter
%and vector-like pairs. \ Each oriented line denotes $g$ bifundamentals.
%\ Vector-like pairs are denoted by dashed lines with the multiplicity
%indicated.}}{\Qlb{qplusvectmatter}}{qplusvectmatter.eps}%
%width 1.9752in;  height 2.9611in;  depth 0pt;  original-width 3.2214in;
%original-height 4.8447in;  cropleft "0";  croptop "1";  cropright "1";
%cropbottom "0";  filename 'QPLUSVECTMATTER.eps';file-properties "NPEU";}} }%
%BeginExpansion
\begin{figure}
[ptb]
\begin{center}
\includegraphics[
height=2.9611in,
width=1.9752in
]%
{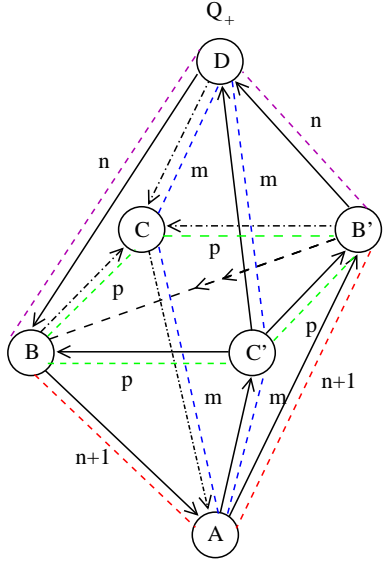}%
\caption{Depiction of the single node extension of the cover of the
MQSM\ which attaches by both chiral matter and vector-like pairs. \ Each
oriented line denotes $g$ bifundamentals. \ Vector-like pairs are denoted by
dashed lines with the multiplicity indicated.}%
\label{qplusvectmatter}%
\end{center}
\end{figure}
%EndExpansion

\subsection{Superpotential Analysis} \label{Superpot}

The analysis of the previous sections establishes that at the level of the
matter content and connectivity of the quiver theory, a periodic cascading
structure may occur in the confining scenario provided all vector-like pairs
except for the Higgs pair lift at each intermediate stage. \ To
demonstrate that this is indeed realized, in Appendix \ref{DUALPOT} we classify all
candidate terms in the superpotential of a given electric theory which could potentially produce a
quadratic term in the superpotential of the dual magnetic theory. \ Due to the fact that
quadratic terms in the magnetic theory can sometimes appear as composite cubic
or quartic operators in the original electric theory variables, we first
present the most general form of the superpotential up to quartic order in the
quiver fields. \ It is a consequence of the topology of the quiver theory that
at nearly all stages of the cascade the dual magnetic theory contains a mass
term for all vector-like matter except for the Higgs pair. \ In fact, we find
that there is only a single stage of the cascade where a candidate mass term
for the Higgs pair can develop.

In order to classify all possible mass terms in the dual magnetic theory
obtained during each stage of the cascade, it is sufficient to determine all
admissible gauge invariant terms up to quartic order for the quiver theories
obtained at intermediate stages of the cascade process. \ To this end, we now
argue that it is enough to only treat the quiver theories of type $a)$ and
$d)$ in figure \ref{fourtypes}. \ Because nodes $b$ and $c$ cannot dualize
when tensor matter is present, nodes $a$ and $d$ always dualize together.
\ Further, because nodes $a$ and $d$ do not share any bifundamental matter,
the number of vector-like pairs attached to either node cannot disappear by
dualizing $a$ or $d$. \ This implies that the dual superpotential given by
dualizing the pair $a$ and $d$ is also independent of the order in which the
cascade proceeds. \ Because dualizing nodes $c$ and $b$ does not introduce any
chiral matter and only reverses the orientation of quiver arrows, it is
therefore enough to determine the form of the dual superpotential for quivers
of type $a)$ and $d)$ in figure \ref{fourtypes}.

As shown in Appendix \ref{DUALPOT}, whereas generically all other vector-like
pairs will develop a mass at some stage of the cascade, there is only one
possible candidate quiver topology which upon dualizing produces a mass term
for the previously massless Higgs pair.

With notation as defined in Appendix \ref{DUALPOT}, in the electric theory corresponding
to the quiver theory $q_{+}$, the superpotential contains the terms:%
\begin{equation}
W_{q_{+}}\supset\lambda_{ij}X_{ca}^{i}X_{a\overline{b}}X_{b\overline{c}}%
^{j}+\widehat{\lambda}_{ij}^{I}X_{ca}^{i}X_{ab}^{I}X_{\overline{bc}}%
^{j}+\varphi_{ijkl}X_{ac}^{i}X_{\overline{cb}}^{j}X_{b\overline{c^{\prime}}%
}^{k}X_{c^{\prime}a}^{l}\text{.}\label{ELECTRICPROBLEM}%
\end{equation}
In addition to the original vector-like Higgs pair between nodes $a$ and $b$,
dualizing node $c$ will produce $g^{2}$ additional vector-like pairs
corresponding to new meson fields. \ Letting $C$ denote the mesons created by
dualizing node $c$, the magnetic dual superpotential of the quiver theory
$cq_{+}$ contains terms of the form:%
\begin{equation}
W_{c(q_{+})}\supset\left[
\begin{array}
[c]{cc}%
X_{\overline{b}a} & C_{\overline{b}a}^{kl}%
\end{array}
\right]  _{1\times\left(  g^{2}+1\right)  }\left[
\begin{array}
[c]{cc}%
\mu_{I} & \mu_{c}\lambda_{ij}\\
\mu_{c}\widehat{\lambda}_{kl}^{I} & \mu_{c}^{2}\varphi_{klij}%
\end{array}
\right]  _{(g^{2}+1)\times\left(  g^{2}+g+1\right)  }\left[
\begin{array}
[c]{c}%
X_{ab}^{I}\\
C_{ab}^{ij}%
\end{array}
\right]  _{\left(  g^{2}+g+1\right)  \times1}
\label{PROBLEM}%
\end{equation}
where the fields $C$ denote meson fields created by dualizing node $c$. \ For
generic values of the couplings, precisely $g$ chiral fields charged in the
fundamental of $b$ and $a$ will remain and \textit{all} vector-like pairs will
develop a mass.

\subsubsection{R-parity}

Although the above analysis demonstrates that in most cases the resulting
vector-like pairs develop a mass, generic values for the couplings produce
phenomenologically undesirable interaction terms. \ Returning to standard
MSSM notation, these include terms of the form:%
\begin{align}
\mu_{i}X_{a\overline{b}}X_{ba}^{i} &  =\mu_{i}H_{u}L^{i}\eol
\widehat{\lambda}_{ij}^{k}X_{ca}^{i}X_{ab}^{k}X_{\overline{bc}}^{j} &
=\widehat{\lambda}_{ij}^{k}Q^{i}L^{k}D^{j}%
\end{align}
which lead to lepton number violating interactions. \ To prevent such terms,
it is customary to require that the Lagrangian density remain invariant under
R-parity.   As already discussed in the context of the Higgsing scenario,
this is compatible with our cascade structure.  This also applies
to the confining scenario.  Nevertheless, for the benefit of the reader
we will discuss the R-parity assignments in the confining cascade scenario in more detail here.
\ In a Lorentz invariant theory, this is equivalent to assigning a
matter parity of $(-1)^{3(B-L)}$ to each superfield of the theory. \ This has
the effect of splitting the $g+1$ bifundamentals:%
\begin{equation}
X_{ba}^{I}\rightarrow X_{ba}^{0}\oplus X_{ba}^{i}\text{.}%
\end{equation}
Due to the fact that this parity assignment corresponds to a real subgroup of
a $U(1)$ group, the matter parity of a field remains the same after Seiberg
dualizing a quiver node. \ All composite operators such as meson fields
therefore also inherit a definite matter parity. \ Returning to the quiver
theory $q_{+}$, the matter parity of the fields are:%
\begin{align}
&\ \ \qquad \qquad R(X_{[\overline{bb^{\prime}}]}^{i})   =\alpha_{i}
\nonumber\\[-5mm]
R(X_{ca}^{i},X_{\overline{bc}}^{i},X_{b\overline{c}}^{i},X_{ba}^{i}%
,X_{(\overline{bb^{\prime}})}^{i})   =-1 & \nonumber \\[-2mm]
& \ \ \qquad \qquad R\left(  X_{d\overline{c}}^{i}\right)    =\beta_{i}\\[-2mm]
R\left(  X_{ab}^{0},X_{a\overline{b}}\right)     =+1 & \nonumber\\[-4mm]
& \ \ \qquad \qquad R\left(  X_{d\overline{b}}^{i}\right)    =\gamma_{j}\nonumber \text{.}%
\end{align}
Scanning through the possible terms of Appendix \ref{DUALPOT}, we find that when
$\alpha_{i}=-1$ and $\beta_{i}=\gamma_{j}$ for all $i$, $j$, the vector-like
pairs of fields studied in the previous section still develop quadratic terms
in the superpotential of the dual magnetic theory. \ This has the consequence that one
of the chiral fields attached to node $d$ has non-trivial lepton number. \ We
shall return to the matter parity of the additional fields in section
\ref{mumu}.

Although imposing matter parity does reduce the number of allowed
superpotential terms, it does not resolve the problem encountered in equation
(\ref{PROBLEM}). \ Indeed, the corresponding quadratic term for the dual meson
fields now takes the form:%
\begin{equation}
R\left(  W_{c(q_{+})}\right)  \supset\left[
\begin{array}
[c]{cc}%
X_{\overline{b}a} & C_{\overline{b}a}^{kl}%
\end{array}
\right]  _{1\times\left(  g^{2}+1\right)  }\left[
\begin{array}
[c]{cc}%
\mu_{0} & \mu_{c}\lambda_{ij}\\
\mu_{c}\widehat{\lambda}_{kl}^{0} & \mu_{c}^{2}\left(  \varphi
_{klij}+\widehat{\varphi}_{ijkl}\right)
\end{array}
\right]  _{(g^{2}+1)\times\left(  g^{2}+1\right)  }\left[
\begin{array}
[c]{c}%
X_{ab}^{0}\\
C_{ab}^{ij}%
\end{array}
\right]  _{\left(  g^{2}+1\right)  \times1} \label{massmatrix}%
\end{equation}
where in the above we have used the fact that the $C^{ij}$ have matter parity
$+1$. \ Note that although the fields $X_{ab}^{i}$ no longer appear in the
corresponding term and are therefore massless, the remaining vector-like
matter between nodes $a$ and $b$ will still generically develop a mass.

\subsection{The Cascading $\mu$ problem and Higgs Regeneration} \label{mumu}

As shown in the previous section, although generic values of the couplings in
the orientifold theory lead to mass terms for most vector-like pairs, at
nearly all stages of the cascade the Higgs pair remains exactly massless.
\ Even so, dualizing node $c$ in the quiver theory $q_{+}$ will generate a
$(g^{2}+1)\times(g^{2}+1)$ mass matrix with rank $g^{2}+1$. \ Such a mass
matrix will lift all vector-like pairs between nodes $a$ and $b$. \ On the
other hand, due to the fact that node $b$ can only dualize when no tensor
matter is present, it follows from the analysis near equation
(\ref{HiggsEquations}) that in order to lift the tensor matter present at node
$b$, the cascade must retain a single nearly massless Higgs pair. \ Once this
pair develops a mass, the results of the previous section establish that while
the cascade can still proceed by dualizing all nodes other than $b$, the rank
of $b$ will remain constant throughout the rest of the cascade.

While the above arguments hold for generic values of the couplings, relations
between couplings in the covering theory can sometimes descend to non-trivial
restrictions on the form of the superpotential in the orientifold theory. \ In
this section we show that while the Higgs pair may disappear from the
massless spectrum at intermediate stages of the cascade, when the couplings
$\varphi_{ijkl}$ of equation (\ref{ELECTRICPROBLEM}) correspond to a
$g^{2}\times g^{2}$ matrix of rank $g^{2}-1$, a new Higgs pair will
automatically regenerate further down the cascade. \ After presenting a
general analysis of how the cascade proceeds as the Higgs pair disappears and
regenerates, we present an explicit realization of this mechanism where the
octahedral symmetry of the covering theory naturally enforces the condition
that the matrix $\varphi_{ijkl}$ is anti-symmetric. \ In this case we find
that the rank of the matrix is generically $g^{2}-1$ only when $g$ is an odd number.

\subsubsection{Higgs Regeneration}

Assuming that all other couplings remain sufficiently generic so that all
other vector-like pairs develop a mass, we now show that when the couplings
$\varphi_{ijkl}$ of equation (\ref{ELECTRICPROBLEM}) correspond to a
$g^{2}\times g^{2}$ matrix of rank
$g^{2}-1$, the Higgs pair between nodes $a$ and $b$ regenerates further down
the cascade. \ For simplicity, we restrict our discussion to matter parity
invariant superpotentials. \ Starting from the quiver theory $q_{+}$, consider
the dualized theory $cq_{+}$. \ It follows from equation (\ref{massmatrix})
that even when $\varphi_{ijkl}$ has rank $g^{2}-1$, the corresponding
quadratic form will still be non-degenerate. \ In this case, all vector-like
pairs between nodes $a$ and $b$ lift, leaving only $g$ oriented lines
corresponding to the lepton doublets. \ Due to the fact that node $b$ still
flows to weak coupling in the IR, the next stage of the cascade must proceed
by dualizing the pair of nodes $a$ and $d$. \ Note that whereas all tensor
matter created at node $c$ lifts as before, the number of arrows attached
between node $b$ and $d$ is now equal to the number between $b$ and $a$.
\ Because these two sets of arrows have opposite orientation, all tensor
matter at node $b$ created by dualizing $a$ and $d$ will generically develop a
mass and lift from the massless spectrum. \ Because similar reasoning applies
to the tensor matter produced at node $c$, it follows that the next stage of
the cascade must proceed by dualizing node $c$.

\begin{figure}
[ptb]
\begin{center}
\includegraphics[
height=1.6in,
width=5.2in
]%
{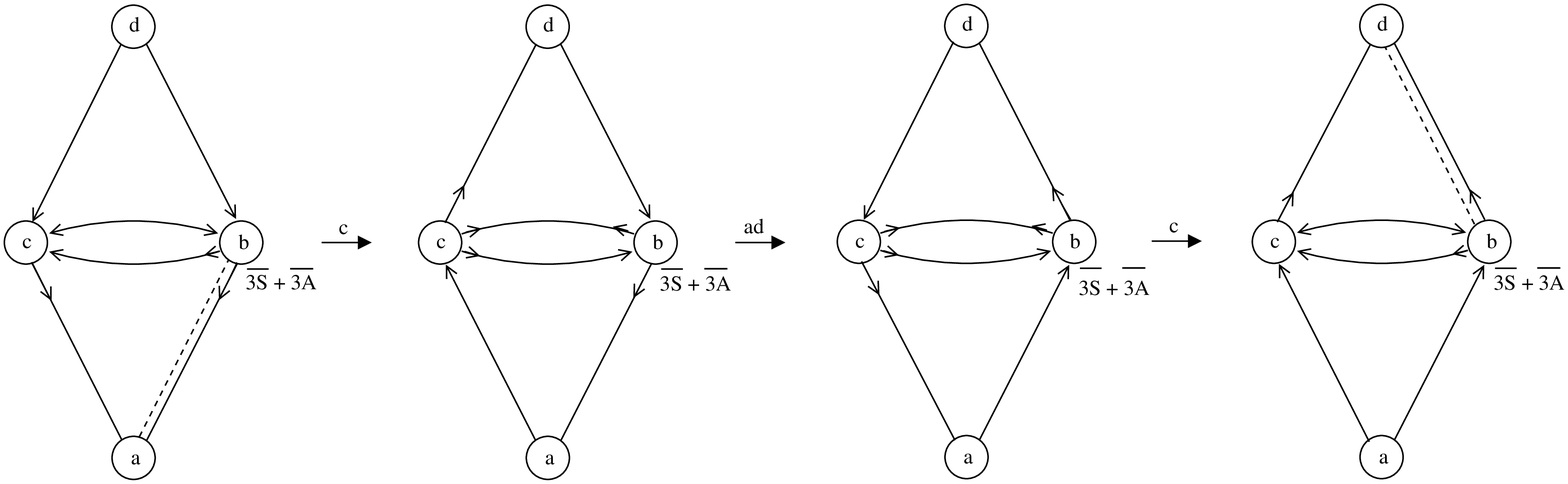}%
\caption{Starting from the quiver theory $q_{+}$, dualizing node $c$
introduces a generic $g^{2}\times g^{2}$ mass term for all vector-like matter
between nodes $a$ and $b$. \ As shown in the rightmost figure, further down
the cascade a Higgs pair will regenerate between nodes $d$ and $b$ when the
mass matrix for the dual meson fields has rank $g^{2}-1$.}%
\label{higgsphoenix}%
\end{center}
\vspace{-2mm}
\end{figure}

In comparison to the original quiver theory $q_{+}$, the quiver theory
$adcq_{+}$ does not possess any vector-like pairs. \ Note in particular that
the r\^{o}les of nodes $a$ and $d$ have also interchanged. \ Indeed, upon
dualizing node $c$, the orientation of lines attached to node $c$ dictate that
no vector-like pairs appear between nodes $a$ and $b$. \ Instead, precisely
$g^{2}$ vector-like pairs develop between nodes $d$ and $b$. \ Assuming that
the matter parity of the superfields $X_{d\overline{c}}^{i}$ and $X_{db}^{i}$
is equal to that of the lepton doublets between nodes $a$ and $b$, it follows
that the quiver theory $adcq_{+}$ contains the terms:%
\begin{equation}
W_{adcq_{+}}\supset\chi_{ijkl}X_{d\overline{c}}^{i}X_{c\overline{b}}%
^{j}X_{bc^{\prime}}^{k}X_{\overline{c^{\prime}}d}^{l}\text{.}%
\end{equation}
The couplings $\chi_{ijkl}$ are the analogue of the couplings $\varphi_{ijkl}$
in the quiver theory $q_{+}$. \ Indeed, upon dualizing node $c$, $\chi_{ijkl}$
corresponds to a $g^{2}\times g^{2}$ quadratic form in the magnetic dual
superpotential. \ Due to the exchange of r\^{o}les for nodes $a$ and $d$, first
suppose that whatever considerations require the matrix $\varphi_{ijkl}$ to
have rank $g^{2}-1$ also apply to the matrix $\chi_{ijkl}$. \ In this case,
the resulting quadratic terms in the superpotential of the dual magnetic
theory $cadcq_{+}$ cause all but one vector-like pair of fields between nodes
$d$ and $b$ to develop a mass. \ Because nodes $a$ and $d$ have simply
exchanged r\^{o}les in the cascade, we therefore conclude that the cascade will
proceed as before. \ Indeed, at the next stage, dualizing the pair of nodes
$a$ and $d$ now generically lifts the tensor matter present at node $b$ so
that node $b$ is free to dualize. \ Iterating through the cascade once more,
it follows that eventually the cascade will return to a quiver theory similar
to $q_{+}$ where the next step of the cascade dualizes node $c$. \ Reversing
the r\^{o}les of the matrices of couplings $\varphi_{ijkl}$ and $\chi_{ijkl}$, in
the next cycle the vector-like matter between nodes $d$ and $b$ generically
develops a mass, but a new massless vector-like pair regenerates between nodes
$a$ and $b$. \ See figure \ref{higgsphoenix} for a sequence of dualities which
exhibits the disappearance and regeneration of the Higgs pair when $\varphi$
and $\chi$ both have rank $g^{2}-1$. \ In order for nodes $a$ and $d$ to
exchange r\^{o}les at intermediate stages of the cascade, the matter attached to
node $a$ must also exchange r\^{o}les with the matter attached to node $d$.
%TCIMACRO{\FRAME{ftbpFU}{4.8118in}{1.4901in}{0pt}{\Qcb{Starting from the quiver
%theory $q_{+}$, dualizing node $c$ introduces a generic $g^{2}\times g^{2}$
%mass term for all vector-like matter between nodes $a$ and $b$. \ As shown in
%the rightmost figure, further down the cascade a Higgs pair will regenerate
%between nodes $d$ and $b$ when the mass matrix for the dual meson fields has
%rank $g^{2}-1$.}}{\Qlb{higgsphoenix}}{higgsphoenix.eps}%
%width 4.8118in;  height 1.4901in;  depth 0pt;  original-width 14.3213in;
%original-height 4.4183in;  cropleft "0";  croptop "1";  cropright "1";
%cropbottom "0";  filename 'higgsphoenix.eps';file-properties "XNPEU";}} }%
%BeginExpansion
%EndExpansion

Alternatively, even when the matrix of couplings $\chi_{ijkl}$ is generic, the
rank constraint on $\varphi_{ijkl}$ will still regenerate a massless Higgs
pair between nodes $a$ and $b$. \ Although dualizing node $c$ of the quiver
theory $adcq_{+}$ in this case produces $g^{2}$ vector-like pairs between
nodes $d$ and $b$ which generically develop a mass, the subsequent dualization
of nodes $a$ and $d$ results in the quiver theory $adcadcq_{+}$. \ But the
topology of this quiver is identical to that of the quiver $q_{+}$ with the
vector-like pair between nodes $a$ and $b$ deleted. \ It now follows by
similar reasoning that dualizing node $c$ regenerates a massless vector-like
pair of fields between nodes $a$ and $b$.

Although originally introduced simply to make the cascade proceed,
regenerating the Higgs pair naturally decouples the $\mu$-term from any
candidate UV completion of the field theory. \ In the sense that generically
the $\mu$-term is expected to be sensitive to Planck scale physics, we can
claim to have partially solved the $\mu$ problem.

\subsubsection{Reflection Symmetry and Path Rules}

The argument of the previous section establishes that when an appropriate
symmetry reduces the ranks of the matrices of couplings $\varphi_{ijkl}$ and
$\chi_{ijkl}$ to $g^{2}-1$, the cascade regenerates the single massless Higgs
pair necessary for other stages of the cascade to proceed. \ Although it is
possible to simply impose a relation on the couplings at some stage of the
cascade, unless this rule derives from a symmetry of some kind, we must
explicitly check that this same symmetry persists further down the cascade.
\ For this reason, we now present an explicit symmetry which reduces the rank
of the corresponding coupling matrix. \ While it is in principle possible
to impose an explicit discrete symmetry of the flavor group $U(3)^{5}$ which
reduces the rank of the matrices of couplings, an arbitrary choice would
likely obscure the underlying geometry of a candidate compactification.  To this end,
we now show that there exists a symmetry of the covering quiver theory such that
the matrices of couplings $\varphi_{ijkl}$ and $\chi_{ijkl}$ will be anti-symmetric.
When $g$ is an odd number, the corresponding matrices will have rank $g^{2}-1$.

Because we expect on general grounds that explicit automorphisms of the quiver
theory will naturally lift to conditions on the geometry probed by a stack of
D-branes, we restrict our analysis to automorphisms of the quiver theory which
interchange the r\^{o}les of the quiver nodes.\footnote{This is manifest in the case of
the brane probe of the conifold studied in \cite{KlebanovStrassler,KlebanovWitten}. \ As a further example, recall that in all quiver
gauge theories given by D-brane probes of toric varieties, each chiral
superfield contributes with opposite sign in precisely two terms of the
superpotential. \ In this case, the resulting F-term relations define toric ideals.}
The automorphisms of the covering theory $Q_{+}$ are generated by two $%
%TCIMACRO{\U{2124} }%
%BeginExpansion
\mathbb{Z}
%EndExpansion
_{2}$ generators $r$ and $R$, where $r$ denotes reflection about the plane cut
by the nodes $A$, $D$, $B$ and $B^{\prime}$ and $R$ denotes rotation by an
angle of $\pi$ about the axis passing through the nodes $A$ and $D$. \ More
precisely, the reflection fixes the bifundamentals $X_{B^{\prime}\overline{B}%
}^{i}$, $X_{B\overline{A}}^{i}$ and $X_{A\overline{B^{\prime}}}^{i}$ and
interchanges all other fields in the obvious fashion. \ In order for the
action of matter parity to remain compatible with the reflection symmetry, at
other stages of the cascade we shall require that all additional meson fields
between $B^{\prime}$ and $B$ remain fixed by the reflection symmetry and that
all additional vector-like pairs of fields between $B$ and $A$ or $B^{\prime}$
and $A$ interchange with a counterpart. \ Whereas the $%
%TCIMACRO{\U{2124} }%
%BeginExpansion
\mathbb{Z}
%EndExpansion
_{2}$ identification of quiver nodes of the orientifold theory corresponds to
identification by the rotation group symmetry, the reflection $r$ does not
descend from the quiver $Q_{+}$ to $q_{+}$. \ Indeed, from the perspective of
the orientifold theory the symmetry $r$ interchanges fields transforming in
different representations of the same gauge group.

Because each piecewise connected path maps to an operator, it is enough to
study the action of $r$ on the algebra of paths with two or more links. \ We
now show that there exists a representation of the $%
%TCIMACRO{\U{2124} }%
%BeginExpansion
\mathbb{Z}
%EndExpansion
_{2}$ action $r$ on the algebra of paths in the quiver theory which imposes
the desired anti-symmetry condition on the matrices of couplings
$\varphi_{ijkl}$ and $\chi_{ijkl}$ while leaving all other couplings
sufficiently generic to lift all remaining vector-like pairs. \ In order to
fully specify the action of the reflection symmetry on the algebra of paths in
the covering theory $Q_{+}$ we shall not require that the starting and ending
point of a given path be the same. \ We claim that a consistent action on the
path algebra of each quiver theory encountered during the cascade is given by
the rules:

\begin{itemize}
\item Path Rule 1: Although invariance under some other symmetry may impose
additional relations on the couplings, no further restriction is imposed on
the coupling constant matrices corresponding to paths which do not map to a
connected path in the image.

\item Path Rule 2: A path/operator with a connected image path which passes
$n$ times perpendicularly through the plane of reflection maps to the image
path/operator weighted by $(-1)^{n}$.
\end{itemize}

In the above, paths which pass perpendicularly through the plane of reflection
are defined as paths which contain a directed or anti-directed (e.g. with all arrows reversed)
subpath of one of the following types:%
\begin{equation}
\{C\rightarrow A\rightarrow C^{\prime},C^{\prime}\rightarrow A\rightarrow
C,C\rightarrow C^{\prime},C^{\prime}\rightarrow C,C\rightarrow D\rightarrow
C^{\prime},C^{\prime}\rightarrow D\rightarrow C\}\text{.} \label{perpdef}%
\end{equation}
The first rule is a necessary condition for a consistent action on the algebra
of paths due to the fact that some gauge invariant paths do not always map to
gauge invariant paths under the action of $r$. \ For example, letting
$\Lambda$ denote an index which runs from $1$ to $2g$, the closed path
$C\rightarrow A\rightarrow B^{\prime}\rightarrow B\rightarrow C$ corresponding
to the operator $\tau_{ij}^{I\Lambda}X_{C\overline{A}}^{i}X_{A\overline
{B^{\prime}}}^{I}X_{B^{\prime}\overline{B}}^{\Lambda}X_{B\overline{C}}^{j}$ in the quiver $Q_{plus}$
does not map to a gauge invariant counterpart. \ Because all meson fields
created during the cascade respect the reflection symmetry of the octahedron,
the above rules remain intact during each stage of the cascade.

Invariance under the action of the reflection symmetry imposes the desired
relation on the matrix of couplings $\varphi_{ijkl}$. \ Indeed, under the
reflection symmetry, the path $C\rightarrow A\rightarrow C^{\prime}\rightarrow
B\rightarrow C$ corresponding to the operator $\varphi_{ijkl}X_{C\overline{A}%
}^{i}X_{A\overline{C^{\prime}}}^{j}X_{C^{\prime}\overline{B}}^{k}%
X_{B\overline{C}}^{l}$ maps to:%
\begin{align}
r\left(  \varphi_{ijkl}X_{C\overline{A}}^{i}X_{A\overline{C^{\prime}}}%
^{j}X_{C^{\prime}\overline{B}}^{k}X_{B\overline{C}}^{l}\right)   &
=-\varphi_{ijkl}X_{C\overline{A}}^{j}X_{A\overline{C^{\prime}}}^{i}%
X_{C^{\prime}\overline{B}}^{l}X_{B\overline{C}}^{k}\\
&  =-\varphi_{jilk}X_{C\overline{A}}^{i}X_{A\overline{C^{\prime}}}%
^{j}X_{C^{\prime}\overline{B}}^{k}X_{B\overline{C}}^{l} \nonumber%
\end{align}
so that invariance of the path implies $\varphi$ is anti-symmetric.

In order to demonstrate that the above symmetry allows the cascade to proceed,
it is enough to show that the relations on all other couplings are
sufficiently mild so as to leave the rest of the cascade intact. \ Because a
similar analysis can be applied for the quiver theory $CC^{\prime}Q_{+}$, it
is enough to treat the relations among couplings in the quiver theories
$Q_{+}$ and $ADQ_{+}$.
Including all possible terms up to quartic order, the most general
superpotential of the quiver theory $Q_{+}$ is:%
\begin{align}
W_{Q_{+}} &  =\mu_{I}X_{B\overline{A}}^{I}X_{A\overline{B}}+\mu_{I}^{\prime
}X_{A\overline{B^{\prime}}}^{I}X_{B^{\prime}\overline{A}}+\lambda
_{ij}X_{C\overline{A}}^{i}X_{A\overline{B}}X_{B\overline{C}}^{j}+\lambda
_{ij}^{\prime}X_{A\overline{C^{\prime}}}^{i}X_{C^{\prime}\overline{B^{\prime}%
}}X_{B^{\prime}\overline{A}}^{j}\eoll
&  +\widehat{\lambda}_{ij}^{I}X_{C\overline{A}}^{i}X_{A\overline{B^{\prime}}%
}^{I}X_{B^{\prime}\overline{C}}^{j}+\widehat{\lambda}_{ij}^{\prime
I}X_{C\overline{A}}^{i}X_{A\overline{B^{\prime}}}^{I}X_{B^{\prime}\overline
{C}}^{j}+\sigma_{\Lambda}^{IJ}X_{A\overline{B^{\prime}}}^{I}X_{B^{\prime
}\overline{B}}^{\Lambda}X_{B\overline{A}}^{J}\eoll
&  +\alpha_{ijkl}X_{D\overline{C}}^{i}X_{C\overline{A}}^{j}X_{A\overline
{C^{\prime}}}^{k}X_{C^{\prime}\overline{D}}^{l}+\gamma_{ij}^{IJ}%
X_{D\overline{B}}^{i}X_{B\overline{A}}^{I}X_{A\overline{B^{\prime}}}%
^{J}X_{B^{\prime}\overline{D}}^{j} \eoll
&  +\tau_{ij}^{I\Lambda}X_{C\overline{A}}^{i}X_{A\overline{B^{\prime}}}%
^{I}X_{B^{\prime}\overline{B}}^{\Lambda}X_{B\overline{C}}^{j}+\tau
_{ij}^{\prime I\Lambda}X_{A\overline{C^{\prime}}}^{i}X_{C^{\prime}%
\overline{B^{\prime}}}^{j}X_{B^{\prime}\overline{B}}^{\Lambda}X_{B\overline
{A}}^{I}\eoll
&  +\beta_{ijk}^{I}X_{D\overline{C}}^{i}X_{C\overline{A}}^{j}X_{A\overline
{B^{\prime}}}^{I}X_{B^{\prime}\overline{D}}^{k}+\beta_{ijk}^{\prime
I}X_{C^{\prime}\overline{D}}^{i}X_{D\overline{B}}^{j}X_{B\overline{A}}%
^{I}X_{A\overline{C^{\prime}}}^{k}\eoll
&  +\varphi_{ijkl}X_{C\overline{A}}^{i}X_{A\overline{C^{\prime}}}%
^{j}X_{C^{\prime}\overline{B}}^{k}X_{B\overline{C}}^{l}+\varphi_{ijkl}%
^{\prime}X_{C\overline{A}}^{i}X_{A\overline{C^{\prime}}}^{j}X_{C^{\prime
}\overline{B^{\prime}}}^{k}X_{B^{\prime}\overline{C}}^{l}\\
&  +\mu_{IJ}^{(1)}(X_{B^{\prime}\overline{A}}X_{A\overline{B^{\prime}}}%
^{I})(X_{B^{\prime}\overline{A}}X_{A\overline{B^{\prime}}}^{J})+\mu_{IJ}%
^{(2)}(X_{A\overline{B}}X_{B\overline{A}}^{I})(X_{A\overline{B}}%
X_{B\overline{A}}^{J})\eoll
&  +\mu_{IJ}^{(3)}(X_{A\overline{B}}X_{B\overline{A}}^{I})(X_{B^{\prime
}\overline{A}}X_{A\overline{B^{\prime}}}^{J})+\kappa_{IJ}^{(1)}X_{B_{1}%
^{\prime}\overline{A_{1}}}X_{A_{1}\overline{B_{2}^{\prime}}}^{I}%
X_{B_{2}^{\prime}\overline{A_{2}}}X_{A_{2}\overline{B_{1}^{\prime}}}^{I}\eoll
&  +\kappa_{IJ}^{(2)}X_{A_{1}\overline{B_{1}}}X_{B\overline{_{1}A_{2}}}%
^{I}X_{A_{2}\overline{B_{2}}}X_{B\overline{_{2}A_{1}}}^{I}+\widehat{\kappa
}_{IJ}X_{B\overline{A}}^{I}X_{A\overline{B^{\prime}}}^{J}X_{B^{\prime
}\overline{A}}X_{A\overline{B}}\nonumber \text{.}%
\end{align}
Invariance under reflection symmetry imposes the following relations
on these couplings:%
\begin{align}
& \qquad \ \ \qquad
\alpha_{ijkl}   =\alpha_{lkji} \nonumber \\[-4.5mm]
\lambda_{ij}  =\widehat{\lambda}_{ij}^{\prime0}& \nonumber
\\[-2mm]
\label{alphasymmetric}
& \qquad \ \ \qquad
\varphi_{ijkl}   =-\varphi_{jilk}\\[-2mm]
\lambda_{ij}^{\prime}   =\widehat{\lambda}_{ij}^{0} & \nonumber \\[-4.5mm]
& \qquad \ \ \qquad
\varphi_{ijkl}^{\prime}  =-\varphi_{jilk}^{\prime}\nonumber%
\end{align}
which descend to the superpotential $W_{q_{+}}$ of the orientifold theory
given by equation (\ref{Wqplus}). \ We note that this symmetry
requires that the Yukawa couplings of the up and down type quarks are exactly
equal at each intermediate stage of the cascade. \ In other words,
%at intermediate stages of the cascade,
the CKM\ matrix of the resulting theory is diagonal. \
Presumably suitably small off diagonal components develop
near the bottom of the cascade.  The observed hierarchy $m_{top} \gg m_{bottom}$ in this scenario must
arise from a proportional hierarchy between the Higgs up and down expectation
values.

Similarly, the most general superpotential of the quiver theory $ADQ_{+}$ is:%
\begin{align}
W_{ADQ_{+}}  &  =\mu_{I}X_{B\overline{A}}^{I}X_{A\overline{B}}+\mu_{I}%
^{\prime}X_{A\overline{B^{\prime}}}^{I}X_{B^{\prime}\overline{A}}\eoll
&  +\alpha_{ijkl}X_{C\overline{D}}^{i}X_{D\overline{C^{\prime}}}%
^{j}X_{C^{\prime}\overline{A}}^{k}X_{A\overline{C}}^{l}+\gamma_{ij}%
^{IJ}X_{B\overline{D}}^{i}X_{D\overline{B^{\prime}}}^{j}X_{B^{\prime}%
\overline{A}}^{I}X_{A\overline{B}}^{J}\eoll
&  +\beta_{ijk}^{I}X_{C\overline{D}}^{i}X_{D\overline{B^{\prime}}}%
^{j}X_{B^{\prime}\overline{A}}^{I}X_{A\overline{C}}^{k}+\beta_{ijk}^{\prime
I}X_{D\overline{C^{\prime}}}^{i}X_{C^{\prime}\overline{A}}^{j}X_{A\overline
{B}}^{I}X_{B\overline{D}}^{k}\\
&  +\mu_{IJ}^{(1)}(X_{B^{\prime}\overline{A}}X_{A\overline{B^{\prime}}}%
^{I})(X_{B^{\prime}\overline{A}}X_{A\overline{B^{\prime}}}^{J})+\mu_{IJ}%
^{(2)}(X_{A\overline{B}}X_{B\overline{A}}^{I})(X_{A\overline{B}}%
X_{B\overline{A}}^{J})\eoll
&  +\mu_{IJ}^{(3)}(X_{A\overline{B}}X_{B\overline{A}}^{I})(X_{B^{\prime
}\overline{A}}X_{A\overline{B^{\prime}}}^{J})+\kappa_{IJ}^{(1)}X_{B_{1}%
^{\prime}\overline{A_{1}}}X_{A_{1}\overline{B_{2}^{\prime}}}^{I}%
X_{B_{2}^{\prime}\overline{A_{2}}}X_{A_{2}\overline{B_{1}^{\prime}}}^{I}\eoll
&  +\kappa_{IJ}^{(2)}X_{A_{1}\overline{B_{1}}}X_{B\overline{_{1}A_{2}}}%
^{I}X_{A_{2}\overline{B_{2}}}X_{B\overline{_{2}A_{1}}}^{I}+\widehat{\kappa
}_{IJ}X_{B\overline{A}}^{I}X_{A\overline{B^{\prime}}}^{J}X_{B^{\prime
}\overline{A}}X_{A\overline{B}}\nonumber \text{.}%
\end{align}
Invariance under the path condition now requires that the $g^{2}\times g^{2}$
matrix of couplings $\alpha_{ijkl}$ must be symmetric.

\begin{figure}
[t]
\begin{center}
\includegraphics[
height=2in,
width=5in
]%
{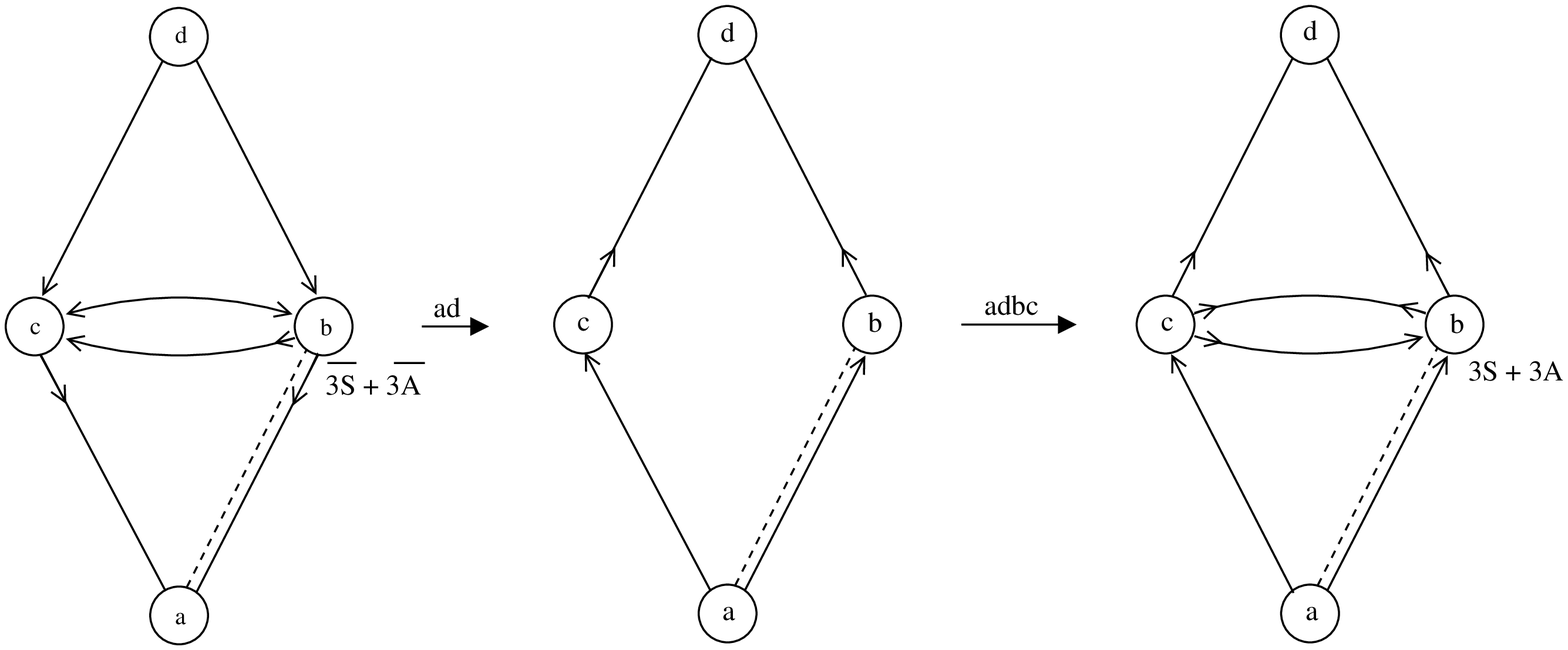}%
\caption{Depiction of the sequence of Seiberg dualities leading from $q_{+}$
to $adbcadq_{+}$. \ At the next stage of the cascade, node $c$ dualizes which
generically eliminates the Higgs pair from the quiver theory.}%
\label{qplussequence}%
\end{center}
\vspace{-4mm}
\end{figure}

\subsection{Candidate Cascade Sequences}\label{CANDIDATES}

As argued previously, there is an essentially unique way to extend the
MQSM\ by a single node so that a cascade can always proceed. \ Even so, the
distinct sequence of Seiberg dualities adopted by a candidate cascade will in
general depend on the initial values of the ranks and coupling constants in
the UV. \ Moreover, while many candidate cascades will eventually repeat back
to the quiver theory $q_{+}$, there are in principle sequences which can fail
to return to the quiver theory $q_{+}$. \ To classify possible cascades, note
that whenever node $c$ dualizes when it is attached to node $b$, it will
either generate or destroy a Higgs pair in the corresponding quiver. \ Letting
$c_{+}$ and $c_{-}$ denote dualizations which respectively create or destroy
such a pair, we note that $c_{+}$ and $c_{-}$ always appear together in the
combination:%
\begin{equation}
I\equiv(ad)c_{+}(ad)c_{-}%
\end{equation}
where as before, a pair of dualizations enclosed by brackets commute.
\ Beginning from the quiver theory $q_{+}$, any candidate sequence of cascades
is necessarily of the form:%
\begin{align}
S_{1} &  \equiv\cdot\cdot\cdot G_{n}^{(i_{n})}\cdot\cdot\cdot IG_{2}^{(i_{2}%
)}IG_{1}^{(i_{1})}q_{+}\eol
S_{2} &  \equiv\cdot\cdot\cdot G_{n}^{(i_{n})}\cdot\cdot\cdot G_{2}^{(i_{2}%
)}IG_{1}^{(i_{1})}Iq_{+}\nonumber%
\end{align}
where each $G_{n}^{(i_{n})}$ corresponds to one of three possible intermediate
sequences:%
\begin{align}
G_{n}^{(1)} &  =(ad)b\eoll
G_{n}^{(2)} &  =(ad)c\\[0mm]
G_{n}^{(3)} &  =(ad)(bc)\nonumber \text{.}%
\end{align}
In order for a subsequence of $I$'s and $G_{i}$'s to eventually repeat back to
$q_{+}$, its conjugate with all arrow directions reversed, or a quiver theory with the r\^{o}les of $a$ and $d$
interchanged, nodes $b$ and $c$ must attach to node $a$ with the same
orientation. \ A similar statement holds for connections to node $d$. \ Due to
the fact that each $I$ and $G^{(3)}$ subsequence preserves the relative
orientation of nodes $b$ and $c$ in terms of how they attach to nodes $a$ and
$d$, only $G^{(1)}$ and $G^{(2)}$ can change this relative orientation. \ We
therefore conclude that a sequence of dualities will only repeat back to
$q_{+}$ or a quiver theory whose connectivity is physically indistinguishable
when the total number of number of $G^{(1)}$ and $G^{(2)}$ dualities is even.
\ To summarize, a sequence beginning at $q_{+}$ will or will not repeat back
to $q_{+}$ or a physically indistinguishable quiver theory when:%
\begin{align}
\text{Repeat}\text{: } &  \#G^{(1)}+\#G^{(2)}\in2%
%TCIMACRO{\U{2124} }%
%BeginExpansion
\mathbb{Z}
%EndExpansion
\eol
\text{Not Repeat} &  \text{: }\#G^{(1)}+\#G^{(2)}\notin2%
%TCIMACRO{\U{2124} }%
%BeginExpansion
\mathbb{Z}
%EndExpansion
\text{.}\nonumber %
\end{align}
This is essentially a consequence of the fact that dualizing node $c$ in
$q_{+}$ destroys the Higgs pair. \ Indeed, assuming the superpotential has
been sufficiently tuned so that dualizing node $c$ does not change the number
of Higgs pairs, it is possible to show that the cascade always eventually
repeats. \ With this in mind, note that although a given sequence may not
return to the quiver theory $q_{+}$, up to the presence of a single Higgs
pair, a similar analysis establishes that the \textit{chiral} matter content
always eventually repeats.

We now consider candidate cascade sequences with additional vector-like pairs present.  As will be shown in more detail in section \ref{IGNITING}, the observed values of the couplings imply that additional vector-like pairs must in fact be added in
order for the cascade to begin dualizing at sub-Planck scales.  Assuming that all
other additional vector-like pairs are always integrated out, the cascade can proceed when an equal number of vector-like pairs attach nodes $a$ to $c$ and $c$ to $d$.  We now show that when the number of vector-like pairs is an even number, this addition will not disrupt the Higgs regeneration mechanism described previously.  To this end, consider dualizing node $c$ of the quiver theory $q_{+}$ with the Higgs pair deleted.  The number of vector-like pairs of mesons created now yields $(m+g)g$ pairs between nodes $a$ and $b$ and $mg$ pairs between nodes $d$ and $b$.  In order for the Higgs regeneration mechanism to proceed as before, we must treat the extra vector-like pairs in a similar fashion to the other bifundamentals between nodes $a$ and $c$.  Further, the mechanism can only work when $(m+g)g$ is an odd number so that $g$ is odd and $m$ is even.  Indeed, because $mg$ is an even number, a generic meson mass matrix will lift all pairs between nodes $d$ and $b$ and leave a single pair between $a$ and $b$.  When $m$ is an odd number, the same mechanism would create a vector-like pair between nodes $d$ and $b$ instead of between nodes $a$ and $b$.  Next dualizing nodes $a$ and $d$, the number of bifundamentals between nodes $b$ and $c$ will now double.  This has the consequence that all subsequent dualizations of node $c$ will always produce an even number of vector-like pairs between nodes $a$ and $b$ and between nodes $d$ and $b$.  We therefore conclude that $m$ must be an even number in order for the Higgs regeneration mechanism to remain intact.

%(XXX Try analyzing the $U(1)_{Y}$ of the model with incorrect hypercharge?
%\ How is it related to the standard model? XXX)%
%TCIMACRO{\FRAME{ftbpFU}{5.1084in}{2.0972in}{0pt}{\Qcb{Depiction of the
%sequence of Seiberg dualities leading from $q_{+}$ to $adbcadq_{+}$. \ At the
%next stage of the cascade, node $c$ dualizes which generically eliminates the
%Higgs pair from the quiver theory.}}{\Qlb{qplussequence}}{qplussequence.eps}%
%width 5.1084in;  height 2.0972in;  depth 0pt;  original-width 10.6692in;
%original-height 4.3613in;  cropleft "0";  croptop "1";  cropright "1";
%cropbottom "0";  filename 'qplussequence.eps';file-properties "XNPEU";}} }%
%BeginExpansion
%EndExpansion

\section{Ultraviolet Unification of Higgsing and Confining
Scenarios}\label{Unification}

%EndExpansion

Although in comparison with the rather intricate cascade combinatorics of the
confining scenario, there is an essentially unique path of dualization for the
Higgsing scenario. \ At intermediate stages, however, the quiver topology of
the confining scenario is nearly identical to that of the Higgsing scenario.
\ Indeed, the only difference between the quiver of the LR\ cascade with a
single Higgs pair and the quiver theory $adq_{+}$ is that the LR\ model has
one additional vector-like pair connecting nodes $d$ and $b$.  We now explain how the two scenarios
unify in the UV of the cascade.

\begin{figure}
[t]
\begin{center}
\includegraphics[
height=2.2009in,
width=3.6521in
]%
{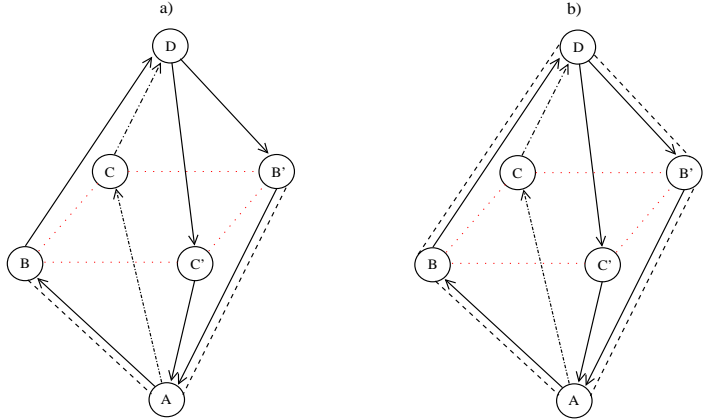}%
\caption{An intermediate stage of the cascade for the covering
quiver in the confining (a) and Higgsing (b) scenarios. \ These two
scenarios unify at higher energy. \ Indeed, the only difference between the
two quivers is the presence (b) or absence (a) of an extra vector-like pair (dashed line).}%
\label{uvunification}%
\end{center}
\vspace{-4mm}
\end{figure}

To this end, note that introducing a $\mu$ term for one of the two vector-like
pairs of the LR\ model would break the $%
%TCIMACRO{\U{2124} }%
%BeginExpansion
\mathbb{Z}
%EndExpansion
_{2}$ symmetry of the model. \ Nevertheless, a cascading structure will remain
intact, although the MQSM\ will now arise via the confining scenario.
\ Conversely, when an ostensibly massive vector-like pair which would normally
be integrated out survives as an anomalously light vector-like pair in the
quiver theory $adq_{+}$, the resulting cascade will now proceed via a
LR\ cascade which connects to the MQSM\ at the bottom of the cascade via the
Higgsing scenario.%
%TCIMACRO{\FRAME{ftbpFU}{3.6521in}{2.2009in}{0pt}{\Qcb{Depiction of an
%intermediate stage of the cascade for the covering quiver in the confining (a)
%and Higgsing (b) scenarios. \ As shown, these two scenarios unify at higher
%energy. \ Indeed, the only difference between the two is the presence (b) or
%absence (a) of an extra vector-like pair of matter which is denoted by a
%dashed line from nodes D to B and nodes D to B'.}}{\Qlb{uvunification}%
%}{uvunification.eps}{\special{ language "Scientific Word";  type "GRAPHIC";
%width 3.6521in;  height 2.2009in;  depth 0pt;  original-width 8.0704in;
%original-height 4.8525in;  cropleft "0";  croptop "1";  cropright "1";
%cropbottom "0";  filename 'UVunification.eps';file-properties "XNPEU";}} }%
%BeginExpansion

\section{The Bottom of the Cascade}\label{Bottom}

In both the Higgsing and confining scenarios, the last steps of the cascade
are in general different from the other intermediate stages due to the
difference in the form of the dual superpotential when a gauge group confines. \ Indeed, just prior to the final step where the four node quiver theory of
the Higgsing scenario transitions to a three node quiver, the
Higgsing scenario contains a $U(1)$ factor at node $b$. \ From the perspective of the
duality cascade, the non-abelian factor at node $b$ has confined without
breaking chiral symmetry. \ In the purely confining scenario, the ranks of the
non-abelian factors at nodes $d$ and $b$ have both depleted to zero. \ This
implies that both gauge group factors have confined. \ In this section
we describe in further detail how the form of the superpotential of
the MQSM arises from the last few duality steps realized at the bottom of the cascade.
We also comment on how the extra node $d$ may play a r\^{o}le in initiating supersymmetry
breaking at the bottom of the cascade.

\subsection{Higgsing Scenario}

The duality cascade described in section \ref{alrsymmcascade} reduces in the infrared
to a supersymmetric, but otherwise minimal, left-right symmetric
extension of the Standard Model with left-right gauge group $USp(2)_L \times USp(2)_R \simeq
SU(2)_L \times SU(2)_R$.  The phenomenology of such left-right symmetric models
has been quite extensively studied in the literature \cite{MohapatraLR,AulakhMelfoLR,AulakhRasinLR,ChakoMohapatra,MohapatraBook}.
It is beyond the scope of this paper to elaborate on the phenomenological strengths and weaknesses
of this class of scenarios.  To this end, we shall limit our discussion to scenarios where the bottom of the cascade realizes a
phenomenological LR symmetric model.
In this discussion, we start in the IR and work our way up towards the UV.
To set the stage, we start with the left-right model itself and how
its superpotential connects with that of the MQSM.
We then proceed to discuss the first two duality cycles.

\subsubsection{Left-Right Symmetry Breaking}

We now discuss how the superpotential of the LR quiver in figure \ref{LRQuiver}a descends to
the superpotential of the MQSM quiver indicated in figure \ref{LRQuiver}b.
The left-right quiver theory has the form of a diamond, with oriented lines corresponding
to the quarks and leptons, and two unoriented lines corresponding to the Higgs fields
that will break the left-right and electro-weak symmetry.
Left-right symmetry breaking may occur when the superpotential for the
Higgs scalars $H_3,H_4$ connecting nodes $b$ and $d$ takes the form given by equation
(\ref{higgsright}).  In one supersymmetric minimum of the corresponding effective potential, the extra Higgs fields attain the
vacuum expectation values:%
\begin{equation}
 H_3 =\left(
       \begin{array}{c}
         a_1 \\
         0 \\
       \end{array}
     \right)\, ,
\qquad H_4 =\left(
       \begin{array}{c}
         0 \\
         a_1 \\
       \end{array}
     \right).
\end{equation}
Thus 3 gauge bosons eat 3 chiral
fields and obtain a mass of order
\begin{equation}
 \Lambda_{LR} \equiv g_{U(1)_b} \, a_1.
\end{equation}
We are also left with an extra singlet
field $S$ corresponding to changing $H_3,H_4$ in the direction of the vev.
Because the left-right symmetric quiver contains no oriented triangles, only quadratic and quartic
terms appear to lowest order:
\begin{eqnarray}\label{WLR}
W_{LR} &=& \mu_L \, H_u H_d + \mu_R\,  H_3 H_4  \\[2mm]
& & + A_{IJKL}\, Q^{I}Q^{J}Q^{K}Q^{L}+B_{IJ}{}^{kl}\, Q^{I}Q^{J}%
X_{k}X_{l}+C^{ijkl}\,X_{i}X_{j}X_{k}X_{l} + \mathcal{O}_{6}.
\nonumber
\end{eqnarray}
In the above, $\mathcal{O}_{6}$ denotes all degree six and higher gauge invariant combinations of fields and
$X$ is a collective label for $L$ or $H$, the fields that are charged under node $b$.
We will discuss in the next subsection how the LR cascade generates this superpotential.
An important point in this regard is that all couplings
in $W_{LR}$ are dimensionful.  This implies that all of their values are expected to be set by the common
the mass scale $\Lambda_1$ of the first step up the cascade.
As we will see, to generate an appropriate seesaw mechanism, the scale $\Lambda_1$ needs to be quite high --  above $10^{11}$ TeV or so.

The MQSM superpotential follows by replacing $H_3,H_4$ by its vev.
%The coefficients in the superpotential depend on the ratios of scales % \begin{equation}\label{ABCscales} A \sim \frac{1}{M_2}, \qquad B \sim \frac{1}{\Lambda_1}, \qquad C \sim \frac{M_2}{\Lambda_1^2}.  \end{equation}
% Here $\Lambda_1$ is the scale of the last Seiberg duality at the bottom of the cascade, and $M_2$ should be close to the scale of the next-to-last Seiberg duality, $M_2 \sim \Lambda_2$.
%The potential for $H_3,H_4$ %\begin{equation} \mu_R \, H_3 H_4 + C \,(H_3 H_4)^2 %\ \sim \ \mu_R\,  S^2 + \ldots \end{equation} % can be arranged such that it has a non-zero minimum at $a_1 \sim \Lambda_1 M_2^{-1/2} \mu_R^{-1/2}$.  All of this is consistent with $\mu_R$ being close to but less than $\Lambda_{LR}$, and about two orders of magnitude between $\Lambda_{LR}$ and $\Lambda_1$ and between $\Lambda_1$ and $M_2$.  We will see later that this also meshes with the evolution of gauge couplings.
The Yukawa couplings arise from the quartic terms
\begin{equation}
\begin{matrix}
  Q_L Q_R H_3 H_d & \to &  a_1 Q_L D H_d, \qquad &  L_L L_R H_3 H_d & \to & a_1 L_L
E H_d  \eol
Q_L Q_R H_u H_4 & \to & {a_1 }
Q_L U H_u, \qquad  &   L_L L_R H_u H_4 & \to & {a_1} L_L N
H_u
\end{matrix}
\end{equation}
These Yukawa couplings are naturally of the right overall magnitude, provided the scale $\Lambda_{LR}$ of left-right symmetry
breaking is sufficiently close to the scale $\Lambda_1$ set by the first Seiberg duality up the cascade.
To reproduce the correct mass spectrum for all the Standard Model particles,
the superpotential of the left-right model must accommodate the required
mass hierarchy. Here we will not address whether this is a natural requirement,
except to note that the model naturally includes
a Majorana mass term for the sterile neutrinos
\begin{equation}
%{\frac{M_2}{\Lambda_1^2}}
L_{R}H_{4}L_{R}H_{4}\rightarrow %{\frac{a_{1}^{2}M_2}{\Lambda_1^2}}
a_1^2 N_{R}N_{R} \, . \nonumber%
\end{equation}
Thus if we choose a high scale for $\Lambda_{LR}$, the theory
automatically incorporates a seesaw mechanism.
Further we get a second contribution to the $\mu$-term:
\begin{equation}
 H_u H_d H_3 H_4 \to {a_1^2}\, H_u H_d \, . \nonumber %+ {a_1}\, H_u H_d S
\end{equation}
In the context of this LR model, the $\mu$ problem is the question why this contribution cancels
to such a large degree of accuracy with $\mu_L$.
%However, note that $S$ should be interpreted
%as the singlet field of the NMSSM, so there is conceivably a dynamical solution
%to the $\mu$-problem in our model.

The superpotential is restricted
to be invariant under the matter parity assignments $R(Q)=R(L)=-1,\ R(H)=+1$.
This invariance forbids the following operators
\begin{equation}
\begin{matrix}
 Q_L Q_R H_3 L_L & \to &  a_1\, Q_L D L_L \qquad \qquad \qquad &
 L_L L_R H_3 L_L & \to &  a_1\, L_L E L_L \eol
 L_L H_3 H_u H_4 & \to & {a_1^2 }\, L_L  H_u %+{a_1 }\, L_L H_u S
\qquad \quad \qquad &
 H_u H_d L_R H_4 & \to & {a_1 }\, H_u H_d N \eol
\end{matrix}
\end{equation}
which confirms that the matter parity of the LR model directly descends to the usual
matter parity of the MSSM. For more details on the phenomenology of the LR model,
we refer to the standard literature on this subject.

\subsubsection{The First Duality: (De)Confinement of Node $b$}

In this subsection we discuss the relation between the superpotential in the LR quiver
and the superpotential in the first step up the cascade.
The cascade terminates (towards the IR) when the $SU(9)$ gauge group on node $b$
confines at a scale $\Lambda_1$. Node $b$ has $N_F=10$, and therefore if $N_c=9$,
it confines without chiral symmetry breaking.  The quiver at this step is as shown in
figure \ref{LRcascade} with $(2N_a,N_b,N_c) = (2,9,3)$.
The gauge coupling for the $SU(9)_b$ is a priori unrelated to that of  $U(1)_b$,
and the scale $\Lambda_1$ is therefore a free parameter in the cascade. As we will
see shortly, it is natural to chose $\Lambda_1$ slightly above the scale $\Lambda_{LR}$
where the left-right symmetry gets restored. We will refer to the theory
above the scale $\Lambda_1$ as the electric theory, and to the one below $\Lambda_1$ as the
magnetic theory.

In the electric theory, we denote the fields charged under node $d$ by $Q$ and the
fields charged under node $b$ by $l,h$.
We impose invariance under the matter parity transformation
$R(Q)=R(l)=-1$ and $R(h)=+1$.
Via the duality and LR symmetry breaking, this descends to the standard
matter parity of the MQSM.
The superpotential up to quartic terms is of the schematic form
\begin{eqnarray}\label{SD1electricsuper}
W & = &  \kappa_L\, h_u h_d + \kappa_R\,  h_3 h_4
\\[2mm]
& &
+ \alpha_{IJKL}\,Q^{I}Q^{J}Q^{K}Q^{L}+\beta_{IJkl}\,Q^{I}%
Q^{J}\mathrm{x}^{k}\mathrm{x}^{l}+\gamma_{ijkl}\,\mathrm{x}^{i}\mathrm{x}%
^{j}\mathrm{x}^{k}\mathrm{x}^{l}%
\nonumber
\end{eqnarray}
where $\mathrm{x}$ is a collective label for the fields charged under node $b$, that is $l$ or $h$.
The quartic couplings are of order $1/M_2$, where $M_2$ will later be related
to scales higher up in the cascade.
Around the scale $\Lambda_1$, the $SU(9)$ gauge sector confines and we must replace the superpotential
(\ref{SD1electricsuper}) by its magnetic dual.
The chiral fields in the dual theory are given by
\begin{equation}
m^{ij}=\mathrm{x}^{i}\mathrm{x}^{j}\,,\qquad\qquad X_{i}={\frac{1}%
{\Lambda_1^{N_{c}-1}}}\;\epsilon_{ij_{1}\ldots j_{N_{c}}}\mathrm{x}^{j_{1}%
}\ldots\mathrm{x}^{j_{N_{c}}}.%
\end{equation}
The dual baryons are the fields of the left-right symmetric theory discussed in the previous
subsection. The dual mesons are additional vector-like matter. We did not consider the mesons
as part of the low energy theory, because for generic superpotential, they
acquire a high scale mass and need to be integrated out to obtain the low energy
superpotential.

For $N_{f}=N_{c}+1$, the magnetic superpotential is generally given
by
\begin{equation}
W={\frac{1}{\Lambda_{1}}}m^{i{j}}X_{i}{X}_{{j}}\;-{\frac{1}{\Lambda
_{1}^{2N_{c}-1}}}\det(m).
\end{equation}
Because the group $USp(2)_{L}\times USp(2)_{R}$ embeds in the $SU(N_{f})\times
SU(N_{f})$ flavor group, the expression $\det(m)$ is gauge invariant and
makes sense. This $\det(m)$ term in the superpotential is of higher order,
and irrelevant for the low energy dynamics of the magnetic model when the
anomalous dimensions of the dual meson fields are suitably small.
The electric superpotential (\ref{SD1electricsuper}) thus descends to the
following magnetic superpotential
\begin{eqnarray}
W & =& \kappa_L\,  m_{ud} + \kappa_R \, m_{34}
\\[2mm]
& &
+ \ \alpha_{IJKL}\,Q^{I}Q^{J}Q^{K}Q^{L}+\beta_{IJkl}\,Q^{I}Q^{J}%
m^{kl}+\gamma_{ijkl}\,m^{ij}m^{kl}+{\frac{1}{\Lambda_{1}}}\,m^{ij}X_{i}X_{j}.
\nonumber
\end{eqnarray}
It is clear that with generic quartic terms in the electric theory, we can
lift all the mesons. Assuming a K\"{a}hler potential $K\sim m\bar{m}%
/\Lambda_{1}^{2}$, the masses of the mesons are of order $\Lambda_{1}%
^{2}/M_{2}$. We assume that this mass is large compared to $\Lambda_{LR}$.
%Although we cannot determine the scale $M_2$ exactly, even with knowledge of the rest of the cascade, we would not expect it to be so large the masses of the mesons
%lie below $\Lambda_{1}$.
We can then safely integrate out the mesons to get the effective
Lagrangian for the LR model. In this way, one obtains
the superpotential as given in equation (\ref{WLR})
% \begin{eqnarray} W_{LR} &=& \mu_L\, H_u H_d + \mu_R\, H_3 H_4 \eol & & + A_{IJKL}\,Q^{I}Q^{J}Q^{K}Q^{L}+B_{IJ}{}^{kl}\,Q^{I}Q^{J}% X_{k}X_{l}+C^{ijkl}\,X_{i}X_{j}X_{k}X_{l} + {\rm higher\ order} \eol \end{eqnarray}% %
where
\begin{eqnarray}
A_{IJKL} =\alpha_{IJKL}+\beta_{IJkl}\beta_{klmn}\gamma
^{klmn}\,, \qquad & & \nonumber \\[-5mm]
& &  \
\mu_L = \gamma^{udud}\frac{ \kappa_L }{\Lambda_1} + \gamma^{ud34}\frac{
\kappa_R }{\Lambda_1} \nonumber\\[-3mm]
\label{wmatch1}
B_{IJ}{}^{kl}  ={\frac{1}{\Lambda_{1}}}\,\beta_{\operatorname{Im}%
Jn}\gamma^{mnkl} \qquad \qquad \qquad \ \ & & \\[-3mm]
  & & \ \mu_R = \gamma^{34ud}\frac{ \kappa_L }{\Lambda_1} +
\gamma^{3434}\frac{ \kappa_R }{\Lambda_1} \nonumber \\[-5mm]
C^{ijkl}   =\;{\frac{1}{\Lambda_{1}^{2}}}\,\gamma^{ijkl}\,. \qquad
\qquad \ \ \ \qquad  \qquad & \nonumber
\end{eqnarray}
%which should be compared with (\ref{WLR}).

%%%

\subsubsection{Second Step: Dualizing Nodes $a$ and $d$}

We now to the relation between the superpotentials before and after
the second step up the cascade. \ In the theory above the scale $\Lambda_1$, the two $USp(2)$ nodes
$a$ and $d$ have a total of $54$ flavors.
This large number implies that both nodes acquire a large $\beta$ function
and hit strong coupling at a scale $\Lambda_2$ that is just a couple of orders
of magnitude larger than $\Lambda_1$. In the next section we shall provide approximate values for these energy scales.
At the strong coupling scale $\Lambda_2$,
both nodes $a$ and $d$ undergo a Seiberg duality. Since we assume that
left-right symmetry is restored, the two nodes dualize simultaneously.  In fact, the presence of additional tensor matter
will greatly accelerate the running of the $b$ and $c$ nodes so that nodes $a$ and $d$ must dualize at very similar scales so that the theory does not prematurely reach a Landau pole at a quiver node with tensor matter.  We shall return to this point in section \ref{CONFEN}.
The theory above $\Lambda_2$ is the new electric theory.
%with gauge group $USp(48)_a \times U(9)_b\times U(3)_c \times USp(48)_d$.
Its quiver diagram is as shown in figure \ref{LRcascade}, with $(2N_a,N_b,N_c)= (48,9,3)$.
The theory valid between $\Lambda_2$ and
$\Lambda_1$ (the previous electric theory) now becomes the new magnetic theory,
describing the effective dynamics of the dual quarks produced by the confinement
transition of the two $USp(48)$ gauge factors.
The electric superpotential is of the form
\begin{equation}
\label{Wstart}
W = \lambda_L\, Z_3 Z_4 + \lambda_R\, Z_u Z_d + a^{IJKL}\, Y_I Y_J Y_K Y_L + b^{ijkl}\, Z_i Z_j Z_k Z_l
+ c^{IJkl}\, Y_I Y_J Z_k Z_l
\end{equation}
where fields charged under node $c$ are denoted as $Y$ and fields charged
under node $b$ are denoted as $Z$. Under matter parity, the vector-like fields $Z_3,Z_4,Z_u,Z_d$
have charge $+1$ and all other fields have charge $-1$.
The scale of the couplings $a,b,c$ is of order $1/M_3$ with $M_3$ the scale
determined by the physics of the next duality step up in the cascade.

The Seiberg duality comes with an extra new scale m, that is not determined
by the scale $\Lambda_{el} = \Lambda_2$ where the electric theory becomes strongly
coupled. This scale m is related to the % electric and
 magnetic strong coupling scale $\Lambda_{mag}$  %s via \cite{IntriligatorPouliot}
%
%$$\Lambda_{el}^{3 N_c/2 + 3  - N_f/2} \Lambda_{mag}^{3(N_f - N_c-2)/2 -N_f/2} = 16 (-1)^{N_f/2 - N_c/2-1}
%\mathrm{m}^{N_f/2},$$
that influences the way the superpotential
feeds through the duality.
The magnetic superpotential is of the form
\begin{eqnarray}
W &=& \lambda_L\, z_{ud} + \lambda_R\, z_{34} + a^{IJKL}\, y_{IJ} y_{KL} + b^{ijkl}\, z_{ij} z_{kl} +
c^{IJkl}\,  w_{Jk}w_{Li}  \eol
& & + \frac{1}{\mathrm{m}}\, y_{ij} Q^i Q^j + \frac{1}{\mathrm{m}}\, z_{ij} \mathrm{x}^{i}\mathrm{x}^j
+ \frac{1}{\mathrm{m}}\, w_{Ij} Q^{I}\mathrm{x}^j +  \frac{1}{\mathrm{m}}\, w_{iJ} \mathrm{x}^i Q^{J}
\end{eqnarray}
The mesons are defined as
\begin{equation}
y_{IJ} = Y_I Y_J, \quad z_{ij} = Z_i Z_j, \quad
w_{Ij} = Y_I Z_j, \quad w_{iJ} = Z_i Y_J.
\end{equation}
The mesons have a mass of order $\mathrm{m}^2/M_3$.
Integrating them out yields
the superpotential given in (\ref{SD1electricsuper})
% \begin{equation} W = \kappa_L\, h_u h_d + \kappa_R\,  h_3 h_4 + \alpha_{IJKL}\,Q^{I}Q^{J}Q^{K}Q^{L}+\beta_{IJkl}\,Q^{I}% Q^{J}\,\mathrm{x}^{k}\mathrm{x}^{l}+\gamma_{ijkl}\,\mathrm{x}^{i}\mathrm{x}% ^{j}\mathrm{x}^{k}\mathrm{x}^{l}% \end{equation}
%
where
\begin{eqnarray}
 \alpha_{IJKL}   = \frac{a_{IJKL}}{\mathrm{m}^2} \qquad & &
 \nonumber \\[-5mm]
& & \qquad \kappa_L = b_{udud}\frac{ \lambda_L }{\mathrm{m}} +
b_{ud34}\frac{ \lambda_R }{\mathrm{m}} \nonumber\\[-2mm]
\beta_{IJkl} = \frac{c_{IJkl}}{\mathrm{m}^2} \ \qquad  & & \\[-2mm]
& & \qquad
\kappa_R = b_{34ud}\frac{ \lambda_L }{\mathrm{m}} + b_{3434}\frac{
\lambda_R }{\mathrm{m}} \nonumber \\[-5mm]
\gamma_{ijkl}  = \frac{b_{ijkl}}{\mathrm{m}^2}. \ \qquad & & \nonumber
\end{eqnarray}
These formulas summarize the matching relations %of the superpotential
at the second duality step. Combined with the matching rules
(\ref{wmatch1}) at the first duality step
and the formulas of left-right
symmetry breaking, they prescribe how the coefficients in the original
superpotential (\ref{Wstart}) descend to various couplings
of the MQSM. It would be worthwhile to investigate this dictionary in
more detail. Undoubtedly one will find that the theory will
need to satsify some stringent constraints. Whether or not these
constraints can be satisfied with reasonable UV initial conditions
 will determine whether this LR cascade scenario has a chance
of being phenomenologically viable.

\subsection{Confining Scenario}
Whereas the combinatorics of the intermediate stages of the confining scenario are more intricate than those of the Higgsing scenario,
the analysis near the bottom of the cascade is comparatively less involved.  Indeed, because the end of the cascade terminates when nodes
b and d confine, it is only necessary to analyze the contributions to the superpotential from terms specific to when a given gauge group confines.  Because each factor must confine without breaking chiral symmetry\footnote{Strictly speaking,
because fields charged under the fundamental of a $USp$ factor are non-chiral, this terminology (while standard) is somewhat imprecise.
  By confinement without chiral symmetry breaking we shall mean a confining theory such that the origin of the classical moduli space also lies within the quantum moduli space.}, the form of the superpotential will contain terms schematically of the form $BM\bar{B}-detM_{b}$ for node $b$ and $PfM_{d}$ for node $d$ where the subscript on $M$ denotes a generic meson field associated with a given gauge group and $B$ denotes a baryon operator.  In these expressions, the number of generations and the non-zero rank of the neighboring flavor groups imply that these terms are of very high degree in comparison to the higher order quartic terms which we have retained in order to analyze how vector-like pairs develop a mass.  Assuming that the cascade soon enters a regime where the above operators are not dangerous irrelevant, we may ignore their contribution to the low energy dynamics of the cascading theory.  At higher stages of the cascade dualizing nodes $b$ and $d$ would produce additional cubic terms in the superpotential.  These terms do not contribute to the mass terms necessary for lifting vector-like pairs of mesons created when nodes $c$ and $a$ dualize for the last time.

\subsubsection{Igniting the Confining Scenario}\label{IGNITING}

Proceeding from the IR\ to the UV, the above analysis demonstrates that in the
Higgsing scenario the energy scale at which the duality cascade begins can be
arranged to lie below the Planck scale. \ In the most minimal version
of the confining scenario, where the extra node $d$ attaches to the MQSM\ by
purely chiral matter, the first few stages of dualization are already far
beyond the Planck scale. \ To establish this, consider the first stage of
dualization from the IR\ to the UV. \ In the confining scenario, this
corresponds to dualizing node $d$ at some energy scale $\Lambda_{d}$. \ As
argued previously, the unique next step must be the dualization of node $a$.
\ Note, however, that because there is no matter connecting nodes $a$ and $d$,
to one loop order the running of the coupling at node $a$ is identical to that
of the MSSM. \ Because the running of the weak coupling of the MSSM\ only
becomes strongly coupled at energy scales far above the Planck scale, we
conclude that as the energy scale increases, gravitational effects will
dominate before node $a$ dualizes.

In order to lower the first scale of dualization for node $a$, the running of
the couplings must accelerate as the theory proceeds to higher energy scales.
\ In general, this will occur when extra matter has been added to node $a$.
\ It is not possible to add this matter in an arbitrary fashion, however,
because at higher energy scales the cascade must still exhibit a periodic
repeating structure.

We find that there are in fact many ways to accelerate the running of
couplings so that the cascade ignites at sub-Planck scales. \ In this section
we present some examples of how this acceleration can be achieved by adding
extra vector-like pairs which can be treated as additional massless degrees of
freedom at sufficiently high energy scales.

Although the cascade requires that each node with finite gauge coupling must
dualize repeatedly during the cascade, no such restriction holds for flavor
groups with arbitrarily weak gauge coupling. \ Indeed, a possible way to
accelerate the cascade corresponds to adding one additional node to the quiver
theory which attaches to node $a$ by some number of vector-like pairs. \ In
order to preserve the repeating structure of the cascade after node $a$
dualizes, node $d$ must connect to the extra node in exactly the same fashion
as node $a$. \ We note that because the nodes with $U$ type group factors
still cannot dualize with tensor matter present, nodes $a$ and $d$ must
dualize sequentially. \ This implies that the combinatorics of the cascade
proceed as before.

In principle, the presence of an extra node runs counter to the
minimality of the cascade. \ As explained in section
\ref{MOREGENERAL}, a very similar cascading structure exists when
extra vector-like matter has been added symmetrically to the quiver
theory. \ A general concern in this context is that an extra
vector-like pair must not develop a mass and lift as the cascade
proceeds. \ Consider for simplicity the cascade with $m$ vector-like
pairs attached between nodes $d$ and $c$ and nodes $a$ and $c$. \ To
establish that such pairs will not develop a mass as the cascade
proceeds, first recall that the Higgs pair generically lifted from
the low energy spectrum because the presence of additional meson
fields between nodes $b$ and $a$ (or $d$ and $b$) mixed sufficiently
with the Higgs so that \textit{all} vector-like pairs generically
lifted from the resulting low energy spectrum. \ In contrast, at no
stage of the cascade do additional vector-like pairs ever appear
between nodes $d$ and $c$ or between $a$ and $c$. \ It now follows
that as opposed to the Higgs pairs, there is no stage of the cascade
which can produce a mass term for the vector-like pairs. \ Finally,
in order to maintain the Higgs regeneration mechanism discussed in
section \ref{mumu} , we must also preserve the condition that the
anti-symmetric matrix $\varphi$ of equation (\ref{ELECTRICPROBLEM})
continues to have an odd number of rows and columns. \ As explained in section \ref{CANDIDATES},
this requires that the number of additional vector-like pairs must be an even
number.

\subsection{Supersymmetry Breaking at the Bottom of the Cascade}\label{HYBRID}

In this brief section we discuss how supersymmetry breaking could in principle occur in the
context of either the Higgsing or confining scenario.
\ Because a supersymmetric spectrum appears necessary for much of the
cascade to proceed, it is natural to expect that supersymmetry breaking must
occur at a scale below the last dualization. \ As we have seen above, a
cascade can only proceed when at least one additional node has been added to
the MQSM. \ In principle, it is possible to imagine that supersymmetry
breaking also takes place at this extra node and only communicates to the rest
of the particles of the MQSM\ through some higher order loop effects. \ This
would be the case, for example, in direct gauge mediation models
\cite{ArkaniRusselMurayDirect,MurayamaDirect,DimopDirect}. \ In the present
context such a realization will typically be phenomenologically
unviable.\footnote{We thank H. Murayama for discussions on this point.}
\ Indeed, in both the Higgsing and confining scenarios, the extra quiver node
only attaches indirectly to the $SU(2)_{L}$ weak sector. \ At the level of
gaugino masses this would effectively produce a large mass splitting between
the gluino and gluons, while producing an essentially supersymmetric spectrum
for the $W$ bosons and winos.

Although it lies beyond the scope of the present paper, it is in principle
possible that supersymmetry breaking may occur due to the cascade landing in a
metastable non-supersymmetric vacuum. \ Indeed, in many cases such vacua exist
when a small mass term has been added to some subset of vector-like matter
\cite{ISS}. \ We note that such a scenario would naturally accommodate the
proposed method to ignite the confining scenario cascade by adding extra
vector-like matter. \ Further, whereas the metastable vacua of \cite{ISS}
preserved R-symmetry and therefore did not allow the gauginos to develop a
mass, generic higher dimension operators will always be present for the
cascade to proceed. \ As noted in \cite{OoguriDirect}, such generic quartic
terms can deform the ISS\ vacuum to an R-symmetry breaking configuration where
gaugino masses are now possible.

\section{Energy Scales}\label{RUNNING}

\subsection{Energy Scales of the Left-Right Cascade}\label{LREN}

The analysis of the previous section demonstrates that compatibility with the Standard Model imposes expectedly
 stringent boundary conditions on the IR region of the cascade.  In this section we study the
 behavior of the cascade at higher energy scales.
In flowing up the cascade, the ranks of the gauge groups increase very fast.
For example, after just five duality steps, the ranks are
$(2N_a,N_b,N_c) = (48,231,141)$ and after six steps they are $(1526, 231,141)$.
This growth triggers an accelerated sequence of duality steps, that accumulate
over an ever shorter range of scale, and the system quickly hits a duality wall.
The duality wall should appear below the Planck scale.  Indeed, the duality wall may be quite close to the Planck scale but can also
be much lower in energy.  The former option would provide a natural explanation for generating appropriate neutrino
masses via a seesaw mechanism, whereas
the latter option is of course more exciting in the context of potential
observations at the LHC.  At any rate, we have no a priori reason to prefer
either choice of energy scale.  For concreteness in the discussion to follow, we shall assume that
this duality wall arises close to the string scale or Planck scale.
For definiteness, let us assume the wall is placed at $10^{15}$ TeV.  This
UV boundary condition essentially determines the energy scales at which
all the lower duality steps take place.  We now give a rough estimate of these scales.

To obtain the scale dependence of the various couplings along the duality cascade,
it does not suffice to use the perturbative formulas.
The exact beta function of an ${\cal N}=1$ supersymmetric gauge theory with gauge group $G$ reads
\begin{eqnarray}
\label{nsvz}
\beta & = &  \frac{d(4\pi /g^2)}{d \log \mu}
\, =\, \frac{3T(G)
- \sum_{i} T(r_{i})(1- \gamma_i)}{2\pi (1-\frac{g^2}{8\pi^2}
 T(G))}
\end{eqnarray}
where the sum runs over all matter multiplets, and $\gamma_i$ denotes their anomalous dimensions.
Here $T(G) = 3N$ for $SU(N)$ and $3(N+1)$ for $USp(2N)$, and
$T(r)=1/2$ for matter in the (anti-)fundamental representation.
The contribution from the anomalous dimensions and from the non-trivial denominator
in (\ref{nsvz}) can be substantial.

\begin{figure}[th]
\begin{center}
%\resizebox{\textwidth}{!}{
\scalebox{.5}{
\includegraphics[width=\textwidth]{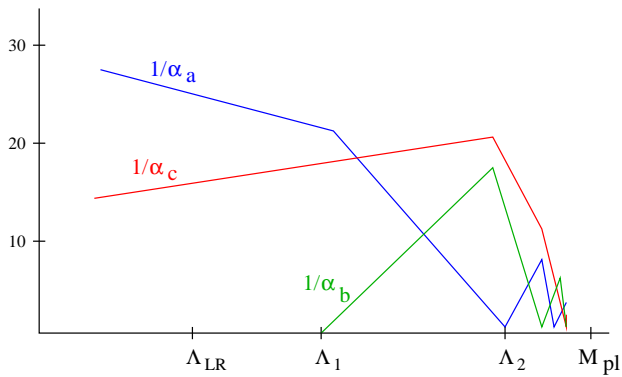}
}
\end{center}
\par
\caption{{A schematic plot of the running couplings of the three non-abelian
gauge groups from $\Lambda_{LR}$ to $M_{Pl}$. In the numeric example, $\Lambda_{LR}
\sim 10^{10}$ TeV. The first two Seiberg dualities occur at scale $\Lambda_1 \sim 10^{11}$ TeV and
$\Lambda_2 \sim 10^{14}$ TeV. The plot shows the acceleration of duality steps and subsequent emergence
of a duality wall close to the Planck scale.  }}%
\label{LRRunning}%
\vspace{-2mm}
\end{figure}

Because we are only interested in a rough estimate of
how fast the duality cascade accelerates in the UV, we shall approximate the above formula
by $\beta = b_0/2\pi$ with
$b_0 = 3N -N_{f}$ for $SU(N)$ with $N_f$ flavors and
$b_0  = 3(N +1)-N_{f}/2$ for  $USp(2N_{c})$ with $N_f$ flavors.\footnote{Here $N_f$ is defined as follows.
%We used the following convention for the number of flavours.
For $SU(N)$, we assume there are an equal number of fundamentals and anti-fundamentals, and $N_{f}$ counts the number
of fundamentals. For $USp(2N)$ the fundamental representation is pseudo-real, and we simply count the total number of fields in this representation.}
Using this estimate, we find that the beta function of node $a$ and $d$ above the scale
$\Lambda_1$ is of order $21/2\pi \sim 3.3$. Assuming that inverse coupling constant $\alpha_a^{-1} = 4\pi/g_a^2$
at the scale $\Lambda_1$ is approximately 25, the second duality takes place at a scale $\Lambda_2$ that is
around 3 orders of magnitude larger than $\Lambda_1$.  A similar estimate shows that
the next scale $\Lambda_3$, where node $b$ again
reaches strong coupling, is at most one order of magnitude above $\Lambda_2$.
The duality wall appears just a little above the scale $\Lambda_{3}$.
In order for the wall to occur at around $10^{15}$ TeV,
we thus find that the scale $\Lambda_1$ must be close to $10^{11}$ TeV or so.
The next scale $\Lambda_{2}$ is then around $10^{14}$ TeV and all other steps take place in the
string regime. Since the left-right symmetry breaking scale needs to be not too far below the first
duality scale $\Lambda_1$, we conclude that a reasonable value for  $\Lambda_{LR}$ is around $10^{10}$ TeV.
A schematic plot of the running of the three non-abelian gauge
couplings over the first five duality stages is shown in figure \ref{LRRunning}.
%%%%

Besides the three non-abelian couplings, the LR cascade also has a gauge coupling
associated with the $U(1)_{B-L}$ gauge symmetry generated by
\begin{equation}
\mathcal{Q}_{B-L}={\frac{1}{2}}(\mathcal{Q}_{U(1)_{c}}-\mathcal{Q}_{U(1)_{b}}).
\end{equation}
This $U(1)_{B-L}$ factor does not actively participate in the Seiberg dualities, and (in our conventions)
its beta function is always negative: $\beta = b_0/2\pi$ with $b_0 =-\sum_{\mathrm{i}}q_{i}^2$.
Given that higher up in the cascade, a large amount of extra matter fields charged under $U(1)_{B-L}$ enter the spectrum, one might
be worried that the corresponding coupling may run into a Landau pole at some sub-Planckian scale. However, %we need to remember that
for the overall $U(1)$ factor inside $U(N)$, fields in the fundamental representation of $SU(N)$
carry $U(1)$ charge $1/N$.  This normalization consistently identifies the $U(1)$ symmetry across Seiberg dualities
because $SU(N)$ baryons will always have $U(1)$ charge $\pm 1$. Hence, in passing through the subsequent cascade steps,
the $U(1)_{B-L}$ charges, and therefore its coupling scale inversely with the rank of the gauge groups.
Thus the cascade naturally avoids the Landau pole.

A disadvantage of the cascade scenario is that it does not predict gauge coupling unification.
Given that the couplings fluctuate wildly near the UV
end of the cascade, any coincidence in the structure of IR
couplings would seem accidental. Even so, it is not unnatural that the
two non-abelian couplings $\alpha_{c}$ and
$\alpha_{a}$ appear to intersect at a scale that is close to
the duality wall. Indeed, both couplings are driven to
weak coupling at the next-to last stage of the cascade.
Gauge coupling unification is the additional coincidence that the
hypercharge coupling meets at the same point. This extra coincidence can perhaps be made natural in slightly less minimal versions of the
LR cascade which connect to the Standard Model via an intermediate Pati-Salam-like
theory, with $U(1)_{B-L}$ and $SU(3)_c$ unified into $SU(4)_c$.  An interesting discussion of gauge coupling unification in D-brane
scenarios can be found in \cite{LustGUTS}.

\subsection{Energy Scales of the Confining Scenario Cascade}\label{CONFEN}

The analysis of the previous section demonstrates that there are a range of
energy scales at which node $b$ can first dualize in proceeding from the IR to the UV.
\ Indeed, once node $b$ deconfines,
the resulting gauge group introduces a large number of additional flavors
which strongly alter the running of the two $USp$ factors at nodes $a$ and $d$.
\ Recall that in the case of the confining scenario the gauge coupling at node
$a$ approached a Landau pole below the Planck scale only after additional
vector-like pairs appeared in the low energy spectrum. \ Turning the
discussion around, if extra vector-like matter appears at suitably low energy
scales, say 1 TeV (perhaps even as a candidate for dark matter) the resulting gauge coupling at node $a$ will suffer from a Landau pole
at a sub-Planckian value. \ In this case, it is especially natural to Seiberg
dualize the theory at energy scales close to this Landau pole.  In this section we will
assume that this is the scale at which new physics will appear.  Depending on the
number of vector-like pairs which appear, the duality wall can be at low or high energy
scales.  For concreteness we keep the duality wall near the Planck scale in the discussion
below.

A priori, lowering the Landau pole of node $a$ would still allow a range of
values over which node $d$ and then node $a$ could dualize in proceeding from the
IR to the UV. \ Note, however, that once node $d$ deconfines a large amount of
additional tensor matter will now alter the running of the gauge coupling at
node $c$. \ This will generically cause the coupling constant for node $c$ to
approach a Landau pole quite rapidly. \ Proceeding from the IR to the UV, the
appearance of this Landau pole is in principle problematic because the Landau
pole for the gauge coupling at node $a$ may in general lie above that of node $c$.

Recall, however, that a gauge coupling which develops a Landau pole
prematurely when additional tensor matter appears cannot arise as part of a
consistent RG\ flow. \ In proceeding from the UV to the IR, nodes with a
sufficient amount of tensor matter do not dualize during the cascade.
\ Nevertheless, the resulting value of the gauge coupling may in general be
too small to match to observed values.

We now show that the scales of dualization for nodes $a$ and $d$ must indeed be
relatively close together so that the Landau pole for the gauge coupling at
node $c$ is at a higher scale. \ As in the previous section, we shall
approximate the running of the couplings by their one loop values. \ To keep
our discussion as general as possible, we shall at first allow the cascade to
ignite via $m$ even vector-like pairs between nodes $d$ and $c$ and nodes $c$ and $a$
and $v$ vector like pairs between nodes $d$ and $E$ and nodes $E$ and $a$, where $E$
denotes a possible extra $U(1)$ flavor node. \ For simplicity, we shall assume
that the mass of these additional vector-like pairs all enter the low energy
spectrum at a scale $\mu=\mu_{new}$. \ At scales $\mu_{new}<\mu<\Lambda_{d}$,
the running of the couplings for the non-abelian gauge group factors is
therefore:%
\begin{align}
{2\pi\alpha}_{c}^{-1}\left(  \mu\right)   &  =\left(  3-2m\right)  \log
\frac{\mu}{\mu_{new}}+3\log\frac{\mu_{new}}{M}+{2\pi\alpha}_{c}^{-1}\left(
M\right) \\
{2\pi\alpha}_{a}^{-1}\left(  \mu\right)   &  =\left(  -1-3m-v\right)
\log\frac{\mu}{\mu_{new}}-\log\frac{\mu_{new}}{M}+{2\pi\alpha}_{a}^{-1}\left(
M\right)
\end{align}
where to first approximation we may identify $M$ with the mass of the Z boson
(91 GeV) and the corresponding couplings as ${2\pi\alpha}_{c}^{-1}\left(
M\right)  \sim53$ and ${2\pi\alpha}_{a}^{-1}\left(  M\right)  \sim181$. \ The
approximate energy scale at which ${\alpha}_{a}$ diverges is therefore:%
\begin{equation}
\Lambda_{a}=\mu_{new}\left(  \frac{\mu_{new}}{M}\right)  ^{-1/(1+3m+v)}%
\exp\left(  \frac{{2\pi\alpha}_{a}^{-1}\left(  M\right)  }{1+3m+v}\right)
\text{.} \label{lambdaA}%
\end{equation}
When $\mu_{new}\sim1$ TeV, the corresponding Landau pole for various values of
$m$ and $v$ are:%
\begin{align}
\Lambda_{a}(m  &  =0,v=5)\sim10^{16}\text{ GeV}\\
\Lambda_{a}(m  &  =2,v=0)\sim10^{14}\text{ GeV}%
\end{align}
so that with only a small amount of additional vector-like matter, the Landau
pole for the $a$ node gauge coupling will be significantly smaller.

Once node $d$ deconfines, a large amount of additional matter will accelerate
the running of the gauge coupling at node $c$ but at one loop order will not
alter that of the gauge coupling at node $a$. \ There are two sources of matter
which accelerate the running of the $c$ node gauge coupling. \ The first is from
meson fields which contain bifundamentals between nodes $d$ and $c$. \ The tensor
matter content once node $d$ deconfines is now $g(1+g+2m)/2$ fields in the
$\overline{A}$ and $g(-1+g+2m)/2$ in the $\overline{S}$ of $U(3)$. \ In
addition, there are $g(g+m)$ fields in the $(\overline{3},\overline{1})$ and
$mg$ in the $(3,\overline{1})$ of $U(3)_{c}\times U(1)_{b}$. \ The second
source of matter corresponds to meson fields which contain bifundamentals
between nodes $d$ and $E$. \ In this case, there are $gv$ fields in the
$(\overline{3},1)$ as well as in the $(\overline{3},\overline{1})$ of
$U(3)_{c}\times U(1)_{E}$. \ Assuming node $a$ has not dualized, the running of
${\alpha}_{c}^{-1}$ accelerates:%
\begin{equation}
{2\pi\alpha}_{c}^{-1}\left(  \mu\right)  = b^{\prime}_{c} \log\frac{\mu}{\Lambda_{d}}+(3-2m)\log\frac{\Lambda_{d}}%
{\mu_{new}}+3\log\frac{\mu_{new}}{M}+{2\pi\alpha}_{c}^{-1}\left(  M\right)
\end{equation}
where $b^{\prime}_{c}=-24-31m-6m^{2}-6v-2mv$ when $g=3$.\footnote{In the formula for $b^{\prime}_{c}$ we have cancelled all vector-like pairs in the dualized quiver theory which have the same matter parity and which generically develop a large mass.  If we simply cancel all vector-like pairs without regard to their matter parity, the numerical factor 24 changes to 21.} \ The corresponding Landau pole is:%
\begin{equation}
{\Lambda}_{c}=\Lambda_{d}\left(  \frac{\Lambda_{d}}{\mu_{new}}\right)
^{-(3-2m)/b^{\prime}_{c}}\left(  \frac{\mu_{new}}{M}\right)
^{-3/b^{\prime}_{c}}\exp\left(  \frac{{-2\pi\alpha}_{c}^{-1}\left(
M\right)  }{b^{\prime}_{c}}\right)  \text{.} \label{lambdaC}%
\end{equation}
Comparing equations (\ref{lambdaA}) and (\ref{lambdaC}), it follows that the
denominators of the fractional exponents are significantly larger in the
contribution to ${\Lambda}_{c}$ so that ${\Lambda}_{c}$ and $\Lambda_{d}$ will
differ by at most only a few orders of magnitude. \ In order to match to the
observed values of ${\alpha}_{a}^{-1}\left(  M\right)  $ and ${\alpha}%
_{c}^{-1}\left(  M\right)  $, the tensor matter created by dualizing node $d$
must quickly pair up with additional matter created by dualizing node $a$ so
that $\Lambda_{a}=\Lambda_{d}$, just as in the left-right symmetric cascade.
\ Indeed, in retrospect it is immediate that the more general Higgsing
scenario cascade will suffer from a similar problem unless nodes $a$ and $d$
dualize at similar energy scales.

When nodes $a$ and $d$ dualize at nearly the same scale, the running of ${\alpha
}_{c}^{-1}\left(  \mu\right)  $ changes to:%
\begin{equation}
{2\pi\alpha}_{c}^{-1}\left(  \mu\right)  = b_{c} \log\frac{\mu}{\Lambda_{d}}+(3-2m)\log\frac{\Lambda_{d}}%
{\mu_{new}}+3\log\frac{\mu_{new}}{M}+{2\pi\alpha}_{c}^{-1}\left(  M\right)
\end{equation}
where $b_{c}=-15-34m-12m^{2}-6v-4mv$ so that the Landau pole for $\Lambda_{c}$ is:%
\begin{equation}
\Lambda_{c}=\Lambda_{d}\left(  \frac{\Lambda_{d}}{\mu_{new}}\right)
^{-(3-2m)/b_{c}}\left(  \frac{\mu_{new}}{M}\right)
^{-3/b_{c}}\exp\left(  \frac{{-2\pi\alpha}_{c}^{-1}\left(
M\right)  }{b_{c}}\right)  \text{.}%
\end{equation}
While there is some freedom in choosing the energy scale $\Lambda_{b}$ at
which node $b$ deconfines, as in the case of the Higgsing scenario, each
subsequent stage of dualization converges rapidly to an accumulation point in
the ultraviolet.

\subsection{The Duality Wall}\label{dualitywall}

The appearance of a duality wall may at first appear
to undermine our rationale for selecting a periodic cascade as a candidate UV extension
of the MSSM.  The wall indeed prevents the cascade from going through
a large number of well-separated duality cycles because each successive duality takes
place in such a short range of scales that the gauge theory gets trapped
in a regime of strong coupling.  We interpret this to mean that a purely field theoretic
interpretation of Seiberg duality becomes ambiguous.
Besides the absence of a weakly coupled limit, the gauge couplings of the system also do
not vary slowly enough to give a meaningful definition of
the infrared degrees of freedom.  In a sense, this is a symptom of the fact that the Planck scale is of finite size.
Indeed, in the limit in which the Planck scale decouples, by suitably tuning the couplings of the field theory the accumulation point
corresponding to the duality wall can be taken to be arbitrarily large.  For this reason it is important to establish that the cascades in question do indeed exhibit a periodic cascading structure which can in principle continue indefinitely.  On the other hand, in realistic applications, the finite size of the Planck scale places an upper bound on the value of the energy scale at the duality wall so that the cascade will in practice rapidly become trapped in a regime of strong coupling.

However, in spite of this complication, the appearance of an accelerated
cascade with a duality wall in fact helps us, in the following sense.
A logarithmic cascade (like the Klebanov Strassler system), while better controlled,
would in fact unfold itself too slowly and fail to produce a large $N$ theory at
sub-Planckian energy. Precisely because the ranks grow quickly in the accelerated
cascade so that the system becomes trapped in a regime of strong coupling, it is
natural to expect a dual description of the ultraviolet behavior of the
cascading gauge theory in terms of a dual closed string theory on a suitable
warped background geometry. Fundamentally, the closed string worldsheets
arise via the planar diagrams of the large $N$ theory.  When the 't Hooft coupling is large,
the direct geometric description in terms of string theory on a classical background
should become better controlled at higher energy scales in the cascade.  Although somewhat different from the perspective we have adopted in this paper, another possibility may be that at large $N$ a theory
of tensionless strings decoupled from gravity, such as little string theory (LST) may also emerge.  Indeed, circumstantial evidence for the appearance of a duality wall in terms of a scale intrinsic to LST has also appeared \cite{FiolWalls}.

\section{Candidate D-brane Configurations}\label{DBRANECONFIGS}

We now discuss candidate D-brane configurations which may realize the cascading theories described above. \ To this end, we first comment on similarities of the single generation model with the del Pezzo 3 brane probe theory. \ As shown in Appendix \ref{Windows}, when $g$ is at least three,
there is no common conformal window for all the quiver nodes of either the Higgsing or confining scenario.
\ We next explain what types of restrictions this imposes on any candidate D-brane configuration which realizes the above cascade.

\subsection{Single Generation Model and Del Pezzo 3 Probe Theory}

\begin{figure}
[ptb]
\begin{center}
\includegraphics[
height=2.7717in,
width=5.1301in
]%
{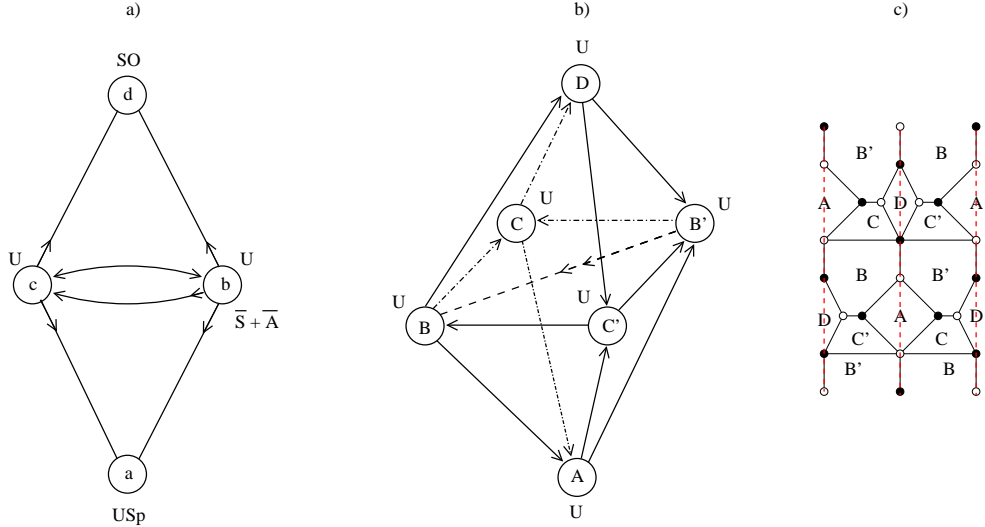}%
\caption{A toric phase of the del Pezzo three brane probe theory
(b), as well as the orientifold theory when a $SO$ factor is present at node
$d$ (a). \ The $\mathbb{Z} _{2}$ action of the orientifold leaves fixed the
dashed red lines of the dimer model (c). \ Deleting the arrows between $B$ and
$B^{\prime}$, the connectivity of the quiver theory is identical to that of an
octahedron.}%
\label{dpthreeandofoldanddimer}%
\end{center}
\end{figure}

Assuming the vector-like Higgs pair has been integrated out by introducing a
suitable mass term, when the number of generations $g=1$, reversing all arrows
attached to the $D$ node of $Q_{+}$ yields a quiver
identical to the del Pezzo 3 brane probe gauge theory. \ See figure
\ref{dpthreeandofoldanddimer} for a depiction of this theory. \ The dimer
model for this theory has been treated in \cite{HananyDimers}. \ Orientifolds
of general dimer models have been treated in \cite{UrangaDimers}. \ Although we omit the details, a similar cascade to that
of the multi-generation model now proceeds provided\footnote{We are assuming
that the orientifolding operation chooses the gauge group type based on compatibility
with the existence of a cascade in the lower theory.  This is natural
if the gravity dual of the cascade exists.} the gauge group is of type
$USp_{a}\times U_{b}\times U_{c}\times SO_{d}$. \ Note, however, that whereas
any node with tensor matter in the multi-generation model would always flow to
weak coupling at low energies, in the single generation model there is far
less tensor matter at nodes $b$ and $c$.%
%TCIMACRO{\FRAME{ftbpFU}{5.1301in}{2.7717in}{0pt}{\Qcb{Depiction of a toric
%phase of the del Pezzo three brane probe theory (b), as well as the
%orientifold theory when a $USp$ factor is present at node d (a). \ The
%$\U{2124} _{2}$ action of the orientifold leaves fixed the dashed red lines of
%the dimer model (c). \ Deleting the arrows between $B$ and $B^{\prime}$, the
%connectivity of the quiver theory is identical to that of an octahedron.}}%
%{\Qlb{dpthreeandofoldanddimer}}{dpthreeandofoldanddimer.eps}%
%width 5.1301in;  height 2.7717in;  depth 0pt;  original-width 10.1832in;
%original-height 5.489in;  cropleft "0";  croptop "1";  cropright "1";
%filename 'DPTHREEandOFOLDandDimer.eps';file-properties "XNPEU";}} }%
%BeginExpansion
%EndExpansion

\subsection{D-Brane Bound States and Non-Conformal Quivers}

General arguments from string theory suggest that at low energies, the
worldvolume theory of a stack of D3-branes probing a geometric singularity
will flow to a non-trivial conformal fixed point. \ For more complicated bound
states of branes, the resulting low energy theory may not possess such a
conformal fixed point. \ In either case, the ranks of the quiver gauge theory
indicate the brane content of the bound state.  Explicit D-brane constructions of
semi-realistic Standard Model vacua have been studied in \cite{UrangaSMthroat}.
Our expectation is that a similar brane construction will realize the cascades studied in this paper.
The analysis of Appendix \ref{Windows} demonstrates
that all of the intermediate quiver theories of both the Higgsing and confining scenario
cascades do not possess a common conformal window for all quiver nodes.  \ This presumably
indicates that some of the nodes should be thought of as composite objects as was
seen in the constructions of \cite{WijnholtVerlindeDPeight,WijnholtGeometry}.
Indeed, it is known that an appropriate Higgsing of the del Pezzo five quiver leads to the MSSM.
One might imagine embedding the MSSM in a del Pezzo five cascade, since
in this case there is in fact a dual supergravity picture. Moreover,
in this case the cascade is approximately conformal, so we can
extend the cascade over many more orders of magnitude than the
cascades we studied previously, which are limited by the existence
of a duality wall.

\section{Distinguishing Features of the MSSM}\label{DISTINGUISH}

While there are in principle many ways in which a given cascade can terminate,
in this section we explain in what sense either cascade scenario distinguishes
the MSSM\ at the bottom of the cascade. \ To this end, consider first the
Higgsing scenario cascade. \ While there is no restriction on the number of
Higgs pairs in this case, a non-trivial feature of the Higgsing scenario is
that the resulting quiver theory at the bottom of the cascade must contain a
$USp(2)$ factor for the weak sector gauge group.  From the perspective of the Standard Model,
this is a consequence of the fact that once the bifundamental matter content of the Standard Model and in particular the multiplicity of $U$ and $D$ type quarks has been specified, cancellation of non-abelian anomalies for the QCD factor requires a $USp(2N)_{a} \times U(N)_{b}$ gauge group structure.  This relation is amplified in the left-right cascade.  \ Indeed, note that at the
final stage of the cascade two of the nodes collapse to a single node. \ In
the breaking process, bifundamentals charged under the $USp_{d}$ node become
the right-handed $U$ and $D$ type quarks. \ In the event that the rank of the
$d$ node gauge group had been larger, additional bifundamental matter would
exist in the Higgsed quiver theory. \ This in turn implies that the gauge
group factor of the extra $d$ node must be $USp(2)$. \ Prior to Higgsing, the non-abelian
anomalies at node $c$ vanish only when the gauge group factor at node $a$ is $USp(2)$.

In the context of the confining scenario, the multiplicity of $U$ and $D$ type quarks relates to the rank of the $USp_{a}$ factor for a different reason. \ Consider a more general class of quiver theories such that nodes $a$ and $d$ attach to additional vector-like pairs. \ Indeed, in order to properly
accelerate the cascade we must allow such vector-like pairs. \ To
maintain the repeating structure of the cascade, these additional pairs must
attach in a symmetric fashion to both nodes $a$ and $d$. \ Recall that at the
bottom of the cascade in the confining scenario, both nodes $d$ and $b$ confine
without breaking chiral symmetry. \ Before nodes $a$ and $d$ dualize for the last
time to reach the quiver theory $q_{+}$, \ the corresponding quiver theory
will be given by $adq_{+}$ and will not contain any tensor matter. \ Even if
node $b$ has already confined, in order to cancel the potential non-abelian
anomaly present at node $c$, the ranks of the $USp$ factors at nodes $a$ and $d$
must be identical. \ As argued previously, node $d$ must confine without
breaking chiral symmetry. \ Prior to confinement, it follows that when the gauge group factor is
$USp(2N_{d})$, precisely $F_{d}=2N_{d}+4$ bifundamentals (counted with
appropriate multiplicity) attach to node $d$. \ Besides the single Higgs pair,
each vector-like pair attached to node $a$ has a symmetric counterpart attached
to node $d$. \ It therefore follows that $F_{a}=2N_{d}+4+2N_{b}$ bifundamentals
attach to node $a$. \ Dualizing node $a$ now yields the gauge group factor
$USp(2N_{b})$ when a single Higgs pair is present. \ Because the $U$ and $D$ type quarks arise as dual meson fields, the number of Higgs pairs also determines their multiplicity. In this way the multiplicity of quarks and the $USp(2)$ gauge group factor both get related to the fact that there is a single Higgs pair. \ Viewed in this light, it is interesting to note that the beta function of the $SU(2)$ gauge coupling switches sign in going from the Standard Model to the MSSM.

Turning the discussion around, note that the gauge group of the MQSM\ can only be reached in the
confining scenario when there is a single Higgs pair. \ Indeed, while required
in order for intermediate stages of the cascade to proceed, the presence of
this pair is also necessary to achieve the correct gauge group assignment at
the bottom of the cascade. \ The precise Higgs regeneration mechanism
discussed previously also requires a particular set of conditions to be met by
the topology of the MQSM. \ For example, RG\ node locking will occur provided
$g\geq3$. \ In the more general scenario where an additional $m$ vector-like
pairs attach between node $c$ and $a$, the analysis of section \ref{CANDIDATES}
demonstrates that the Higgs regeneration mechanism will proceed only when $m$ is an
even number and $g$ is an odd number.  In other words, the number of generations
must be odd in order for the cascade to proceed.  It would be interesting to establish
whether further restrictions require $g$ to be exactly three.

\section{Discussion}\label{DISCUSSION}

If the Standard Model or some supersymmetric extension thereof is realized on
a stack of D-branes, general considerations from string theory greatly
restrict possible ways in which extra matter can enter the low energy
spectrum. \ In the context of D-brane constructions probing local geometries,
it is also natural to treat the large $N$ limit of such configurations.
When the ranks of the gauge groups change so as to perturb the system away from a
conformal fixed point, such theories will undergo RG\ flow and will
typically undergo a duality cascade. \ In this paper we have considered an
explicit minimal realization of this paradigm where the connectivity of the
cascading quiver must periodically repeat to its original form such that the
MSSM\ appears at the very bottom of the duality cascade. \ We have presented
two generic ways in which a four node quiver can cascade to the
MSSM\ depending on whether at the final stages the extra node Higgses or
confines. \ In both cases, we have found a rather rigid structure which in fact
unifies in the ultraviolet. \ Indeed, at higher energies the only difference
between the Higgsing scenario and the confining scenario corresponds to the
presence of an additional vector-like pair. \ Higgsing to the MQSM represents
the most minimal cascading quiver which can descend to the MQSM. \ It is intriguing
that the left-right symmetric extension of the MSSM naturally
comes equipped with a cascading structure. \ In the context of the confining
scenario we have found an intricate sequence of dualities which in fact
require a single Higgs pair in order for the cascade to proceed. \ Moreover,
we have also presented a general mechanism whereby this Higgs pair can appear
and disappear as the cascade proceeds to lower energies. \ This has the
appealing feature that the minimal confining cascade scenario implies a
naturally light Higgs pair.

Much of the above discussion must of course be tempered by phenomenological
considerations. \ While the phenomenology of the Higgsing scenario has been
studied in the context of left-right symmetric extensions of the Standard
Model, it would be interesting to determine the precise distinction between
such partial unification schemes and cascading scenarios. \ By comparison with the Higgsing scenario, the matter content and running of the
couplings is far less flexible in the confining scenario. \ Due to the
intriguing prediction of a single Higgs pair, it would be interesting to see
whether further phenomenological considerations apply. \ As some
possibilities, we now discuss both right-handed neutrinos and CKM\ matrix
elements in the MSSM.

Whereas right-handed neutrinos naturally appear at the last stage of the
Higgsing scenario as matter transforming in the adjoint of node $b$, in the confining scenario
these fields can appear as dual mesons charged under the adjoint of node $b$. \ This would suggest that in either scenario, one natural possibility is to have the
Majorana mass of the right-handed neutrinos near energy scales set by the bottom of the cascade.

It is also of interest to determine whether the cascading scenario
naturally imposes any restrictions on the Yukawa couplings of the
MSSM. \ Such constraints translate into restrictions on the mixing
between generations of the right-handed quarks known as the CKM
matrix. \ While this is essentially a generic parameter in the
context of the Higgsing scenario, for the confining scenario, the
explicit method of Higgs regeneration discussed in section
\ref{mumu} required that all terms of the superpotential remain
invariant
under a general $%
%TCIMACRO{\U{2124} }%
%BeginExpansion
\mathbb{Z}
%EndExpansion
_{2}$ reflection symmetry of the oriented covering theory. \ In the
orientifold theory, this descends to the condition that the up and down type
Yukawa couplings are in fact identical. \ At zeroth order, this would imply
that the CKM matrix is diagonal. \ Note that no similar relation
exists for the mixing of neutrinos. This structure for the mixings is, to
leading order what one would expect based on the observed values of the CKM
and MNS mixing matrices.

Perhaps the most outstanding issue in this context is how supersymmetry is broken near the
bottom of the cascade. \ While we have proposed some general possibilities
where supersymmetry breaking communicates via the extra node, these scenarios
appear to produce a splitting between the W bosons and winos which is too small. \ It
would be interesting to see whether other minimal cascading scenarios can
naturally accomodate supersymmetry breaking.

Proceeding from the IR\ to the UV, we have found that the ranks of the gauge
group factors participating in the cascade rapidly increase. \ This leads to a
large increase in the number of field degrees of freedom in the corresponding
gauge theory. \ It is tempting to speculate that this increase in the
number of degrees of freedom has some relation to other phenomena in string
theory where a large profusion of states appear at a critical temperature,
such as in the Hagedorn phase transition.

The appearance of the duality wall appears unavoidable and may lead to potentially interesting
consequences for cosmology.  We now briefly speculate on some possibilities.  At higher and higher
temperatures an increasing number of matter fields appear.  In a dual
gravity description at least one extra spatial dimension will appear to decompactify at higher energy scales.
Said differently, as the universe cools, such a cosmology will compactify the extra dimensions.  Roughly
speaking the compactification process would likely begin at high scales where the wall starts to appear.  Near the energy scale where the final dualization occurs, the resulting theory would then be best described as an an effective four dimensional theory.  In view of the concrete gauge theory realization of this growth in the number of degrees of freedom at higher energy scales, it would be interesting to study such issues in further detail.

\section*{Acknowledgements}

We thank M. Aganagic, N. Arkani-Hamed, D. Berenstein, S. Franco, H. Murayama, S. Pinansky, M. Ro\v{c}ek, N. Seiberg and M. J. Strassler
for helpful discussions. \ We also thank the Stony
Brook physics department and the fifth Simons workshop in Mathematics and
Physics for their hospitality while this project was initiated. JJH also thanks the theory group at Caltech for hospitality
during the final stages of this work.\ The work of JJH\ and CV is supported in part by NSF grants PHY-0244821 and DMS-0244464.\ The research of JJH\ is also supported by an NSF\ Graduate Fellowship.\ The work of HV is supported in
part by NSF grant PHY-0243680. \ The research of MW is supported by a Marie Curie Fellowship.

\appendix
\section*{Appendices}
\section{Partial Classification of Cascading Quivers}\label{CovercascadeA}

In this Appendix we partially classify candidate cascade paths for
single node extensions of the covering theory of the MQSM by purely chiral
matter which can descend to a cascade in the orientifold theory. \ To keep our
discussion as general as possible we shall allow $n$ Higgs pairs but otherwise
no additional vector-like matter.\ \ Assigning gauge group ranks compatible
with the $%
%TCIMACRO{\U{2124} }%
%BeginExpansion
\mathbb{Z}
%EndExpansion
_{2}$ orientifold symmetry, we find that cancelling all non-abelian anomalies
greatly restricts the ways in which the extra node can attach to the MQSM.
\ Restricting to this class of quiver topologies, we next consider candidate
cascades. \ We find that unless the arrows of the covering theory have
multiplicity $g$, a cascade in the covering theory will not properly descend
to the orientifold theory.

When the number of generations $g$ is greater than one there is a unique way
to attach one additional node by purely chiral matter to the covering quiver
of the MQSM so that the resulting quiver admits a cascade which repeats after
finitely many steps. \ Let the signed integers $n_{DA}$, $n_{DB}$,
$n_{DB^{\prime}}$, $n_{DC}$, $n_{DC^{\prime}}$ denote the number of
bifundamentals in the covering theory charged in the fundamental
representation of node $D$, where a negative number indicates that the
orientation of the arrow is reversed. \ The $%
%TCIMACRO{\U{2124} }%
%BeginExpansion
\mathbb{Z}
%EndExpansion
_{2}$ symmetry of the quiver imposes the constraint:%
\begin{equation}
n_{DB} =-n_{DB^{\prime}}\qquad\qquad n_{DC} =-n_{DC^{\prime}}\text{.}%
\end{equation}
Due to the fact that the cascade must repeat, we may assume that the ranks of
all gauge groups factors are greater than one. \ The non-abelian anomalies of
the covering theory gauge groups vanish when:%
\begin{align}
n_{DA}  &  =0\nonumber\\
n_{DC}N_{D}-2gN_{B}+gN_{A}  &  =0\\
n_{DB}N_{D}-2gN_{B}+gN_{A}  &  =0\nonumber
\end{align}
where in the above we have used the fact that the $%
%TCIMACRO{\U{2124} }%
%BeginExpansion
\mathbb{Z}
%EndExpansion
_{2}$ symmetry of the quiver imposes the constraints $N_{B}=N_{B^{\prime}}$
and $N_{C}=N_{C^{\prime}}$. \ This implies:%
\begin{equation}
n_{DB}=n_{DC}\text{.}%
\end{equation}
We therefore conclude that for a given positive integer $\alpha=\left\vert
n_{DB}\right\vert $, there are two candidate single node extensions of the
covering quiver theory which we denote by $Q_{+}$ and $Q_{-}$ for $n_{DB}$
respectively positive or negative (see figure \ref{qminusandqplus}).

We now argue that when $g>1$, the quiver theory $Q_{-}$ does not admit a
sequence of Seiberg dualities leading to a repeating cascade structure. \ To
this end, we show that each candidate cascade only proceeds a finite number of
steps. \ Because too much bifundamental matter connects $B$ to $B^{\prime}$,
the only admissible Seiberg dual quiver theories are $AQ_{-}$, $DQ_{-}$ and
$CC^{\prime}Q_{-}$. \ First consider the quiver theories $AQ_{-}$ and $DQ_{-}$.
\ Because no bifundamentals connect $A$ to $D$ and because there exist
bifundamentals connecting the pair $BB^{\prime}$ as well as the pair
$CC^{\prime}$, the next stage of the cascade is $ADQ_{-}$. \ Due to the
presence of massless vector-like matter connecting $A$ to $B$ as well as $A$
to $B^{\prime}$, dualizing node $A$ yields $(g+n)^{2}$ meson fields
connecting $B$ to $B^{\prime}$ and $n^{2}$ meson fields connecting $B^{\prime
}$ to $B$. \ Integrating out all vector-like pairs other than those between
$A$ and $B$ and $A$ and $B^{\prime}$, the net number of bifundamentals from
$C$ to $C^{\prime}$ and the number from $B$ to $B^{\prime}$ is:%
\begin{align}
n_{CC^{\prime}}  &  =g^{2}+\alpha^{2}>0\nonumber\\[0mm]
n_{BB^{\prime}}  &  =g^{2}+\alpha^{2}+2g(n-1)>0\text{.}%
\end{align}
We therefore conclude that for $g>1$ a $%
%TCIMACRO{\U{2124} }%
%BeginExpansion
\mathbb{Z}
%EndExpansion
_{2}$ symmetric cascade cannot proceed.

Next consider the quiver theory $CC^{\prime}Q_{-}$. \ The next stage of
dualization now leads uniquely to the quiver theory $ADCC^{\prime}Q_{-}$. \ As
in the previous case, the cascade cannot proceed because the number of
bifundamentals from $C^{\prime}$ to $C$ and the number from $B$ to $B^{\prime
}$ is:%
\begin{align}
n_{C^{\prime}C}  &  =g^{2}+\alpha^{2}>0\nonumber\\[0mm]
n_{BB^{\prime}}  &  =g^{2}+\alpha^{2}+2g(n-1)>0\text{.}%
\end{align}
Hence, the quiver theory $Q_{-}$ does not admit a periodically repeating cascade.

We now show that the remaining candidate quiver theory $Q_{+}$ only admits a
repeating cascade which descends to the orientifold theory when $\alpha=g$.
\ To begin, first consider a cascade which dualizes node $A$ or $D$ of the
quiver theory $Q_{+}$. \ This produces a quiver theory with bifundamentals
between the pair $B$ and $B^{\prime}$ and the pair $C$ and $C^{\prime}$ so
that the next stage of dualization always leads to the quiver theory $ADQ_{+}%
$. \ Due to the different direction of orientation for all arrows attached to
node $D$, the number of bifundamentals from $C^{\prime}$ to $C$ and the number
from $B$ to $B^{\prime}$ is:%
\begin{align}
n_{C^{\prime}C}  &  =\alpha^{2}-g^{2}\\
n_{BB^{\prime}}  &  =g^{2}-\alpha^{2}+2g(n-1)\text{.}\nonumber
\end{align}
It follows that the cascade can only proceed provided $\alpha=mg$ for some
positive integer $m$ so that:%
\begin{align}
n_{C^{\prime}C}  &  =g^{2}(m^{2}-1)\label{CC}\\
n_{BB^{\prime}}  &  =g^{2}(1-m^{2})+2g(n-1)\text{.}\nonumber
\end{align}
When $m>1$, the resulting cascade cannot proceed unless $n_{BB^{\prime}}$
vanishes. \ In this case:%
\begin{equation}
n=\frac{g\left(  m^{2}-1\right)  }{2}+1>1\text{.} \label{HiggsPairs}%
\end{equation}

To proceed further, it is most efficient to descend to the orientifold theory
and ask whether the above assignment of Higgs pairs eliminates all tensor
matter. \ A priori, the additional node $D$ can descend to either a $SO$ or
$USp$ type factor. \ It follows from the general analysis of Seiberg duality
for $USp$ and $SO$ type factors given in section \ref{Dualizing} that dualizing the pair of
nodes $a$ and $d$ in the orientifold theory $q_{+}$ generates additional
tensor matter at node $b$. \ Letting $\varepsilon=+1$ (resp. $-1$) when node
$d$ corresponds to an $SO$ (resp. $USp$) type factor, the total amount of
tensor matter at node $b$ is:
\begin{align}
\text{Node b} &  \text{: }g\overline{S_{b}}+g\overline{A_{b}}+\frac
{\alpha(\alpha+\varepsilon)}{2}\overline{S_{b}}+\frac{\alpha(\alpha
-\varepsilon)}{2}\overline{A_{b}}+\frac{n(n+1)}{2}\overline{A_{b}}%
+\frac{n(n-1)}{2}\overline{S_{b}}\label{UNCANCONE}\\
&  +\frac{\left(  g+n+1\right)  (g+n)}{2}S_{b}+\frac{\left(  g+n-1\right)
(g+n)}{2}A_{b}\label{UNCANCTWO}\\
&  =\text{ \ }\frac{1}{2}\left(  g^{2}+g(2n-3)-\alpha\left(  \alpha
+\varepsilon\right)  \right)  S_{b}\label{SSIMP}\\
&  +\frac{1}{2}\left(  g^{2}+g(2n-1)+\alpha\left(  -\alpha+\varepsilon\right)
\right)  A_{b}\label{ASIMP}%
\end{align}
where in the above, lines (\ref{UNCANCONE}) and (\ref{UNCANCTWO}) denote all tensor matter created by dualizing nodes $a$ and $d$ and lines
(\ref{SSIMP}) and (\ref{ASIMP}) denote the total amount after all vector-like pairs have been integrated out.
If all vector-like pairs do indeed cancel so that node $b$ is free to dualize, the two coefficients multiplying $S_{b}$ and $A_{b}$ in
lines (\ref{SSIMP}) and (\ref{ASIMP}) respectively must separately vanish. \ We note in passing that
because the multiplicity of each type of tensor matter is divisible by $g$, the
corresponding node will indeed \textquotedblleft RG lock\textquotedblright\ if
the amount of uncancelled tensor matter is non-zero. \ Subtracting the two
coefficients yields the condition:%
\begin{equation}
-g-\alpha\varepsilon=0\text{.}\label{constraint}%
\end{equation}
Because $\alpha$ is a positive integer, the above equality will hold only when $\alpha=g$
and $\varepsilon=-1$ so that the gauge group factor at node $d$ is of $USp$ type.

To complete our analysis, we now establish that the existence of a repeating
cascade structure for the quiver $CC^{\prime}Q_{+}$ which properly descends to
the orientifold also requires $\alpha=g$. \ The essential point is that after
cancelling all vector-like pairs, dualizing the pair $CC^{\prime}$ only
re-orients arrows attached to this pair of nodes and otherwise leaves the
quiver theory unchanged. \ As in the quiver theory $ADQ_{+}$ and its
orientifold descendant, the tensor matter of the theory $adcq_{+}$ can only
cancel when equation (\ref{constraint}) holds.

\section{Dual Superpotentials}\label{DUALPOT}

As argued in section \ref{Superpot}, because each stage of the confining
cascade scenario must eventually pass through the quiver theory $adq_{+}$,
$q_{+}$ or a similar quiver theory with the orientations of some arrows
reversed, it is enough to analyze only these theories and their immediate
Seiberg duals. \ In this Appendix we collect the explicit expressions for the
superpotential of all possible duals of the quiver theories $adq_{+}$ and
$q_{+}$. \ We perform this analysis to explicitly demonstrate that generic
cubic and quartic terms of the electric theory will indeed lift nearly all
vector-like pairs in the dual magnetic theory. \ In the absence of
restrictions on the form of the couplings, this analysis corresponds to
classifying all possible paths in the quiver theory with a given number of
link fields. \ In the presence of symmetries of the string theory or field
theory, there will be further restrictions on the form of the couplings. \ For
this reason, it is also important to establish that a given symmetry does not
forbid (resp. does forbid) a given vector-like pair from developing a mass.
\ Indeed, in the minimal realization of the cascading scenario which
terminates via confinement, we must establish that the massless Higgs pair
does not develop a $\mu$ term at some further stage of the cascade.  While we shall present this analysis for quiver theories where the superpotential is generically not invariant under matter parity, by suitably restricting the form of the coupling constant matrices, a similar discussion holds in this case as well.

We now classify all terms of the superpotential of degree four or less in the
quiver theory $adq_{+}$. \ By inspection, this quiver is identical to figure
\ref{fourtypes}$d)$. \ Further, the absence of closed triangles in the theory
implies that all gauge invariant operators are of even degree in the quiver
fields. \ Labeling the various chiral superfields by the variable $X$, we
shall distinguish which gauge group a given field is charged under as well as
whether it is a fundamental or anti-fundamental index by the subscript of the
field. To avoid cluttering the notation too much, we shall suppress the presence
of the appropriate $\epsilon$ tensors necessary for contracting all $USp$ indices.
 \ While this notation admittedly obscures the connection of the $X$
fields to the content of the MQSM, it will prove convenient in discussing the
magnetic dual superpotentials. \ The superpotential for the quiver theory
$adq_{+}$ is:%
\begin{align}
W_{adq_{+}}  &  =\mu_{I}X_{ab}X_{\overline{b}a}^{I}+\alpha_{ijkl}X_{dc}%
^{i}X_{\overline{c}a}^{j}X_{a\overline{c^{\prime}}}^{k}X_{c^{\prime}d}%
^{l}\eoll & +\beta_{ijk}^{I}X_{dc}^{i}X_{\overline{c}a}^{j}X_{a\overline{b}%
}^{I}X_{bd}^{k}+\gamma_{ij}^{IJ}X_{db}^{i}X_{\overline{b}a}^{I}X_{a\overline
{b^{\prime}}}^{J}X_{b^{\prime}d}^{j}\eoll  &  +\mu_{IJ}\left(  X_{ab}%
X_{\overline{b}a}^{I}\right)  \left(  X_{a^{\prime}b^{\prime}}X_{\overline
{b^{\prime}}a^{\prime}}^{J}\right) \\
&  +\kappa_{IJ}X_{ab}X_{\overline{b}a^{\prime}}^{I}X_{a^{\prime}b^{\prime}%
}X_{\overline{b^{\prime}}a}^{J}+\widehat{\kappa}_{IJ}X_{ab}X_{\overline
{b}a^{\prime}}^{I}X_{a^{\prime}\overline{b^{\prime}}}^{J}X_{b^{\prime}%
a}+\widehat{\mathcal{O}}_{6}  & \nonumber
\end{align}
where in the above, the indices $i,j,k,l$ run from $1,...,g$, $I$ and $J$ run
from $I=0,1,...,g$ and $\widehat{\mathcal{O}}_{6}$ denotes all degree six and higher
gauge invariant combinations of fields.

Dualizing the pair of nodes $a$ and $d$ in $adq_{+}$ produces the magnetic
dual superpotential:%
\begin{align}
W_{ad(adq_{+})}  &  =W_{adq_{+}}|_{M={\mu}^{-1}XY}+W_{meson}=\mu_{a}\mu
_{I}A_{\overline{b}b}^{I}\eoll & +\mu_{a}\mu_{d}\alpha_{ijkl}\left(
A_{[\overline{cc^{\prime}}]}^{(jk)}+A_{(\overline{cc^{\prime}})}%
^{[jk]}\right)  \left(  D_{[cc^{\prime}]}^{(jk)}+D_{(cc^{\prime})}%
^{[jk]}\right)  \eoll  &  +\mu_{a}\mu_{d}\beta_{ijk}^{I}A_{\overline{cb}}%
^{jI}D_{bc}^{ik}+\mu_{a}\mu_{d}\gamma_{ij}^{IJ}\left(  A_{[\overline
{bb^{\prime}}]}^{(IJ)}+A_{(\overline{bb^{\prime}})}^{[IJ]}\right)  \left(
D_{[bb^{\prime}]}^{(jk)}+D_{(bb^{\prime})}^{[jk]}\right)  \eoll & +\mu_{a}%
^{2}\mu_{IJ}\left(  A_{\overline{b}b}^{I}\right)  \left(  A_{\overline
{b^{\prime}}b^{\prime}}^{J}\right)  +\mu_{a}^{2}\kappa_{IJ}A_{\overline
{b}b^{\prime}}^{I}A_{\overline{b^{\prime}}b}^{J}+\mu_{a}^{2}\widehat{\kappa
}_{IJ}A_{[\overline{bb^{\prime}}]}^{(IJ)}A_{[b^{\prime}b]}\\
&  +\left(  D_{[cc^{\prime}]}^{(ij)}+D_{(cc^{\prime})}^{[ij]}\right)
X_{\overline{c^{\prime}}d}^{j}X_{d\overline{c}}^{i}+\left(  D_{[bb^{\prime}%
]}^{(ij)}+D_{(bb^{\prime})}^{[ij]}\right)  X_{\overline{b^{\prime}}d}%
^{j}X_{d\overline{b}}^{i}+D_{bc}^{ij}X_{\overline{c}d}^{j}X_{d\overline{b}%
}^{i}\eoll & +\left(  A_{[\overline{cc^{\prime}}]}^{(jk)}+A_{(\overline
{cc^{\prime}})}^{[jk]}\right)  X_{c^{\prime}a}^{j}X_{ac}^{i}+\left(
A_{[\overline{bb^{\prime}}]}^{(IJ)}+A_{(\overline{bb^{\prime}})}%
^{[IJ]}\right)  X_{b^{\prime}a}^{J}X_{ab}^{I}\eoll  &  +A_{[bb^{\prime}%
]}X_{\overline{b^{\prime}}d}X_{d\overline{b}}+A_{b\overline{b^{\prime}}}%
^{I}X_{b^{\prime}a}^{I}X_{a\overline{b}}+A_{b\overline{c}}^{i}X_{ca}%
^{i}X_{a\overline{b}}+A_{\overline{bc}}^{iI}X_{ca}^{i}X_{ab}^{I}%
+\widehat{\mathcal{O}}_{4}\text{.}  & \nonumber
\end{align}
\newline In the above, the fields $A$ and $D$ denote dual meson fields
generated by dualizing the nodes $a$ and $d$, respectively. \ The
symmetrization and anti-symmetrization of a pair of indices indicates the
number of such meson fields. \ Assuming that the term $\mu_{I}$ has been set
to zero, we note that for generic values of the couplings, no linear terms are
present in the superpotential. \ By inspection of the degree two terms of
$W_{ad(adq_{+})}$, we note that all of the $A$ tensor matter at node $c$ pairs
with $D$ tensor matter at node $c$ via the $\alpha$ couplings. \ On the other
hand, the quadratic term for the $A$ tensor matter at node $b$ is schematically
of the form:%
\begin{align*}
W_{ad(adq_{+})}  &  \supset\left[  D^{(ij)},A\right]  _{1\times g(g+1)/2+1}%
\left[  M^{symm}\right]  _{g(g+1)/2+1\times(g+1)(g+2)/2}\left[  A^{(IJ)}%
\right]  _{(g+1)(g+2)/2\times1}\eol & +\left[  D^{[ij]}\right]  _{1\times
g(g-1)/2}\left[  M^{asymm}\right]  _{g(g-1)/2\times(g+1)g/2}\left[
A^{[IJ]}\right]  _{(g+1)g/2\times1}%
\end{align*}
where the $M$'s denote a generic matrix and the subscripts for each square
bracket indicates the size of the corresponding matrix. \ We therefore
conclude that for generic couplings, precisely $g$ two index symmetric and
anti-symmetric tensor matter fields at node $b$ remain massless. \ Further, we
note that the remaining $A$ fields in the adjoint representation at node $b$
develop a mass due to the couplings $\mu_{IJ}$ and $\kappa_{IJ}$. \ Finally,
it follows from the general form of the coupling $\beta$ that only $g^{2}$ of
the $A_{\overline{cb}}^{jI}$ develop a mass by pairing with $D$ fields. \ This
again leaves precisely $g$ massless fields. \ Note, however, that when the
linear term proportional to $\mu_{I}A_{\overline{b}b}^{I}$ is absent, it is
consistent to set the vevs of all fields to zero. \ In this case, there are no
quadratic terms of the form $X_{a\overline{b}}X_{ab}^{I}$. \ In other words,
once set to zero, the $\mu$-term for the Higgs pair remains zero.

Dualizing node $b$ produces the magnetic dual superpotential:%
\begin{align}
W_{b(adq_{+})}  &  =\mu_{b}\mu_{I}B_{aa}^{I}+\mu_{b}^{2}\gamma_{ij}^{IJ}%
B_{da}^{iI}B_{ad}^{Jj}+\mu_{b}^{2}\mu_{IJ}\left(  B_{aa}^{I}\right)  \left(
B_{a^{\prime}a^{\prime}}^{J}\right)  \eoll & +\mu_{b}^{2}\kappa_{IJ}%
B_{aa^{\prime}}^{I}B_{a^{\prime}a}^{J}+\mu_{b}^{2}\widehat{\kappa}%
_{IJ}B_{aa^{\prime}}^{I}B_{a^{\prime}a}^{J}\eoll  &  +\mu_{b}\beta_{ijk}%
^{I}X_{dc}^{i}X_{\overline{c}a}^{j}B_{ad}^{kI}+B_{ad}^{iI}X_{d\overline{b}%
}^{i}X_{ba}^{I}+B_{aa^{\prime}}^{I}X_{a^{\prime}b}^{I}X_{\overline{b}a}\\
&  +B_{aa^{\prime}}^{I}X_{a^{\prime}\overline{b}}X_{ba}^{I}+\alpha
_{ijkl}X_{dc}^{i}X_{\overline{c}a}^{j}X_{a\overline{c^{\prime}}}%
^{k}X_{c^{\prime}d}^{l}+\widehat{\mathcal{O}}_{4}\text{.}  & \nonumber
\end{align}
We note that the meson fields $B$ between nodes $a$ and $d$ all develop a mass
due to the $\gamma$ couplings and the $B$ adjoint fields at node $a$ all
develop a mass due to $\mu,\kappa$ and $\widehat{\kappa}$. \ As before, the
dual theory does not contain a generalized $\mu$-term.

A similar analysis establishes that dualizing node $c$ produces the
superpotential:%
\begin{align}
W_{c(adq_{+})}  &  =\mu_{I}X_{ab}X_{\overline{b}a}^{I}+\mu_{c}^{2}%
\alpha_{ijkl}C_{da}^{ij}C_{ad}^{kl}+C_{ad}^{ij}X_{d\overline{c}}^{i}X_{ca}%
^{j}\eoll & +\mu_{c}\beta_{ijk}^{I}C_{da}^{ij}X_{a\overline{b}}^{I}X_{bd}%
^{k}+\gamma_{ij}^{IJ}X_{db}^{i}X_{\overline{b}a}^{I}X_{a\overline{b^{\prime}}%
}^{J}X_{b^{\prime}d}^{j}\eoll  &  +\mu_{IJ}\left(  X_{ab}X_{\overline{b}a}%
^{I}\right)  \left(  X_{a^{\prime}b^{\prime}}X_{\overline{b^{\prime}}%
a^{\prime}}^{J}\right) \\
&  +\kappa_{IJ}X_{ab}X_{\overline{b}a^{\prime}}^{I}X_{a^{\prime}b^{\prime}%
}X_{\overline{b^{\prime}}a}^{J}+\widehat{\kappa}_{IJ}X_{ab}X_{\overline
{b}a^{\prime}}^{I}X_{a^{\prime}\overline{b^{\prime}}}^{J}X_{b^{\prime}%
a}+\widehat{\mathcal{O}}_{4}  & \nonumber
\end{align}
so that all of the meson fields $C$ between nodes $a$ and $d$ develop a mass
due to the $\alpha$ couplings.

To summarize, the above analysis establishes that a $\mu$ term will never
appear by dualizing a node in the theory $adq_{+}$. \ We now show that whereas
dualizing the pair $ad$ in $q_{+}$ also does not produce a $\mu$ term, such a
term does appear upon dualizing node $c$.

We begin by classifying all terms of the superpotential of degree four or less
in the quiver theory $q_{+}$: \
\begin{align}
W_{q_{+}} &
  =\mu_{I}X_{a\overline{b}}X_{ba}^{I}
\eoll &
  +\lambda_{ij}%
  X_{ca}^{i}X_{a\overline{b}}X_{b\overline{c}}^{j}+\widehat{\lambda}_{ij}%
  ^{I}X_{ca}^{i}X_{ab}^{I}X_{\overline{bc}}^{j}+\sigma_{i}^{IJ}X_{ab}%
  ^{I}X_{(\overline{bb^{\prime}})}^{i}X_{b^{\prime}a}^{J}+\rho_{i}^{IJ}%
  X_{ab}^{I}X_{[\overline{bb^{\prime}}]}^{i}X_{b^{\prime}a}^{J}
\eoll &
  +\alpha_{ijkl}X_{d\overline{c}}^{i}X_{ca}^{j}X_{ac^{\prime}}^{k}%
  X_{\overline{c^{\prime}}d}^{l}+\beta_{ijk}^{I}X_{d\overline{c}}^{i}X_{ca}%
  ^{j}X_{ab}^{I}X_{\overline{b}d}^{k}+\gamma_{ij}^{IJ}X_{d\overline{b}}%
  ^{i}X_{ba}^{I}X_{ab^{\prime}}^{J}X_{\overline{b^{\prime}}d}^{j}\eoll &
  +\tau_{ijk}^{I}X_{ac}^{i}X_{\overline{c}b}^{j}X_{(\overline{bb^{\prime}})}%
  ^{k}X_{b^{\prime}a}^{I}+\omega_{ijk}^{I}X_{ac}^{i}X_{\overline{c}b}%
  ^{j}X_{[\overline{bb^{\prime}}]}^{k}X_{b^{\prime}a}^{I}
\eoll &  +\varphi
  _{ijkl}X_{ac}^{i}X_{\overline{cb}}^{j}X_{b\overline{c^{\prime}}}%
  ^{k}X_{c^{\prime}a}^{l}
\label{Wqplus}\\
  &  +\mu_{IJ}\left(  X_{ab}X_{\overline{b}a}^{I}\right)  \left(  X_{a^{\prime
  }b^{\prime}}X_{\overline{b^{\prime}}a^{\prime}}^{J}\right)  +\kappa_{IJ}%
  X_{ab}X_{\overline{b}a^{\prime}}^{I}X_{a^{\prime}b^{\prime}}X_{\overline
  {b^{\prime}}a}^{J}
\eoll &   +\widehat{\kappa}_{IJ}X_{ab}X_{\overline
  {b}a^{\prime}}^{I}X_{a^{\prime}\overline{b^{\prime}}}^{J}X_{b^{\prime}%
  a}+\widehat{\mathcal{O}}_{5}\text{.} \nonumber
\end{align}
Note in particular that the chiral fields attached to node $d$ first
contribute at quartic order. \ Due to the fact that node $b$ contains too much
tensor matter, any candidate cascade must proceed by dualizing the pair of
nodes $ad$ or the node $c$. \ Dualizing $ad$ yields the superpotential:%
\begin{align}
W_{ad(q_{+})} &  =\mu_{a}\mu_{I}A_{b\overline{b}}^{I}+\mu_{a}\lambda
_{ij}A_{c\overline{b}}^{i}X_{b\overline{c}}^{j}+\mu_{a}\widehat{\lambda}%
_{ij}^{I}A_{cb}^{iI}X_{\overline{bc}}^{j}\eoll & +\mu_{a}\sigma_{i}%
^{IJ}X_{(\overline{bb^{\prime}})}^{i}A_{(b^{\prime}b)}^{[JI]}+\mu_{a}\rho
_{i}^{IJ}X_{[\overline{bb^{\prime}}]}^{i}A_{[b^{\prime}b]}^{(JI)}\eoll &
+\mu_{a}^{2}\mu_{IJ}\left(  A_{b\overline{b}}^{I}\right)  \left(
A_{b^{\prime}\overline{b^{\prime}}}^{J}\right)  +\mu_{a}^{2}\kappa
_{IJ}A_{\overline{b}b^{\prime}}^{I}A_{\overline{b^{\prime}}b}^{J}+\mu_{a}%
^{2}\widehat{\kappa}_{IJ}A_{[\overline{bb^{\prime}}]}^{(IJ)}A_{[b^{\prime}%
b]}\eoll & +\mu_{a}\mu_{d}\alpha_{ijkl}\left(  A_{[cc^{\prime}]}%
^{(jk)}+A_{(cc^{\prime})}^{[jk]}\right)  \left(  D_{[\overline{c^{\prime}c}%
]}^{(jk)}+D_{(\overline{c^{\prime}c})}^{[jk]}\right)  +\mu_{a}\mu_{d}%
\beta_{ijk}^{I}A_{cb}^{jI}D_{\overline{bc}}^{ki}\eoll &  +\mu_{a}\mu_{d}%
\gamma_{ij}^{IJ}\left(  A_{[bb^{\prime}]}^{(IJ)}+A_{(bb^{\prime})}%
^{[IJ]}\right)  \left(  D_{[\overline{b^{\prime}b}]}^{(ji)}+D_{(\overline
{b^{\prime}b})}^{[ji]}\right)   &  & \label{adtwoW}\\
&  +\mu_{a}\tau_{ijk}^{I}X_{\overline{c}b}^{j}X_{(\overline{bb^{\prime}})}%
^{k}A_{b^{\prime}c}^{Ii}+\mu_{a}\omega_{ijk}^{I}X_{\overline{c}b}%
^{j}X_{[\overline{bb^{\prime}}]}^{k}A_{b^{\prime}c}^{Ii}\eoll & +\mu
_{a}\varphi_{ijkl}X_{\overline{cb}}^{j}X_{b\overline{c^{\prime}}}^{k}\left(
A_{[c^{\prime}c]}^{(li)}+A_{(c^{\prime}c)}^{[li]}\right)  \eoll &
+A_{b\overline{b^{\prime}}}^{I}X_{b^{\prime}a}^{I}X_{a\overline{b}}+\left(
D_{[\overline{bb^{\prime}}]}^{(ij)}+D_{(\overline{bb^{\prime}})}%
^{[ij]}\right)  X_{b^{\prime}d}^{i}X_{db}^{j}\eoll & +\left(  A_{[bb^{\prime
}]}^{(IJ)}+A_{(bb^{\prime})}^{[IJ]}\right)  X_{\overline{b^{\prime}}a}%
^{I}X_{a\overline{b}}^{J}+A_{[\overline{bb^{\prime}}]}X_{b^{\prime}a}%
X_{ab}\eoll &  +\left(  D_{[\overline{cc^{\prime}}]}^{(ij)}+D_{(\overline
{cc^{\prime}})}^{[ij]}\right)  X_{c^{\prime}d}^{i}X_{dc}^{j}+\left(
A_{[cc^{\prime}]}^{(ij)}+A_{(cc^{\prime})}^{[ij]}\right)  X_{\overline
{c^{\prime}}a}^{I}X_{a\overline{c}}^{J}\eoll &   +D_{\overline{bc}}%
^{ij}X_{cd}^{j}X_{db}^{i}+A_{c\overline{b}}^{i}X_{ba}X_{a\overline{c}}%
^{i}+A_{cb}^{iI}X_{\overline{b}a}^{I}X_{a\overline{c}}^{i}+\widehat
{\mathcal{O}}_{3}\text{.}\nonumber
\end{align}
By inspection, when $\mu_{I}$ vanishes, all vector-like pairs other than the
Higgs pair develop a mass. \ As before, we therefore conclude that
once the generalized $\mu$-term has been set to zero, for generic values of
the couplings it will not regenerate.

The final way in which the quiver theory $q_{+}$ can dualize is via node $c$.
\ In fact, it can be shown by analyzing all possible candidate paths that each
candidate cascade must eventually dualize node $c$ in either the quiver theory
$q_{+}$ or its conjugate with all arrow directions reversed. \ Dualizing node $c$ yields the superpotential:%
\begin{align}
W_{c(q_{+})} &  =\mu_{I}X_{a\overline{b}}X_{ba}^{I}+\mu_{c}\lambda
_{ij}X_{a\overline{b}}C_{ba}^{ji}+\mu_{c}\widehat{\lambda}_{ij}^{I}X_{ab}%
^{I}C_{\overline{b}a}^{ji}\eoll & +\mu_{c}^{2}\varphi_{ijkl}C_{a\overline{b}%
}^{ij}C_{ba}^{kl}+\mu_{c}^{2}\alpha_{ijkl}C_{da}^{ij}C_{ad}^{kl}\eoll &
+\sigma_{i}^{IJ}X_{ab}^{I}X_{(\overline{bb^{\prime}})}^{i}X_{b^{\prime}a}%
^{J}+\rho_{i}^{IJ}X_{ab}^{I}X_{[\overline{bb^{\prime}}]}^{i}X_{b^{\prime}%
a}^{J}+\mu_{c}\beta_{ijk}^{I}C_{da}^{ij}X_{ab}^{I}X_{\overline{b}d}^{k}+\eoll
&    +\mu_{c}\tau_{ijk}^{I}C_{ab}^{ij}X_{(\overline{bb^{\prime}})}%
^{k}X_{b^{\prime}a}^{I}+\mu_{c}\omega_{ijk}^{I}C_{ab}^{ij}X_{[\overline
{bb^{\prime}}]}^{k}X_{b^{\prime}a}^{I}\label{cqfive}\\
&  +C_{ad}^{ij}X_{dc}^{i}X_{\overline{c}a}^{j}+C_{a\overline{b}}^{ij}%
X_{bc}^{i}X_{\overline{c}a}^{j}+C_{ab}^{ij}X_{\overline{b}c}^{i}%
X_{\overline{c}a}^{j}\eoll & +\gamma_{ij}^{IJ}X_{d\overline{b}}^{i}X_{ba}%
^{I}X_{ab^{\prime}}^{J}X_{\overline{b^{\prime}}d}^{j}+\mu_{IJ}\left(
X_{ab}X_{\overline{b}a}^{I}\right)  \left(  X_{a^{\prime}b^{\prime}%
}X_{\overline{b^{\prime}}a^{\prime}}^{J}\right)  \eoll &  +\kappa_{IJ}%
X_{ab}X_{\overline{b}a^{\prime}}^{I}X_{a^{\prime}b^{\prime}}X_{\overline
{b^{\prime}}a}^{J}+\widehat{\kappa}_{IJ}X_{ab}X_{\overline{b}a^{\prime}}%
^{I}X_{a^{\prime}\overline{b^{\prime}}}^{J}X_{b^{\prime}a}+\widehat
{\mathcal{O}}_{3}\text{.} &  & \nonumber
\end{align}
Just as generic values of the $\alpha$ couplings induce mass term for all of
the $C$ mesons between $a$ and $d$, generic values of the couplings $\mu$,
$\lambda$, $\widehat{\lambda}$, and $\varphi$ will also give a mass to
\textit{all} vector-like pairs between nodes $a$ and $b$. \ Indeed, the
corresponding quadratic term schematically takes the form:%
\begin{equation}
W_{c(q_{+})}\supset\left[
\begin{array}
[c]{cc}%
X_{\overline{b}a} & C_{\overline{b}a}^{kl}%
\end{array}
\right]  _{1\times\left(  g^{2}+1\right)  }\left[
\begin{array}
[c]{cc}%
\mu_{I} & \mu_{c}\lambda_{ij}\\
\mu_{c}\widehat{\lambda}_{kl}^{I} & \mu_{c}^{2}\varphi_{klij}%
\end{array}
\right]  _{(g^{2}+1)\times\left(  g^{2}+g+1\right)  }\left[
\begin{array}
[c]{c}%
X_{ab}^{I}\\
C_{ab}^{ij}%
\end{array}
\right]  _{\left(  g^{2}+g+1\right)  \times1}\label{problematicmassterm}%
\end{equation}
\newline so that for generic values of the couplings, precisely $g$ chiral
fields charged in the fundamental of $b$ and $a$ will remain.

\section{Conformal Windows}\label{Windows}

In this Appendix we demonstrate that in both the Higgsing and confining scenario cascades, when
$g\geq3$, there does not exist an assignment of ranks in the resulting quiver
gauge theory such that the theory flows to a conformal fixed point. \ To
establish this, we show that there is no overlap in the conformal window for
each quiver node. \ On the other hand, we shall also demonstrate that when $g\leq2$ and the number of
vector-like pairs has been suitably adjusted, there exist intermediate stages of the
cascade which admit a common conformal window for each quiver node. \ Our aim
in this regard is not to present an exhaustive list of variants, but to simply
indicate some possibilities for future investigation.

To establish the absence of a conformal window when $g\geq3$, recall that the
conformal window for $SU(N)$ SQCD with $F$ flavors is:%
\begin{equation}
3N>F>\frac{3}{2}N\text{.}%
\end{equation}
Similarly, the conformal window for $USp(2N)$ SQCD\ with $F$ quarks
transforming in the fundamental is:%
\begin{equation}
3(2N+2)>F>3(N+1)\text{.}%
\end{equation}
It follows from the arguments below equation (\ref{betaone}) that any quiver node with enough
matter transforming in two index representations will fall outside the
conformal window. \ The classification of figure \ref{fourtypes}
implies that when $g\geq3$, the only candidate quivers which do not contain
any tensor matter are of type d). \ This indicates the absence of a conformal
fixed point. \ Indeed, suppose to the contrary that at some stage of either
the Higgsing or confining scenario cascade one of the intermediate quiver
topologies admitted a conformal fixed point for appropriate rank assignments.
\ Seiberg dualizing the resulting theory would produce a dual description of
the same conformal fixed point. \ In general, one of the dual theories would
contain far too much tensor matter for the associated quiver node to lie
within a conformal window.

A possible caveat to the above argument is that all quiver nodes in question
may lie at strong coupling so that the meaning of a strict Seiberg duality
where the flavor groups are only weakly gauged may not apply. \ While much of
the analysis of the paper assumes that it is still possible to dualize the
resulting quiver theory, to fully establish the absence of a common conformal
window, we now give a direct demonstration that no conformal window exists for
quiver topologies where no tensor matter is present.

As explained in section \ref{Unification}, when no tensor matter is present at any node, the
only difference between the confining and Higgsing cascade scenarios is the
absence or presence\ of an additional vector-like pair between nodes $d$ and $b$.
\ To keep our discussion as general as possible, let $n$ denote the number of
Higgs pairs between nodes $a$ and $b$ and $n^{\prime}$ the number between $d$ and $b$.
\ Because variants of the confining scenario also require additional
vector-like pairs in order to properly accelerate the running of couplings, we
shall also allow additional vector-like pairs of flavors which we denote by
$F_{i}$ where $i$ indicates the node in question. \ All non-abelian anomalies
vanish when the ranks of the two $USp$ factors are identical. \ The four
conformal windows for the quiver theory with gauge group $USp(2N_{d}%
)\times U(N_{b})\times U(N_{c})\times USp(2N_{d})$ are then:%
\begin{align}
\text{Node a}  &  \text{: }3(2N_{d}+2)>(g+2n)N_{b}+gN_{c}+F_{a}>3(N_{d}%
+1)\label{confone}\\
\text{Node b}  &  \text{: }3N_{b}>(2g+2n+2n^{\prime})N_{d}+F_{b}>\frac{3}%
{2}N_{b}\label{conftwo}\\
\text{Node c}  &  \text{: }3N_{c}>2gN_{d}+F_{c}>\frac{3}{2}N_{c}%
\label{confthree}\\
\text{Node d}  &  \text{: }3(2N_{d}+2)>(g+2n^{\prime})N_{b}+gN_{c}%
+F_{d}>3(N_{d}+1)\text{.} \label{conffour}%
\end{align}
Adding the leftmost inequalities of lines (\ref{conftwo}) and (\ref{confthree}%
) yields the condition:%
\begin{equation}
3(N_{b}+N_{c})>(4g+2n+2n^{\prime})N_{d}+F_{b}+F_{c}\geq(4g+2n+2n^{\prime
})N_{d}\text{.}%
\end{equation}
On the other hand, adding the leftmost inequalities of lines (\ref{confone})
and (\ref{conffour}) yields:%
\begin{equation}
6(2N_{d}+2)>(2g+2n+2n^{\prime})N_{b}+2gN_{c}+F_{a}+F_{d}\geq2g(N_{b}%
+N_{c})\text{.}%
\end{equation}
Combining the above inequalities yields the further condition:%
\begin{equation}
6(2N_{d}+2)>2g(N_{b}+N_{c})>\frac{2g}{3}(4g+2n+2n^{\prime})N_{d}%
\end{equation}
or:%
\begin{equation}
18>(4g^{2}+2gn+2gn^{\prime}-18)N_{d}\text{.}%
\end{equation}
For general $N_{d}$ and $n,n^{\prime}\geq0$, the above inequality is never
satisfied when $g\geq3$. \ This establishes the absence of a common conformal
window for all quiver nodes. \ A similar argument establishes the absence of a
common conformal window when $g\geq2$ and at least one of either $n$ or
$n^{\prime}$ is one.

When $g\leq2$, integrating out or introducing additional vector-like pairs in
an intermediate quiver theory of a cascade can yield a common conformal window
for all quiver nodes. \ Although our aim is not to present an exhaustive
classification of such rank assignments, we shall give two representative
examples of this possibility. \ Consider again the intermediate quiver theory
with no tensor matter. \ When $g=2$ and $n=n^{\prime}=F_{i}=0$ for all $i,$
the corresponding quiver theory admits a common conformal window for all
quiver nodes. \ For example, the rank assignments: $N_{b}=N_{c}=1.4N_{d}$
falls within the common conformal window defined by lines (\ref{confone}%
)-(\ref{conffour}). \ As a final example, consider the covering quiver theory
$AQ_{+}$ with $g=1$. \ Deleting all vector-like pairs, the resulting quiver
theory with all ranks equal falls within the common conformal window. \ These
examples suggest that even when $g\geq3$, a description in terms of a bound
state configuration of D-branes exists.

\pagebreak

\bibliographystyle{ssg}
\bibliography{smcascade}

\end{document}